\pdfoutput=1
\newcommand*{\ATLASLATEXPATH}{}
\documentclass[cernpreprint, PAPER, atlasdraft=false, texlive=2016, UKenglish]{\ATLASLATEXPATH atlasdoc}
 
\usepackage[block=none]{\ATLASLATEXPATH atlaspackage}
\usepackage{\ATLASLATEXPATH atlasbiblatex}
 
\usepackage{\ATLASLATEXPATH atlasphysics}
 
\graphicspath{{logos/}{figures/}}
 
\usepackage{ANA-TOPQ-2017-07-PAPER-defs}
 
\AtlasTitle{Search for top-quark decays $t \to Hq$ with 36 fb$^{-1}$ of $pp$ collision data at $\sqrt{s}=13~\tev$ with the ATLAS detector}
 
\AtlasAbstract{
A search for flavour-changing neutral current decays of a top quark into an up-type quark ($q=u, c$) and the
Standard Model Higgs boson, $t\to Hq$, is presented.
The search is based on a dataset of $pp$ collisions at $\sqrt{s}=13~\tev$ recorded in 2015 and 2016 with the ATLAS detector at the
CERN Large Hadron Collider and corresponding to an integrated luminosity of 36.1 fb$^{-1}$.
Two complementary analyses are performed to search for top-quark pair events in which one top quark decays into $Wb$ and the other top quark decays into $Hq$,
and target the $H \to b\bar{b}$ and $H \to \tau^+\tau^-$ decay modes, respectively.
The high multiplicity of $b$-quark jets, or the presence of hadronically decaying $\tau$-leptons, is exploited in the two analyses respectively.
Multivariate techniques are used to separate the signal from the background, which is dominated by top-quark pair production.
No significant excess of events above the background expectation is found, and 95\% CL upper limits on the $t\to Hq$ branching ratios are derived.
The combination of these searches with ATLAS searches in diphoton and multilepton final states
yields observed (expected) 95\% CL upper limits on the $t\to Hc$ and $t\to Hu$ branching ratios of $1.1 \times 10^{-3}$ ($8.3 \times 10^{-4}$)
and $1.2 \times 10^{-3}$ ($8.3 \times 10^{-4}$), respectively.
The corresponding combined observed (expected) upper limits on the $|\lambda_{tcH}|$ and $|\lambda_{tuH}|$ couplings are 0.064 (0.055) and 0.066 (0.055), respectively.
}
 
\author{The ATLAS Collaboration}
 
\AtlasRefCode{TOPQ-2017-07}

\AtlasJournal{JHEP}
\AtlasJournalRef{JHEP 05 (2019) 123}
\AtlasDOI{10.1007/JHEP05(2019)123}
 
\PreprintIdNumber{CERN-EP-2018-295}

\AtlasCoverSupportingNote{Search for $\ttbar \to WbHq$, $H \to b\bar{b}$}{https://cds.cern.ch/record/2257631}
\AtlasCoverSupportingNote{Search for $\ttbar \to WbHq$, $H \to \tau^+\tau^-$}{https://cds.cern.ch/record/2273683}
\AtlasCoverSupportingNote{Combination of $\ttbar \to WbHq$ searches}{https://cds.cern.ch/record/2312520/}
 
\AtlasCoverCommentsDeadline{4 November 2018}
 
\AtlasCoverAnalysisTeam{
$H \to b\bar{b}$: Trisha Farooque, Davide Gerbaudo, Aurelio Juste, Nicola Orlando,  \\ Cheng Peng,  Laura Pereira,
Yulia Rodina, Yanjun Tu,  Lo\"ic Val\'ery, \\ Tal Van Daalen, Jia-Shian Wang, Daiki Yamaguchi; \\
$H \to \tau^+\tau^-$: Xin Chen, Antonio De Maria, Boyang Li, Ligang Xia, Gang Zhang; \\
Combination: Peter Onyisi, Harish Potti}
 
\AtlasCoverEdBoardMember{Ketevi Assamagan~(chair), Garabed Halladjian, Fabrice Hubaut, Michele Weber~(chair)}

\AtlasCoverEgroupEditors{atlas-TOPQ-2017-07-editors@cern.ch}
 
\AtlasCoverEgroupEdBoard{atlas-TOPQ-2017-07-editorial-board@cern.ch}

\hypersetup{pdftitle={ATLAS document},pdfauthor={The ATLAS Collaboration}}
 
\begin{document}
 
\maketitle

 
\section{Introduction}
\label{sec:intro}
 
Following the observation of the Higgs boson by the ATLAS and CMS experiments~\cite{Aad:2012tfa,Chatrchyan:2012ufa} at
the Large Hadron Collider (LHC), a comprehensive programme of measurements of its properties is underway.
An interesting possibility is the presence of flavour-changing neutral-current (FCNC) interactions
between the Higgs boson, the top quark, and a $u$- or $c$-quark, $tqH$ ($q=u,c$). Since the Higgs boson is lighter than the top quark~\cite{Aad:2015zhl},
such interactions would manifest themselves as FCNC top-quark decays~\cite{Agashe:2013hma}, $t\to H q$.
In the Standard Model (SM), such decays are suppressed relative to the dominant $t\to Wb$ decay mode, since $tqH$
interactions are forbidden at the tree level and suppressed even at higher orders in the perturbative expansion due to the
Glashow--Iliopoulos--Maiani (GIM) mechanism~\cite{Glashow:1970gm}.  As a result, the SM predictions for the $t \to Hq$ branching
ratios ($\BR$) are exceedingly small, $\BR(t\to Hu) \sim 10^{-17} $ and $\BR(t\to Hc) \sim 10^{-15}$~\cite{Eilam:1990zc,Mele:1998ag,AguilarSaavedra:2004wm,Zhang:2013xya}, making them undetectable in the foreseeable future.
In contrast, large enhancements of these branching ratios are possible in some scenarios beyond the SM.
Examples include quark-singlet models~\cite{AguilarSaavedra:2002kr}, two-Higgs-doublet models (2HDM) of type I, with explicit flavour conservation,
and of type II, such as the minimal supersymmetric SM (MSSM)~\cite{Bejar:2000ub, Guasch:1999jp,Cao:2007dk,Cao:2014udj}, supersymmetric models
with R-parity violation~\cite{Eilam:2001dh}, composite Higgs models with partial compositeness~\cite{Azatov:2014lha},
or warped extra dimensions models with SM fermions in the bulk~\cite{Azatov:2009na}.
In these scenarios, branching ratios can be as high as $\BR(t\to Hq) \sim 10^{-5}$.
An even larger branching ratio of  $\BR(t\to Hc) \sim 10^{-3}$ can be reached in 2HDM without explicit flavour conservation (type III),
since a tree-level FCNC coupling is not forbidden by any symmetry~\cite{Cheng:1987rs,Baum:2008qm,Chen:2013qta,Chiang:2015cba,Crivellin:2015hha,Botella:2015hoa, Gori:2017tvg,Chiang:2017fjr}.
While other FCNC top couplings ($tq\gamma$, $tqZ$, $tqg$) are also enhanced in these scenarios beyond the SM,
the largest enhancements are typically found for the $tqH$ couplings, and in particular the $tcH$ coupling~\cite{Agashe:2013hma}.
 
Searches for $t \to Hq$ decays have been performed by the ATLAS and CMS collaborations, taking advantage of the large samples
of top-quark pair ($\ttbar$) events collected in proton-proton ($pp$) collisions at centre-of-mass energies of $\sqrt{s}=7~\tev$ and $8~\tev$~\cite{Aad:2014dya,Aad:2015pja,Khachatryan:2016atv} during Run~1 of the LHC, as well as at $\sqrt{s}=13~\tev$~\cite{Aaboud:2017mfd,Aaboud:2018pob,Sirunyan:2017uae} using early Run~2 data.
In these searches, one of the top quarks is required to decay into $Wb$, while the other top quark decays into $Hq$, yielding $\ttbar \to WbHq$.\footnote{
In the following, $WbHq$ is used to denote both $W^+b H\bar{q}$ and its charge conjugate, $HqW^- \bar{b}$. Similarly,
$WbWb$ is used to denote $W^+b W^- \bar{b}$.}  The Higgs boson is assumed to have a mass of $m_H=125~\gev$ and to decay as predicted by
the SM. The simplifying assumption of SM-like Higgs boson branching ratios is motivated by the fact that measurements of the flavour-diagonal Higgs boson couplings by the ATLAS and CMS collaborations are in agreement with the SM prediction within about 10\%~\cite{Khachatryan:2016vau,Sirunyan:2018koj}.
Furthermore, typical beyond-the-SM scenarios that predict significant enhancements to $\BR(t\to Hq)$, also predict modifications to the Higgs boson branching ratios at the few percent level or below, well beyond the current experimental precision.
Some of the most sensitive single-channel searches have been performed in the $H\to\gamma\gamma$ decay mode, which
has a small branching ratio of $\BR(H\to \gamma\gamma)\simeq 0.2\%$, but benefits from having a very small background contamination
and excellent diphoton mass re\-so\-lu\-tion.
Searches targeting signatures with two same-charge leptons or three leptons (electrons or muons), generically referred to as multileptons,
are able to exploit a branching ratio that is significantly larger for the $H \rightarrow WW^*, \tau\tau$ decay modes than for the $H \rightarrow \gamma\gamma$ decay mode,
and are also characterised by relatively small backgrounds.
Finally, searches have also been performed exploiting the dominant Higgs boson decay mode, $H\to b\bar{b}$, which has a branching ratio
of $\BR(H\to b\bar{b})\simeq 58\%$. Compared with Run~1, the Run~2 searches benefit from the increased $\ttbar$ cross section at $\sqrt{s}=13~\tev$,
as well as the larger integrated luminosity.
Using 36.1 fb$^{-1}$ of data at $\sqrt{s}=13~\tev$, the ATLAS Collaboration has derived upper limits at 95\% confidence level (CL) of
$\BR(t\to Hc)<0.22\%$ using $H\to \gamma\gamma$ decays~\cite{Aaboud:2017mfd}, and of $\BR(t\to Hc)<0.16\%$ based on
multilepton signatures resulting from
$H \to WW^*$, $H\to \tau^+\tau^-$ in which both $\tau$-leptons decay leptonically, or $H \to ZZ^*$~\cite{Aaboud:2018pob}.
These upper limits are derived assuming that $\BR(t\to Hu)=0$. Similar upper limits are obtained for $\BR(t\to Hu)$ if $\BR(t\to Hc)=0$.
The CMS Collaboration has performed a search using
$H\to b\bar{b}$ decays~\cite{Sirunyan:2017uae} with 35.9 fb$^{-1}$ of data at $\sqrt{s}=13~\tev$, resulting
in upper limits of $\BR(t\to Hc)<0.47\%$ and $\BR(t\to Hu)<0.47\%$, in each case neglecting the other decay mode.
Compared with previous searches, the search in Ref.~\cite{Sirunyan:2017uae} considers in addition the contribution to the signal from
$pp \to tH$ production~\cite{Greljo:2014dka}.

The searches presented in this paper are focussed on fermionic decay modes of the Higgs boson. Therefore, they help to complete the
ATLAS experiment's programme of searches for $t \to Hq$ decays based on $pp$ collision
data at $\sqrt{s}=13~\tev$ recorded in 2015 and 2016. The corresponding integrated luminosity is 36.1 fb$^{-1}$.
Two analyses are performed, searching for $\ttbar \to WbHq$ production (ignoring $pp \to tH$ production) and targeting the
$H \to b\bar{b}$ and $H \to \tau^+\tau^-$ decay modes, which this paper refers to as ``$\Hbb$ search'' and ``$\Htautau$ search'', respectively.
The $\Hbb$ search selects events with one isolated electron or muon from the $W \to \ell\nu$ decay, and multiple jets, several
of which are identified with high purity as originating from the hadronisation of $b$-quarks.
The $\Htautau$ search selects events with two $\tau$-lepton candidates, at least one of which decays hadronically, as well as multiple jets.
The latter requirement aims to select events with a hadronically decaying $W$ boson, since this allows an improved reconstruction of the
event kinematics.
 
Both searches employ multivariate techniques to discriminate between the signal and the background on the basis of their different kinematics.
These two searches are combined with previous ATLAS searches in the diphoton and multilepton final states using the same dataset~\cite{Aaboud:2017mfd,Aaboud:2018pob}, and bounds are set on $\BR(t\to Hc)$ and $\BR(t\to Hu)$, as well as on the corresponding non-flavour-diagonal Yukawa couplings.
The combination is performed after verifying the overall consistency of the results obtained by the different searches, which exploit very different
experimental signatures and thus are affected by different backgrounds and related systematic uncertainties.
By combining all searches, the expected sensitivity is improved by about a factor of two relative to the most sensitive individual results.

 
\section{ATLAS detector}
\label{sec:detector}
 
The ATLAS detector~\cite{PERF-2007-01} at the LHC covers almost the entire solid angle around the collision point,\footnote{ATLAS
uses a right-handed coordinate system with its origin at the nominal interaction point (IP) in the
centre of the detector.
The $x$-axis points from
the IP to the centre of the LHC ring, 
the $y$-axis points upward,
and the $z$-axis coincides with the axis of the beam pipe.
Cylindrical coordinates ($r$,$\phi$) are used
in the transverse plane, $\phi$ being the azimuthal angle around the beam pipe. The pseudorapidity is defined in
terms of the polar angle $\theta$ as $\eta = - \ln \tan(\theta/2)$.
Angular distance is measured in units of $\Delta R\equiv \sqrt{(\Delta\eta)^2+(\Delta\phi)^2}$.} and
consists of an inner tracking detector surrounded by a thin superconducting solenoid producing a
2~T axial magnetic field, electromagnetic and hadronic calorimeters, and a muon spectrometer
incorporating three large toroid magnet assemblies with eight coils each. The inner detector contains a high-granularity silicon pixel detector,
including the 
insertable B-layer~\cite{IBL1,IBL2,Abbott:2018ikt}, installed in 2014,
and a silicon microstrip tracker,
together providing a precise reconstruction of tracks of charged particles in the pseudorapidity range $|\eta|<2.5$.
The inner detector also includes a transition radiation tracker that provides tracking and electron identification for $|\eta|<2.0$.
The calorimeter system covers the pseudorapidity range $|\eta| < 4.9$.
Within the region $|\eta|< 3.2$, electromagnetic (EM) calorimetry is provided by barrel and
endcap high-granularity lead/liquid-argon (LAr) sampling calorimeters,
with an additional thin LAr presampler covering $|\eta| < 1.8$,
to correct for energy loss in material upstream of the calorimeters.
Hadronic calorimetry is provided by 
a steel/scintillator-tile calorimeter,
segmented into three barrel structures within $|\eta| < 1.7$, and two copper/LAr hadronic endcap calorimeters.
The solid angle coverage is completed with forward copper/LAr and tungsten/LAr calorimeter modules
optimised for electromagnetic and hadronic measurements, respectively.
The calorimeters are surrounded by a muon spectrometer within a magnetic field provided by air-core toroid magnets
with a bending integral of about 2.5 Tm in the barrel and up to 6 Tm in the endcaps.
The muon spectrometer measures the trajectories of muons with $|\eta|<2.7$ using multiple layers of high-precision tracking chambers,
and is instrumented with separate trigger chambers covering $|\eta|<2.4$.
A two-level trigger system~\cite{Aaboud:2016leb}, consisting of a hardware-based level-1 trigger followed by
a software-based high-level trigger, is used to reduce the event rate to a maximum of around 1 kHz for offline storage.
 
 
\section{Event reconstruction}
\label{sec:objects}
 
The event reconstruction is affected by multiple $pp$ collisions in a single bunch crossing and by collisions
in neighbouring bunch crossings, referred to as pile-up.
Interaction vertices from the $pp$ collisions are reconstructed from at least two tracks
with transverse momentum ($\pt$) larger than $400~\mev$ that are consistent with originating from the
beam collision region in the $x$--$y$ plane. If more than one primary vertex candidate is found, the
candidate whose associated tracks form the largest sum of squared $\pt$~\cite{ATL-PHYS-PUB-2015-026}
is selected as the hard-scatter primary vertex.
 
Electron candidates~\cite{ATLAS-CONF-2016-024,Aaboud:2018ugz} are reconstructed from energy
clusters in the EM calorimeter that are matched to reconstructed tracks in the inner detector;
electron candidates in the transition region between the EM barrel and endcap calorimeters
($1.37 < |\eta_{\textrm{cluster}}| < 1.52$) are excluded.
In the $\Hbb$ ($\Htautau$) search, electron candidates are required to have $\pt>30~(15)~\gev$ and
$|\eta_{\textrm{cluster}}| < 2.47$, and to satisfy tight (medium) likelihood-based identification
criteria~\cite{ATLAS-CONF-2016-024} based on calorimeter, tracking and combined variables that provide
separation between electrons and jets.
 
Muon candidates~\cite{Aad:2016jkr} are reconstructed by matching track segments in 
different layers of the muon spectrometer to tracks found in the inner detector;
the resulting muon candidates are re-fitted using the complete track information from both detector systems.
In the $\Hbb$ ($\Htautau$) search, muon candidates are required to have $\pt>30~(10)~\gev$ and $|\eta|<2.5$
and to satisfy medium identification criteria~\cite{Aad:2016jkr}.
 
Electron (muon) candidates are matched to the primary vertex by requiring that the significance of their transverse impact parameter, $d_0$,
satisfies $|d_0/\sigma(d_0)|<5\,(3)$, where $\sigma(d_0)$ is the measured uncertainty in $d_0$,
and by requiring that their longitudinal impact parameter, $z_0$, satisfies $|z_0 \sin\theta|<0.5$~mm.
To further reduce the background from non-prompt leptons, photon conversions and hadrons, lepton candidates are also required to be isolated
in the tracker and in the calorimeter.
A track-based lepton isolation criterion is defined by calculating the quantity $I_R = \sum \pt^{\textrm{trk}}$, where
the scalar sum includes all tracks (excluding the lepton candidate itself) within the cone defined by $\Delta R<R_{\textrm{cut}}$ around the 
direction of the lepton.  The value of $R_{\textrm{cut}}$ is the smaller of $r_{\textrm{min}}$ and $10~\gev/\pt^\ell$, where
$r_{\textrm{min}}$ is set to 0.2 (0.3) for electron (muon) candidates, and $\pt^\ell$ is the lepton $\pt$.
The $\Hbb$ search requires lepton candidates to satisfy $I_R/\pt^\ell < 0.06$, while the $\Htautau$ search
makes $\pt$-dependent requirements on $I_R/\pt^\ell$. Additionally, the $\Htautau$ search requires leptons to
satisfy a calorimeter-based isolation criterion: the sum of the transverse energy within a cone of size
$\Delta R<0.2$ around the lepton, after subtracting the contributions
from pile-up and the energy deposit of the lepton itself, is required to be less than a $\pt$-dependent
fraction of the lepton energy.
 
Candidate jets are reconstructed with the anti-$k_t$ algorithm~\cite{Cacciari:2008gp,Cacciari:2005hq} with a
radius parameter $R=0.4$, as implemented in the \fastjet\ package~\cite{Cacciari:2011ma}.
Jet reconstruction in the calorimeter starts from topological clustering~\cite{Aad:2016upy} of individual calorimeter cells calibrated to the electromagnetic energy scale.
The reconstructed jets are then calibrated to the particle level by the application of a jet energy scale
derived from simulation and in situ corrections based on $\sqrt{s}=13~\tev$ data~\cite{Aaboud:2017jcu}.
The calibrated jets used in the $\Hbb$ search are required to have $\pt > 25~\gev$ and $|\eta| < 2.5$,
while the $\Htautau$ search uses jets with $\pt > 30~\gev$ and $|\eta| < 4.5$.
Jet four-momenta are corrected for pile-up effects using the jet-area method~\cite{Cacciari:2008gn}.
 
Quality criteria are imposed to reject events that contain any jets arising from non-collision sources
or detector noise~\cite{ATLAS-CONF-2015-029}.  To reduce the contamination due to jets originating from pile-up interactions,
additional requirements are imposed on the jet vertex tagger (JVT)~\cite{Aad:2015ina} output for jets with $\pt<60~\gev$ and $|\eta| < 2.4$,
or on the forward JVT~\cite{Aaboud:2017pou} output for jets with $\pt<50~\gev$ and $|\eta| > 2.5$.
 
Jets containing $b$-hadrons are identified ($b$-tagged) via an algorithm~\cite{Aad:2015ydr,ATL-PHYS-PUB-2016-012}
that uses multivariate techniques to combine information about the impact parameters of displaced tracks and the  topological properties
of secondary and tertiary decay vertices reconstructed within the jet. For each jet, a value for the multivariate $b$-tagging discriminant is
calculated. In the $\Htautau$ search, a jet is considered $b$-tagged if this value is above the threshold corresponding to
an average 70\% efficiency to tag a $b$-quark jet, with a light-jet\footnote{Light-jet refers to a jet originating from the hadronisation of a light quark
($u$, $d$, $s$) or a gluon.} rejection factor of about 380 and a charm-jet rejection factor of about 12, as determined for jets with
$\pt >20~\gev$ and $|\eta|<2.5$ in simulated $\ttbar$ events. In contrast, the $\Hbb$ search employs a tighter $b$-tagging requirement,
corresponding to an average efficiency of 60\% to tag a $b$-quark jet, and light-jet and charm-jet rejection factors of about 1500 and 34, respectively.
 
Hadronically decaying $\tau$-lepton ($\had$) candidates are reconstructed from energy clusters in the calorimeters and
associated inner-detector tracks~\cite{ATL-PHYS-PUB-2015-045}. Candidates are required to have either one or three associated tracks,
with a total charge of $\pm 1$. Candidates are required to have $\pt > 25~\gev$ and $|\eta|<2.5$, excluding the EM calorimeter's transition region.
A boosted decision tree (BDT) discriminant~\cite{Breiman:1984jka,Friedman:2002we,Freund:1997xna} using calorimeter- and tracking-based variables is used to identify $\had$ candidates and reject
jet backgrounds. Three working points labelled loose, medium and tight are defined, and correspond to different $\had$ identification efficiency
values, with the efficiency designed to be independent of $\pt$. The $\Htautau$ search uses the medium
working point for the nominal selection, while the loose working point is used for background estimation.
The medium working point has a combined reconstruction and identification efficiency of 55\% (40\%) for one-prong (three-prong) $\had$
decays~\cite{ATLAS-CONF-2017-029}, and an expected rejection factor against light-jets of 100~\cite{ATL-PHYS-PUB-2015-045}.
Electrons that are reconstructed as one-prong $\had$ candidates are removed via a BDT trained to reject electrons.
Any $\had$ candidate that is also $b$-tagged is rejected.

Overlaps between reconstructed objects are removed sequentially. In the $\Hbb$ search, firstly, electron candidates that lie
within $\Delta R = 0.01$ of a muon candidate are removed to suppress contributions from muon bremsstrahlung.
Overlaps between electron and jet candidates are resolved next, and finally, overlaps between remaining jet candidates
and muon candidates are removed. Energy clusters from identified electrons are not excluded during jet reconstruction.
In order to avoid double-counting of electrons as jets, the closest jet whose axis is within ${\Delta}R = 0.2$ of an electron
is discarded. If the electron is within ${\Delta}R = 0.4$ of the axis of any jet after this initial removal, the jet is retained and  the electron is removed.
The overlap removal procedure between the remaining jet candidates and muon candidates is designed to remove those muons
that are likely to have arisen in the decay of hadrons and to retain the overlapping jet instead.
Jets and muons may also appear in close proximity when the jet results from high-$\pt$ muon bremsstrahlung,
and in such cases the jet should be removed and the muon retained. Such jets are characterised by having very
few matching inner-detector tracks. Selected muons that satisfy $\Delta R(\mu,{\textrm{jet}}) < 0.04+10~\gev/\pt^\mu$ are rejected
if the jet has at least three tracks originating from the primary vertex; otherwise the jet is removed and the muon is kept.
The overlap removal procedure in the $\Htautau$ search is similar to that of the $\Hbb$ search, except that the
first step is the removal of $\had$ candidates within $\Delta R=0.2$ of electrons or muons, and the last step is the
removal of jets whose axis lies within $\Delta R=0.2$ of the leading (highest-$\pt$) $\had$ candidate or the two leading $\had$ candidates (depending on the
search channel). In addition, the muon--jet overlap removal is slightly different:
if a muon lies within $\Delta R = 0.2$ of the axis of a jet, the jet is removed if either it has fewer than three tracks originating from the
primary vertex or it has a small $\pt$ compared with that of the muon (the $\pt$ of the jet is less than 50\% of the $\pt$ of the muon,
or the scalar sum of the $\pt$ of the tracks associated with the jet is less than 70\% of the $\pt$ of the muon).

The missing transverse momentum $\mpt$ (with magnitude $\met$) is defined as the negative vector sum of the
$\pt$ of all selected and calibrated objects in the event, including a term to account for momentum from soft particles
in the event which are not associated with any of the selected objects.
This soft term is calculated from inner-detector tracks matched to the selected primary vertex to make it more resilient to
contamination from pile-up interactions~\cite{Aaboud:2018tkc}.

 
\section{Data sample and event preselection}
\label{sec:data_presel}
 
Both searches are based on a dataset of $pp$ collisions at $\sqrt{s}=13~\tev$ with 25 ns bunch spacing collected in 2015 and 2016, corresponding to an integrated luminosity of $36.1~\ifb$.
Only events recorded with a single-electron trigger, a single-muon trigger, or a di-$\tau$ trigger under stable beam conditions
and for which all detector subsystems were operational are considered.
The number of $pp$ interactions per bunch crossing in this dataset ranges from about 8 to 45, with an average of 24.
 
Single-electron and single-muon triggers with low $\pt$ thresholds and lepton isolation requirements are combined in a logical OR
with higher-threshold triggers but with a looser identification criterion and without any isolation requirement.
The lowest $\pt$ threshold used for muons is 20 (26)~\gev\ in 2015 (2016), while for electrons the threshold is 24 (26)~\gev.
For di-$\tau$ triggers, the $\pt$ threshold of the leading (trailing) $\had$ candidate is 35 (25)~\gev.
In both searches, events satisfying the trigger selection are required to have at least one primary vertex candidate.
 
Events selected by the $\Hbb$ search are recorded with a single-electron or single-muon trigger and
are required to have exactly one electron or muon that matches, with $\Delta R < 0.15$, the lepton reconstructed by the trigger.
Furthermore, at least four jets are required, of which at least two must be $b$-tagged.
 
In the $\Htautau$ search, events are classified into $\lephad$ and $\hadhad$ channels depending on the
multiplicity of selected leptons. Events in the $\lephad$ channel are recorded with a single-electron or single-muon trigger
and are required to have exactly one selected electron or muon and at least one $\had$ candidate.
The selected electron or muon is required to match, with $\Delta R < 0.15$, the lepton reconstructed by the trigger
and to have a $\pt$ exceeding the trigger $\pt$ threshold by 1~\gev\ or 2~\gev\ (depending on the lepton trigger and
data-taking conditions). In addition, its electric charge is required to be of opposite sign to that of the leading $\had$ candidate.
Events in the $\hadhad$ channel are recorded with a di-$\tau$ trigger, and are required to have at least two $\had$ candidates and
no selected electrons or muons. The two leading $\had$ candidates are required to have charges of opposite sign.
In addition, in both $\Htautau$ search channels, trigger matching for $\had$ candidates, at least three jets and exactly one $b$-tagged jet are required.
 
The above requirements apply to the reconstructed objects defined in Section~\ref{sec:objects}.
These requirements, which ensure a negligible overlap between the $\Hbb$ and $\Htautau$ searches,
are referred to as the preselection and are summarised in Table~\ref{tab:preselection}.
 
\begin{table*}[t!]
\caption{\small{Summary of preselection requirements for the $\Hbb$ and $\Htautau$ searches.
The leading and trailing $\had$ candidates are denoted by $\tau_{\mathrm{had,1}}$ and $\tau_{\mathrm{had,2}}$ respectively.}}
\begin{center}
\begin{tabular}{l|c|cc}
\toprule\toprule
\multicolumn{4}{c}{Preselection requirements} \\
\midrule
Requirement &  $\Hbb$ search & \multicolumn{2}{c}{$\Htautau$ search} \\
& & $\lephad$ channel & $\hadhad$ channel \\
\midrule
Trigger & single-lepton trigger & single-lepton trigger & di-$\tau$ trigger  \\
Leptons  & =1 isolated $e$ or $\mu$ & =1 isolated $e$ or $\mu$ & no isolated $e$ or $\mu$ \\
& -- & $\geq$1 $\had$ & $\geq$2 $\had$ \\
Electric charge ($q$) & -- & $q_\ell \times q_{\tau_{\mathrm{had,1}}} < 0$ & $q_{\tau_{\mathrm{had,1}}} \times q_{\tau_{\mathrm{had,2}}} < 0$ \\
Jets  &  $\geq$4 jets & $\geq$3 jets & $\geq$3 jets \\
$b$-tagging & $\geq$2 $b$-tagged jets & =1 $b$-tagged jets & =1 $b$-tagged jets  \\
\bottomrule\bottomrule
\end{tabular}
\label{tab:preselection}
\end{center}
\end{table*}
 
\section{Signal and background modelling}
\label{sec:signal_background_model}
 
Signal and most background processes are modelled using Monte Carlo (MC) simulation.
After the event preselection, the main background is $\ttbar$ production, often in association with jets, denoted by $\ttbar$+jets in the following.
Small contributions arise from single-top-quark, $W/Z$+jets, multijet and diboson ($WW,WZ,ZZ$) production, as well as from the associated
production of a vector boson $V$ ($V=W,Z$) or a Higgs boson and a $\ttbar$ pair ($\ttbar V$ and $\ttbar H$). All backgrounds
with prompt leptons, i.e.\ those originating from the decay of a $W$ boson, a $Z$ boson, or a $\tau$-lepton,
are estimated using samples of simulated events and initially normalised to their theoretical cross sections.
In the simulation, the top-quark and SM Higgs boson masses are set to $172.5~\gev$ and $125~\gev$, respectively,
and the Higgs boson is allowed to decay into all SM particles with branching ratios calculated using \textsc{Hdecay}~\cite{Djouadi:1997yw}.
Backgrounds with non-prompt electrons or muons, with photons or jets misidentified as electrons, or with jets misidentified as $\had$ candidates,
generically referred to as fake leptons, are estimated using data-driven methods.
The background prediction is further improved during the statistical analysis by performing a likelihood
fit to data using several signal-depleted analysis regions, as discussed in Sections~\ref{sec:strategy_Hbb} and~\ref{sec:strategy_Htautau}.
 
\subsection{Simulated signal and background processes}
\label{sec:simulations}
 
Samples of simulated $\ttbar \to WbHq$ events were generated with the next-to-leading-order (NLO) generator\footnote{In the following,
the order of a generator should be understood as referring to the order in the strong coupling constant at which the matrix-element calculation
is performed.} {\amcatnlolong}~2.4.3~\cite{Alwall:2014hca}  (referred to in the following as {\amcatnlo}) with the NNPDF3.0 NLO~\cite{Ball:2014uwa} parton distribution function (PDF) set and interfaced to {\pythia} 8.212~\cite{Sjostrand:2007gs} with the NNPDF2.3 LO~\cite{Ball:2012cx} PDF set for the modelling of parton showering, hadronisation, and the underlying event.
The A14~\cite{ATLASUETune4} set of tuned parameters in {\pythia} controlling the description of multiparton interactions and
initial- and final-state radiation, referred to as the tune, was used.
The signal sample is normalised to the same total cross section as used for the inclusive $t\bar{t}\to WbWb$ sample (see discussion below) and
assuming an arbitrary branching ratio of $\BR_{\mathrm{ref}}(t\to Hq)=1\%$.
The case of both top quarks decaying into $Hq$ is neglected in the analysis given the existing upper limits on $\BR(t \to Hq)$ (Section~\ref{sec:intro}).
 
The nominal sample used to model the $\ttbar$ background was generated with the NLO generator {\powheg}~v2 \cite{Frixione:2007nw,Nason:2004rx,Frixione:2007vw,Alioli:2010xd} using the NNPDF3.0 NLO PDF set. The {\powheg} model parameter $h_{\textrm{damp}}$, which controls
matrix element to parton shower matching and effectively regulates the high-$\pt$ radiation, was set to 1.5 times the top-quark mass.
The parton showers, hadronisation, and underlying event were modelled by {\pythia}~8.210 with the NNPDF2.3 LO PDF set in combination with the A14 tune.
Alternative $\ttbar$ simulation samples used to derive systematic uncertainties are described in Section~\ref{sec:syst_bkgmodeling}.
The generated $\ttbar$ samples are normalised to a theoretical cross section of $832^{+46}_{-51}$~pb,
computed using \textsc{Top++}~v2.0~\cite{Czakon:2011xx} at next-to-next-to-leading order (NNLO),
including resummation of next-to-next-to-leading logarithmic (NNLL) soft gluon
terms~\cite{Cacciari:2011hy,Baernreuther:2012ws,Czakon:2012zr,Czakon:2012pz,Czakon:2013goa}.
 
The $\ttbar$ background selected by the $\Hbb$ search is enriched in $\ttbar$+heavy-flavour production, and thus requires a more sophisticated
treatment than provided by the nominal $\ttbar$ sample; this treatment is briefly outlined below. A detailed discussion can be found in Ref.~\cite{Aaboud:2017rss}.
The simulated $\ttbar$ events are categorised depending
on the flavour content of additional particle jets not originating from the decay of the $\ttbar$ system.
Events labelled as either \ttbin\ or \ttcin\ are generically referred to in the following as $\ttbar$+HF events, where HF stands for heavy flavour.
The remaining events are labelled as $\ttbar$+light-jets events, including those with no additional jets.
A finer categorisation of \ttbin\ events is considered for the purpose of applying further corrections and
assigning systematic uncertainties associated with the modelling of heavy-flavour production in different event topologies~\cite{Aaboud:2017rss}.
In particular, the \ttbin\ events are reweighted to an NLO prediction in the four-flavour (4F) scheme
of \ttbin\ production including parton showering~\cite{Cascioli:2013era}, based on {\ShOLlong}~\cite{Gleisberg:2008ta, Cascioli:2011va} (referred to as {\ShOL}
in the following) using the CT10 4F PDF set.  This reweighting is performed in such a way that the inter-normalisations of the \ttbin\ categories are at NLO accuracy,
while preserving the \ttbin\ cross section of the nominal $\ttbar$ sample.
This reweighting is also applied to the alternative $\ttbar$ samples that are used to study systematic uncertainties.
 
Samples of single-top-quark events corresponding to the $t$-channel production mechanism were generated with the
{\powheg}~v1~\cite{Frederix:2012dh} generator, using the 4F scheme  for the NLO matrix-element calculations
and the fixed 4F \textsc{CT10}f\textsc{4}~\cite{Lai:2010vv} PDF set.
Samples corresponding to the $tW$- and $s$-channel production mechanisms were generated
with {\powheg}~v1 using the CT10 PDF set. Overlaps between the $\ttbar$ and $tW$ final states were avoided by using
the diagram removal scheme~\cite{Frixione:2005vw}.
The parton showers, hadronisation and the underlying event were modelled using {\pythia}~6.428~\cite{Sjostrand:2006za}
with the CTEQ6L1~\cite{Pumplin:2002vw,Nadolsky:2008zw} PDF set
in combination with the Perugia 2012 tune~\cite{Skands:2010ak}.
The single-top-quark samples are normalised to the approximate NNLO theoretical cross
sections~\cite{Kidonakis:2011wy,Kidonakis:2010ux,Kidonakis:2010tc}.
 
Samples of $W/Z$+jets events were generated with the {\sherpa}~2.2.1~\cite{Gleisberg:2008ta} generator.
The matrix element was calculated for up to two partons at NLO and up to four partons at LO using
\textsc{Comix}~\cite{Gleisberg:2008fv} and \textsc{OpenLoops}~\cite{Cascioli:2011va}. The matrix-element calculation
is merged with the {\sherpa} parton shower~\cite{Schumann:2007mg} using the ME+PS@NLO prescription~\cite{Hoeche:2012yf}.
The PDF set used for the matrix-element calculation is NNPDF3.0 NNLO~\cite{Ball:2014uwa} with a dedicated parton shower tuning developed for {\sherpa}.
Separate samples were generated for different $W/Z$+jets categories using filters for a $b$-jet
($W/Z$+$\geq$1$b$+jets), a $c$-jet and no $b$-jet ($W/Z$+$\geq$1$c$+jets), and with a veto on $b$- and $c$-jets
($W/Z$+light-jets), which are combined into the inclusive $W/Z$+jets samples.
Both the $W$+jets and $Z$+jets samples are normalised to their respective inclusive NNLO theoretical
cross sections calculated with \textsc{FEWZ}~\cite{Anastasiou:2003ds}.
 
Samples of $WW/WZ/ZZ$+jets events were generated with {\sherpa}~2.2.1 using the CT10 PDF set
and include processes containing up to four electroweak vertices.
In the case of $WW/WZ$+jets ($ZZ$+jets) the matrix element was calculated for zero (up to one) additional partons
at NLO and up to three partons at LO using the same procedure as for the $W/Z$+jets samples.
The final states simulated require one of the bosons to decay leptonically and the other hadronically.
All diboson samples are normalised to their NLO theoretical cross sections provided by {\sherpa}.
 
Samples of $\ttbar V$ and $\ttbar H$ events were generated with {\amcatnlo}~2.2.1, using NLO matrix elements and the NNPDF3.0 NLO PDF set,
and interfaced to {\pythia}~8.210 with the NNPDF2.3 LO PDF set and the A14 tune.
Instead, the $\ttbar V$ samples used in the $\Hbb$ search are based on LO matrix elements computed for up to two additional partons
using the NNPDF3.0 NLO PDF set, and merged using the CKKW-L approach~\cite{Lonnblad:2001iq}.
The $\ttbar V$ samples are normalised to the NLO cross section computed with {\amcatnlo}, while the $\ttbar H$ sample is normalised using
the NLO cross section recommended in Ref.~\cite{deFlorian:2016spz}.

All generated samples, except those produced with the {\sherpa}~\cite{Gleisberg:2008ta} event generator,
utilise \textsc{EvtGen}~1.2.0~\cite{Lange:2001uf} to model the decays of heavy-flavour hadrons.
To model the effects of pile-up, events from minimum-bias interactions were generated using {\pythia}~8.186~\cite{Sjostrand:2007gs}
in combination with the A2 tune~\cite{ATL-PHYS-PUB-2011-014},
and overlaid onto the simulated hard-scatter events according to the luminosity profile of the recorded data.
The generated events were processed through a simulation~\cite{Aad:2010ah} of the ATLAS detector geometry and response
using \textsc{Geant4}~\cite{Agostinelli:2002hh}. A faster simulation, where the full \textsc{Geant4} simulation of
the calorimeter response is replaced by a detailed parameterisation of the shower shapes~\cite{FastCaloSim},
was adopted for some of the samples used to estimate systematic uncertainties in background modelling.
Simulated events were processed through the same reconstruction software as the data, and corrections were applied so that the object identification
efficiencies, energy scales and energy resolutions match those determined from data control samples.
 
\subsection{Backgrounds with fake leptons}
\label{sec:fakeleptons}
 
\subsubsection{Fake electrons and muons}
In the $\Hbb$ search, the background from multijet production (multijet background in the following) contributes to the selected
data sample via several production and misreconstruction mechanisms.
In the electron channel, it consists of non-prompt electrons (from semileptonic $b$- or $c$-hadron decays) as well as
misidentified photons (from a conversion of a photon into an $e^+e^-$ pair) or jets with a high fraction of
their energy deposited in the EM calorimeter.  In the muon channel, the multijet background originates mainly from
non-prompt muons.  The multijet background normalisation and shape are estimated directly from data by using the matrix method
technique~\cite{Aad:2010ey,ATLAS-CONF-2014-058}, which exploits differences in lepton identification and isolation properties between
prompt leptons and leptons that are either non-prompt or result from the misidentification of photons or jets.
 
\subsubsection{Fake $\tau$-lepton candidates}
\label{sec:faketaus}
In the $\Htautau$ search, the background with one or more fake $\had$ candidates mainly arises from $\ttbar$ or
multijet production, depending on the search channel, with $W$+jets production contributing to a lesser extent.
Studies based on the simulation show that, for all the above processes, fake $\had$ candidates primarily result from the
misidentification of light-quark jets, with the contribution from $b$-quarks and gluon jets playing a subdominant role.
It is also found that the fake rate decreases for all jet flavours as the $\had$ candidate $\pt$ increases.
 
This background is estimated directly from data by defining control regions (CR) enriched in fake $\had$ candidates via loosened $\had$ requirements or flipped charge. These CRs do not overlap with the main search regions (SRs), discussed in Section~\ref{sec:strategy_Htautau}. The CR selection requirements are analogous to those used to define the different SRs, except that the leading (trailing) $\had$ candidate
in the $\lephad$ ($\hadhad$) channel is required to fail the medium $\had$ identification but pass the loose identification, or the two $\had$ candidates have the same charge.
 
The fake $\had$ background prediction in a given SR is modelled by the distribution (referred to as the fake $\had$ template) derived from data in the corresponding CR. The fake $\had$ template is defined as the data distribution from which the contributions from the simulated backgrounds with real $\had$ candidates, originating primarily from
$W(\to \tau\nu)$+jets and $Z(\to \tau\tau)$+jets, are subtracted. In the $\lephad$ channel, simulation studies indicate that the fake $\had$ background composition is consistent between the SR and the CR, and dominated by $\ttbar$ production. In the $\hadhad$ channel, the fake $\had$ background is expected to be dominated by multijet production. However, simulation studies indicate that the contribution of $\ttbar$ events to the fake $\had$ background is higher in the SR than in the CR. Therefore, an appropriate number of simulated $\ttbar$ events with fake $\had$ candidates in the CR is added to the fake $\had$ template to match the fake $\had$ background composition in the SR.
In both the $\lephad$ and $\hadhad$ channels, the fake $\had$ template in each SR is initially normalised to the estimated fake $\had$ background yield,
defined as the data yield minus the contributions from the simulated backgrounds with real $\had$ candidates (assuming no signal contribution).
During the statistical analysis, the normalisation of the fake $\had$ background in each SR is allowed to vary freely in the fit to data, as discussed in Section~\ref{sec:results_Htautau}.
 
 
 
\section{Strategy for the $\Hbb$ search}
\label{sec:strategy_Hbb}
 
This section presents an overview of the analysis strategy adopted in the $\Hbb$ search, which
closely follows that of the previous search performed on the Run 1 dataset~\cite{Aad:2015pja}.
 
\subsection{Event categorisation}
\label{sec:event_categorisation}
 
Given that the $W\to\ell\nu$ and $H\to b\bar{b}$ decay modes are chosen, the $\ttbar \to WbHq$ signal
is expected to have four jets in the final state, three of them originating from $b$-quarks, which
can be effectively exploited to suppress the background.
Additional jets can also be present because of initial- or final-state radiation.
However, the use of the 60\% $b$-tagging efficiency operating point, characterised by a low mistag rate for
$c$- and light-jets, results in both the $\Hc$ and $\Hu$ signals having a similar $b$-tag multiplicity distribution,
with a very small fraction of events having four or more $b$-tagged jets.
 
In order to optimise the sensitivity of the search, the selected events are categorised into different analysis
regions depending on the number of jets (4, 5 and $\geq$6) and on the number of $b$-tagged jets (2, 3 and $\geq$4).
Therefore, a total of nine analysis regions are considered:
(4j, 2b), (4j, 3b), (4j, 4b), (5j, 2b), (5j, 3b), (5j, $\geq$4b), ($\geq$6j, 2b), ($\geq$6j, 3b), and ($\geq$6j, $\geq$4b),
where ($n$j, $m$b) indicates $n$ selected jets and $m$ $b$-tagged jets.
 
The overall rate and composition of the $\ttbar$+jets background strongly depends on the jet and $b$-tag
multiplicities, as illustrated in Figure~\ref{fig:Hbb_Summary}.
Regions with exactly two $b$-tagged jets are dominated by $\ttbar$+light-jets, while regions with
at least four $b$-tagged jets are dominated by \ttbin. Intermediate compositions are found in regions with exactly three
$b$-tagged jets.  Most of the $\ttbar$+light-jets background events in these regions have a $b$-tagged charm jet from the hadronic $W$ boson
decay, in addition to the two $b$-jets from the top-quark decays.
 
In the regions with four or five jets and exactly three $b$-tagged jets, which dominate the sensitivity of this search,
the selected signal events have a $H \to b\bar{b}$ decay in more than 97\% of the events.
The other regions have significantly lower signal-to-background ratios, but they are used to improve
the $\ttbar$+jets background prediction and constraining the related systematic uncertainties
through a likelihood fit to data.
Because of a somewhat larger fraction of $\Hc$ signal in the regions with exactly three $b$-tagged jets,
resulting from the higher mistag rate for $c$-jets than for light-jets,
this analysis is expected to have slightly better sensitivity to a $\Hc$ signal than to a $\Hu$ signal.
 
\begin{figure*}[t]
\begin{center}
\includegraphics[width=0.55\textwidth]{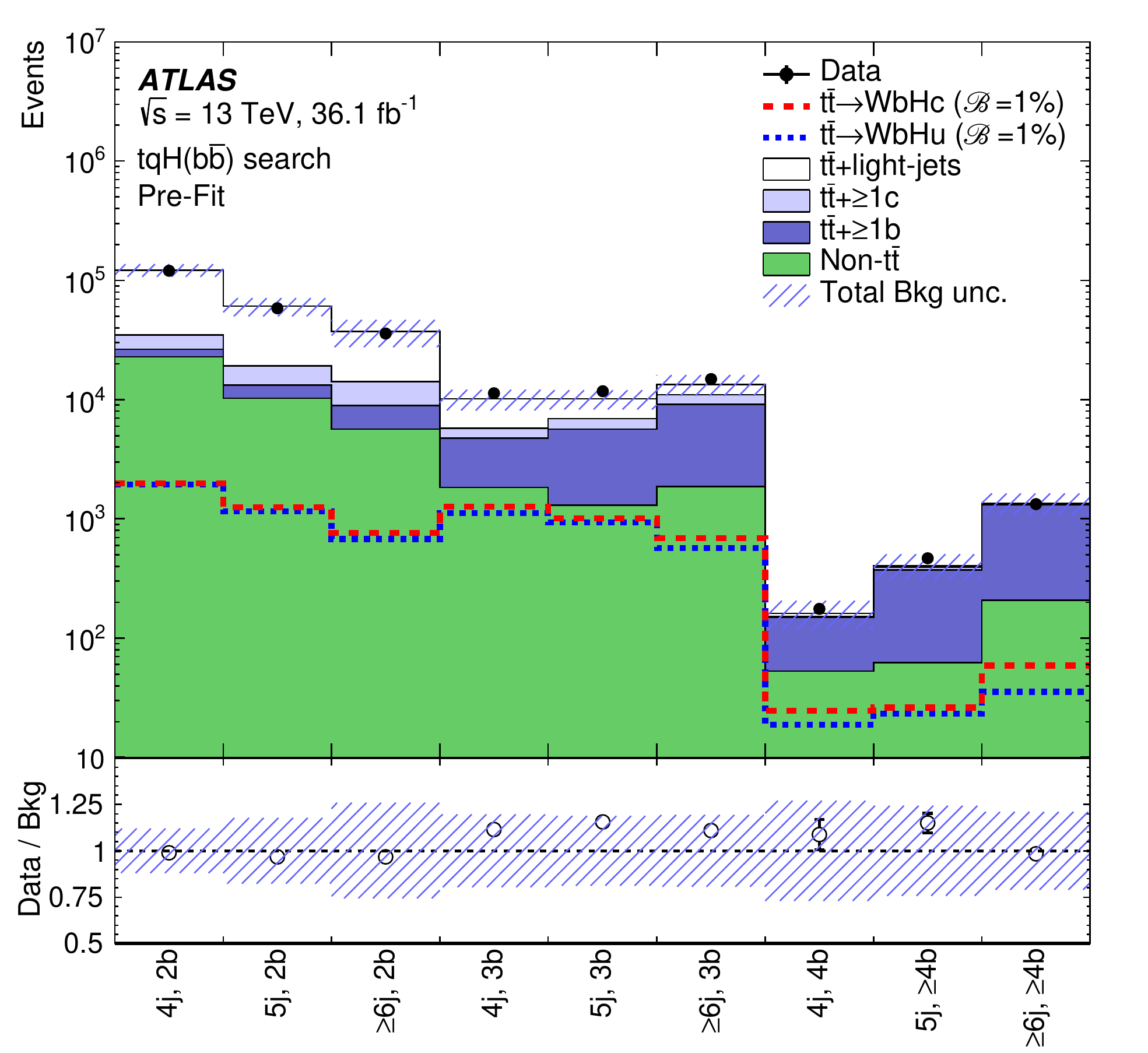}
\caption{$\Hbb$ search: Comparison between the data and predicted background for the event yields in each of the analysis regions considered
before the fit to data (``Pre-Fit''). All events satisfy the preselection requirements, whereas those with exactly two $b$-tagged jets are
in addition required to have a value of the likelihood discriminant above 0.6 (see Section~\ref{sec:likelihood_discriminant}).
Backgrounds are normalised to their nominal cross sections.
The small contributions from $W/Z$+jets,  single-top-quark, diboson and multijet backgrounds are combined into a single background source
referred to as ``Non-$\ttbar$''.
The expected $\Hc$ and $\Hu$ signals (dashed histograms) are shown separately normalised to $\BR(t\to Hq)=1\%$.
The bottom panel displays the ratio of data to the SM background (``Bkg'') prediction.
The hashed area represents the total uncertainty of the background, excluding the normalisation uncertainty of the $\ttbin$ background,
which is determined via a likelihood fit to data.}
\label{fig:Hbb_Summary}
\end{center}
\end{figure*}
 
\subsection{Likelihood discriminant}
\label{sec:likelihood_discriminant}
 
After event categorisation, the signal-to-background ratio is insufficient even in the best cases to achieve sensitivity, and a suitable
discriminating variable between signal and background needs to be constructed in order to improve the sensitivity of the search.
Since both signal and background result from the $\ttbar$ decay,
their discrimination is a challenge and it is based on a few measured quantities.
The most prominent features are the different resonances present in the decay (the Higgs boson in the case
of the $\Hq$ signal and a hadronically decaying $W$ boson in the case of the $\ttbar \to WbWb$ background), and the different flavours of the
jets forming those resonances. However, the large number of jets in the final state causes ambiguities in the calculation
of these kinematic variables to discriminate signal events from background events.
 
This search uses a likelihood (LH) discriminant similar to that developed in Ref.~\cite{Aad:2015pja}.
The LH variable for a given event is defined as:
\begin{equation*}
L(\mathbf{x}) = \frac{P^\textrm{sig}(\mathbf{x}) }{P^\textrm{sig}(\mathbf{x}) +P^\textrm{bkg}(\mathbf{x}) },
\label{eq:D}
\end{equation*}
where $P^\textrm{sig}(\mathbf{x}) $ and $P^\textrm{bkg}(\mathbf{x})$ represent the probability density functions (pdf) of a given event under
the signal hypothesis ($\ttbar \to WbHq$) and under the background hypothesis ($\ttbar \to WbWb$), respectively.
Both $P^\textrm{sig}$ and $P^\textrm{bkg}$ are functions of $\mathbf{x}$, representing the four-momentum vectors of all final-state particles at the reconstruction level:
the lepton, the missing transverse momentum, and the selected jets in a given analysis region.
The value of the multivariate $b$-tagging discriminant for each jet is also included in $\mathbf{x}$.
As in Ref.~\cite{Aad:2015pja}, $P^\textrm{sig}$ and $P^\textrm{bkg}$ are approximated as a product of one-dimensional pdfs
over the set of two-body and three-body invariant masses that correspond to the expected resonances in the event (the leptonically decaying $W$ boson,
the Higgs boson or the hadronically decaying $W$ boson, and the corresponding parent top quarks) and averaged over all possible parton--jet matching combinations.
Combinations are weighted using the per-jet multivariate $b$-tagging discriminant value to suppress the impact from
parton--jet assignments that are inconsistent with the correct flavour of the parton candidates.
The invariant masses are computed from the reconstructed lepton, missing transverse momentum, and jets. After a suitable transformation of the three-body invariant masses
(see Ref.~\cite{Aad:2015pja}), all considered invariant mass variables are largely uncorrelated, thus making possible the factorisation of $P^\textrm{sig}$ and $P^\textrm{bkg}$
as discussed above.
 
Two background hypotheses are considered, corresponding to the dominant backgrounds in
the analysis: $\ttbar$+light-jets and \ttbin. Thus, $P^\textrm{bkg}$ is computed as the average of
the pdfs for the two hypotheses, weighted by their relative fractions found in simulated $\ttbar$+jets events, which depend
on the analysis region considered. Furthermore, in a significant fraction of $\Hq$ simulated events (about 40--50\% in regions with exactly three $b$-tagged jets),
the light-quark jet from the hadronic top-quark decay is not among the selected jets.
Similarly, in about 30--40\% (50--90\%) of simulated $\ttbar$+light-jets ($\ttbin$) background events in regions with exactly three $b$-tagged jets,
the light-quark jet originating from the $W$ boson decay is also not selected. Thus, the calculation of $P^\textrm{sig}$ and
$P^\textrm{bkg}$ also includes an additional hypothesis to account for this topology, again weighted by the corresponding fractions.
In this case, the invariant masses involving the missing jet are computed using the highest-$\pt$ jet not matched
to a decay product from the $\ttbar$ system.

Figure~\ref{fig:Hbb_extravars_4j3b} shows a comparison between data and prediction in the most sensitive analysis region, (4j, 3b),
for several kinematic variables associated with the reconstructed lepton, jets, and missing transverse momentum.
The distributions shown correspond to the lepton $\pt$, the $\met$, the scalar sum of the transverse momenta of
the jets, and the invariant mass distribution of the two $b$-tagged jets with lowest $\Delta R$ separation.
The variables displayed do not correspond directly to those used internally in the evaluation the LH discriminant, as to build them it is necessary
to select a particular signal or background hypothesis and a jet permutation. Instead, these distributions are shown to demonstrate that
a good description of the data by the background prediction is observed in several kinematic variables related to the information used in
the LH discriminant construction.
 
Figure~\ref{fig:LHD} compares the shape of the LH discriminant distribution between the $\Hc$ and $\Hu$ signals and the
$t\bar{t}\to WbWb$ background in each of the analysis regions considered.
Since this analysis has higher expected sensitivity to a $\Hc$ signal than to a $\Hu$ signal, in order to allow probing
of the $\BR(t\to Hu)$ versus $\BR(t\to Hc)$ plane, the LH discriminant optimised for $\Hc$ is used for both
decay modes. It was verified that using the $\Hc$ discriminant for the $\Hu$ search does not result in a significant sensitivity loss.
 
\begin{figure*}[htbp]
\begin{center}
\subfloat[]{\includegraphics[width=0.40\textwidth]{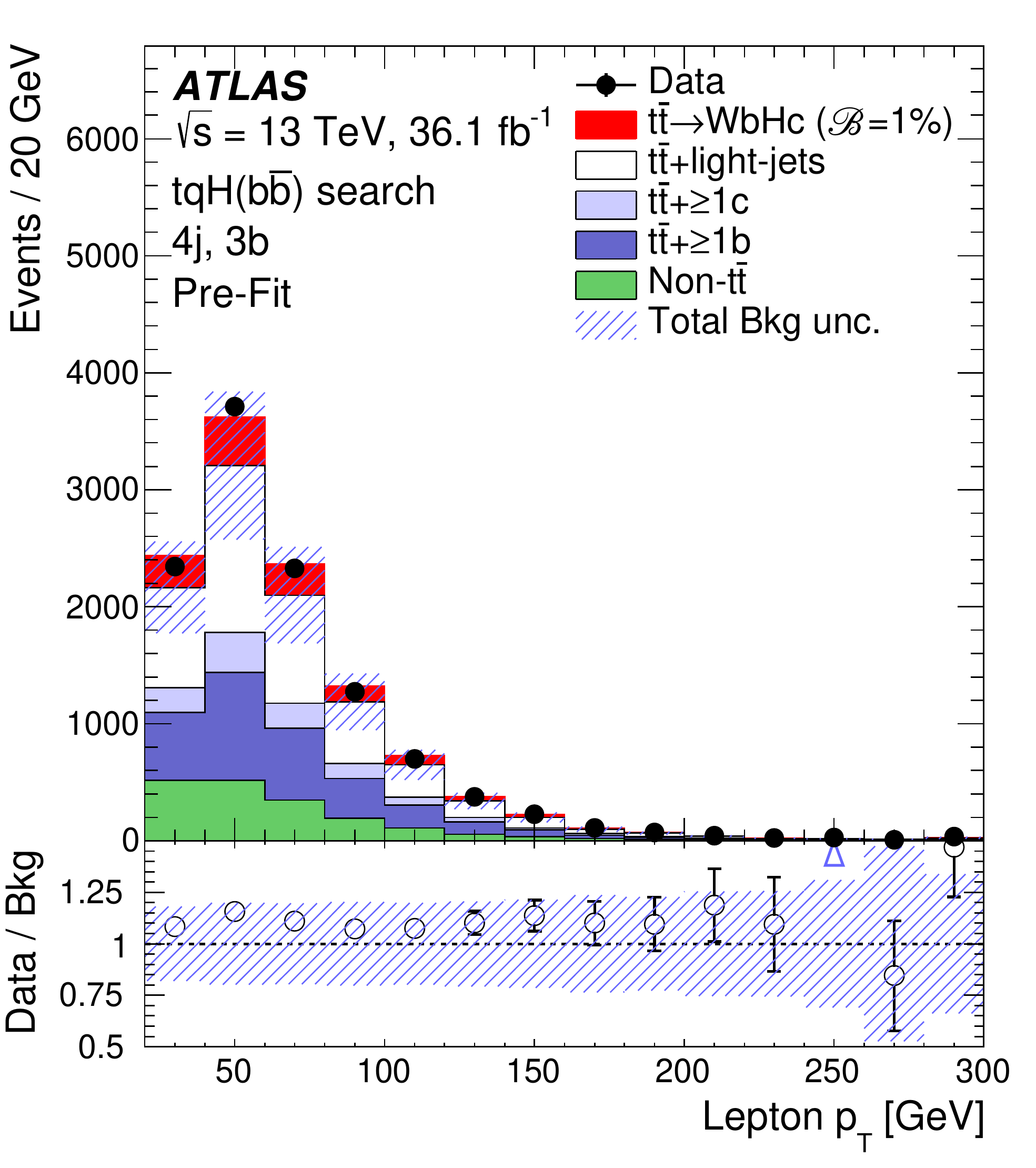}}
\subfloat[]{\includegraphics[width=0.40\textwidth]{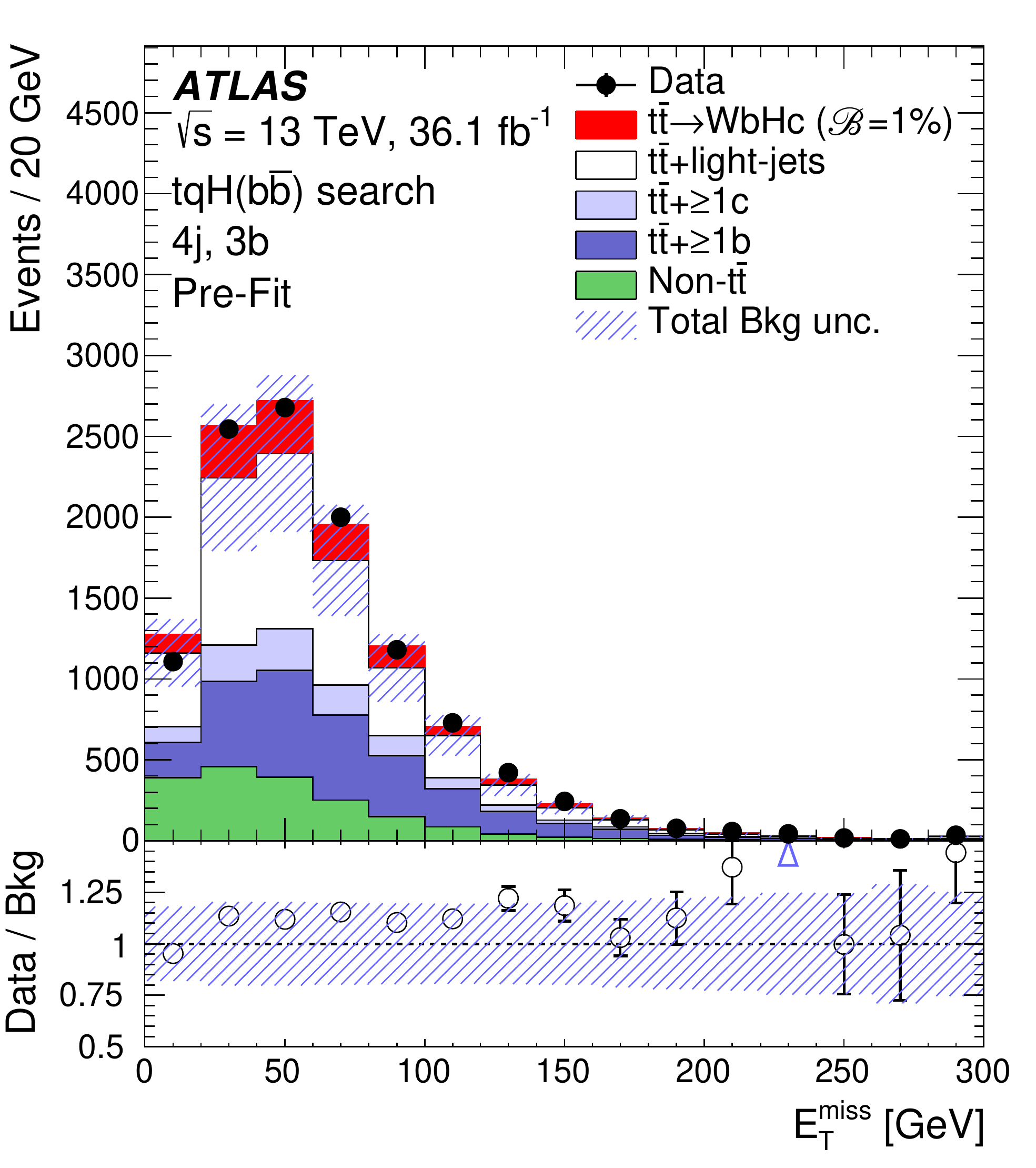}} \\
\subfloat[]{\includegraphics[width=0.40\textwidth]{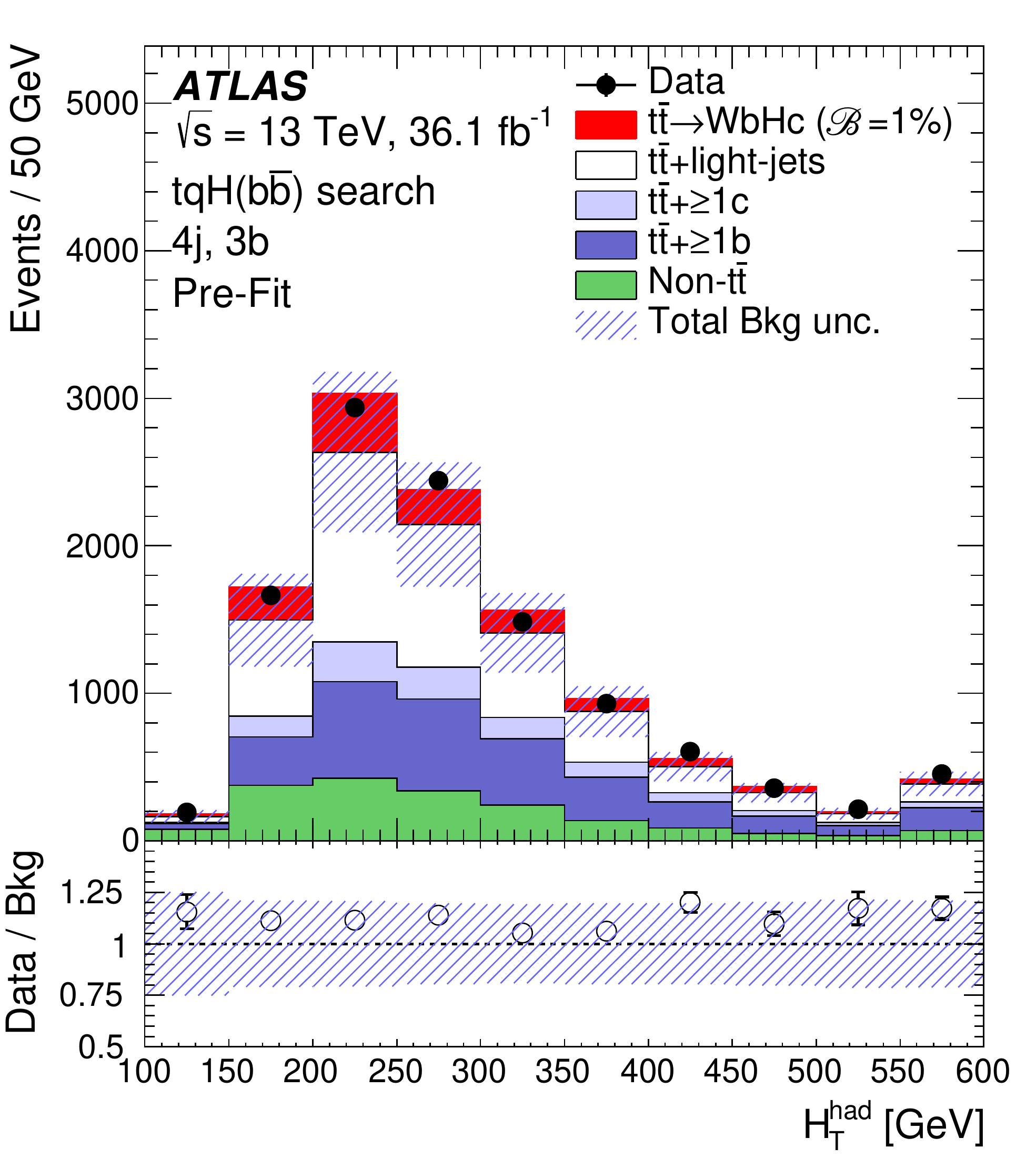}}
\subfloat[]{\includegraphics[width=0.40\textwidth]{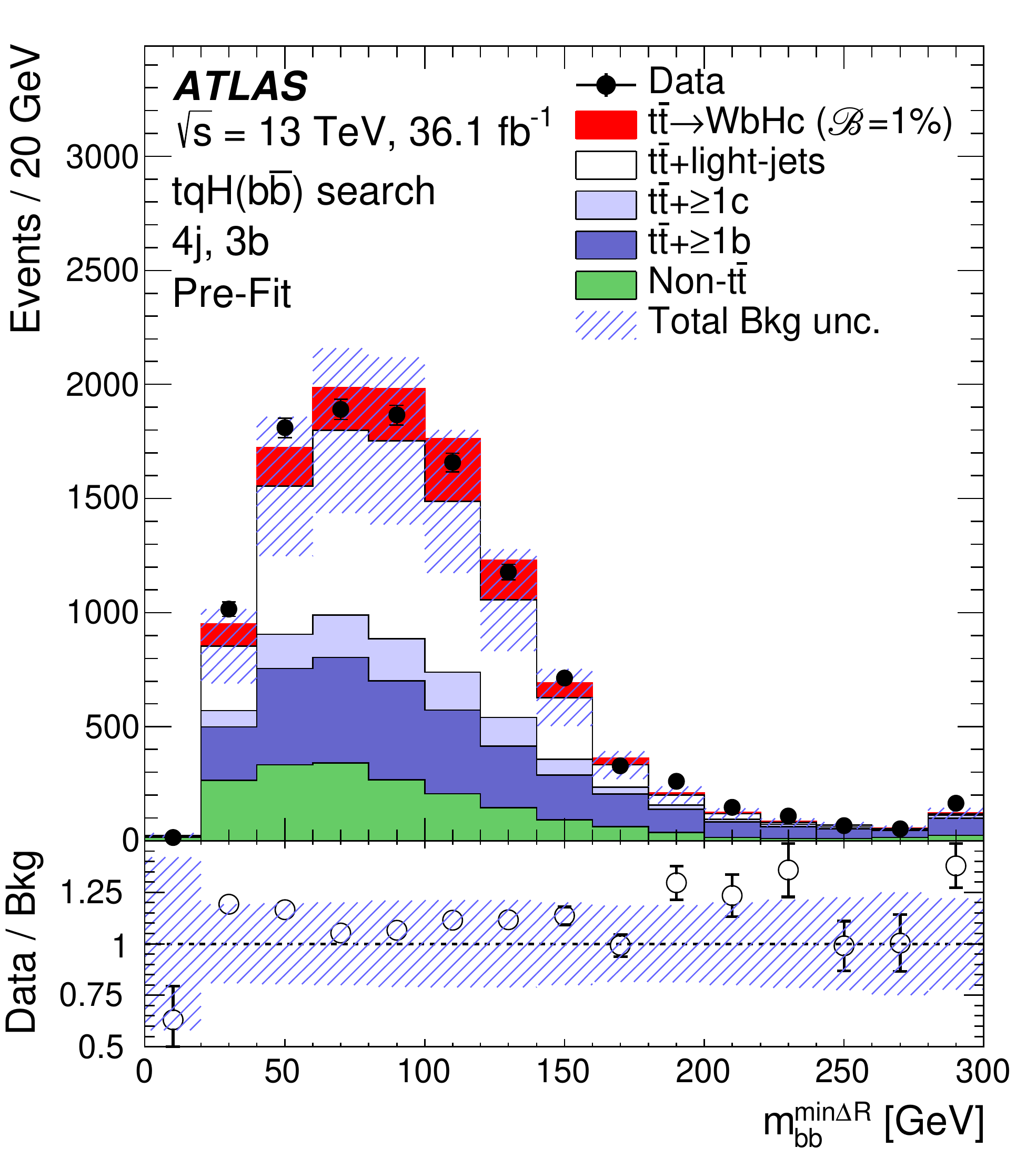}} \\
\caption{\small{$\Hbb$ search: Comparison between the data and predicted background after preselection for several kinematic
distributions in the (4j, 3b) region before the fit to data (``Pre-Fit'').
The distributions are shown for (a) lepton $\pt$, (b) $\met$, (c) scalar sum of the transverse momenta of
the jets ($\hthad$), and (d) the invariant mass of the two $b$-tagged jets with lowest
$\Delta R$ separation ($\mbb$).
The small contributions from $\ttbar V$, $\ttbar H$, single-top-quark, $W/Z$+jets, diboson, and multijet backgrounds are combined
into a single background source referred to as ``Non-$\ttbar$''.
The expected $\Hc$ signal (solid red) corresponding to $\BR(t\to Hc)=1\%$ is also shown,
added to the background prediction.
The last bin in all figures contains the overflow.
The bottom panel displays the ratio of data to the SM background (``Bkg'') prediction.
The blue triangles indicate points that are outside the vertical range of the figure.
The hashed area represents the total uncertainty of the background, excluding the normalisation uncertainty of the $\ttbin$ background,
which is determined via a likelihood fit to data.}}
\label{fig:Hbb_extravars_4j3b}
\end{center}
\end{figure*}
 
\begin{figure*}[htbp]
\begin{center}
\subfloat[]{\includegraphics[width=0.33\textwidth]{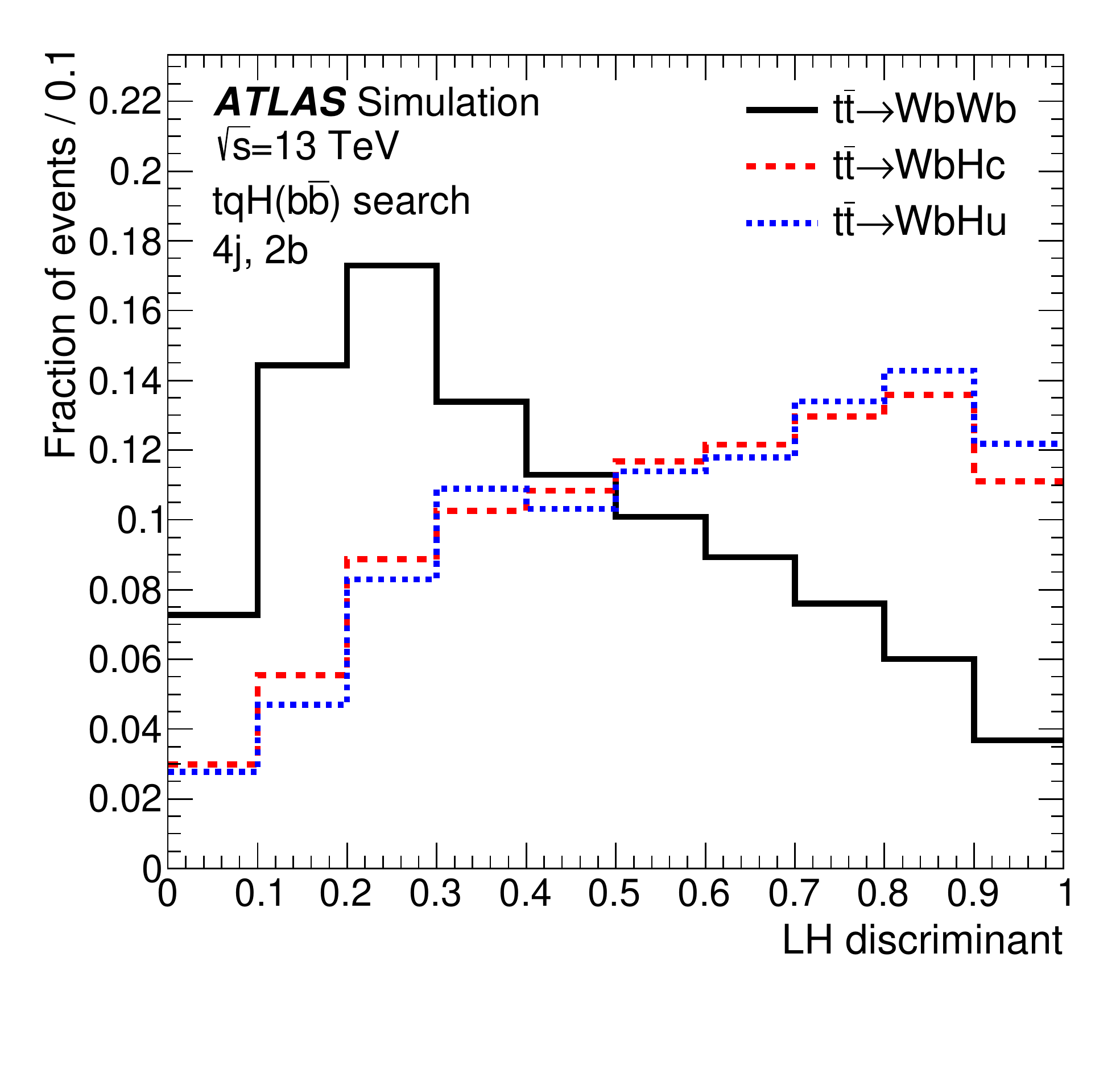}}
\subfloat[]{\includegraphics[width=0.33\textwidth]{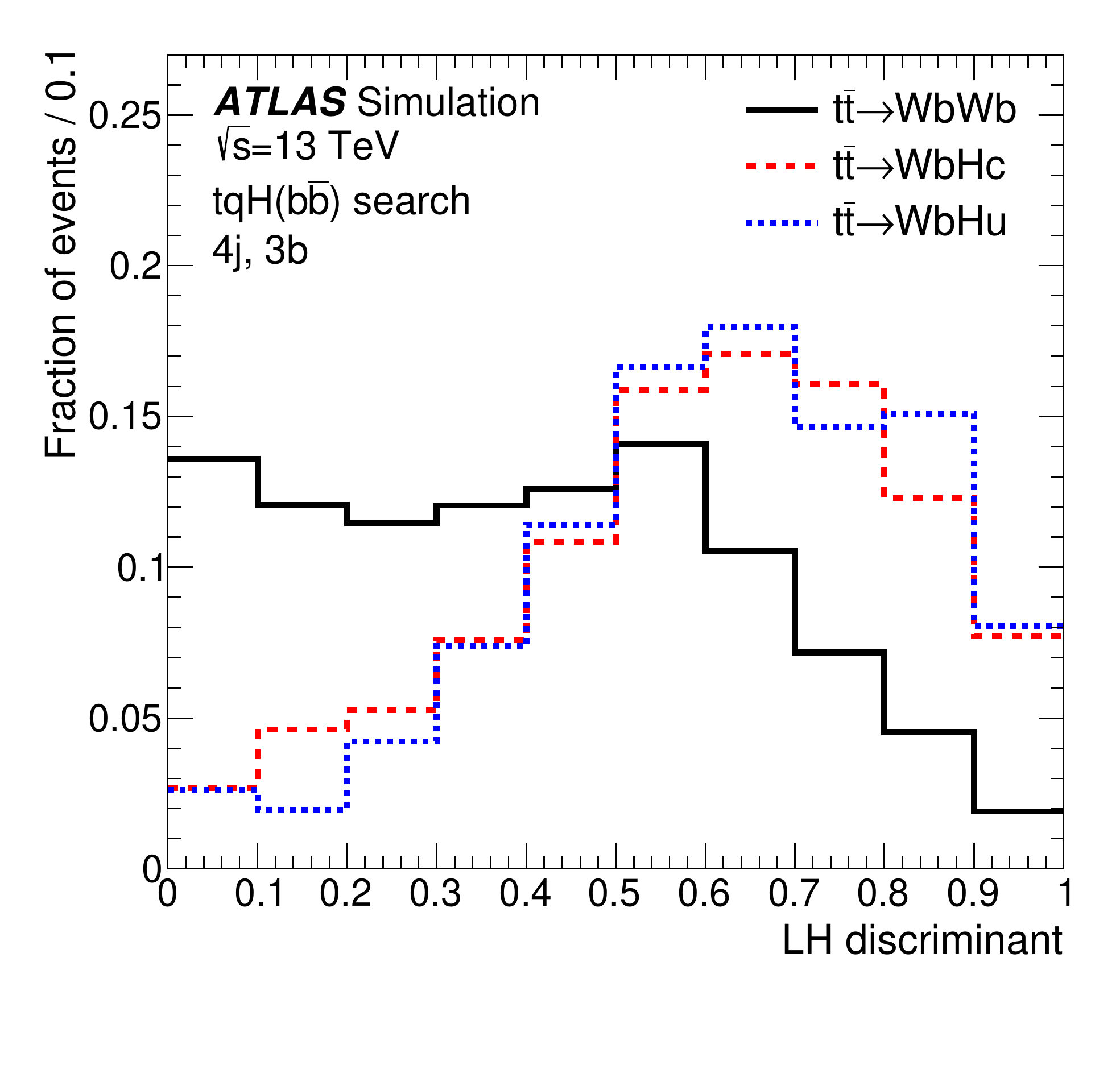}}
\subfloat[]{\includegraphics[width=0.33\textwidth]{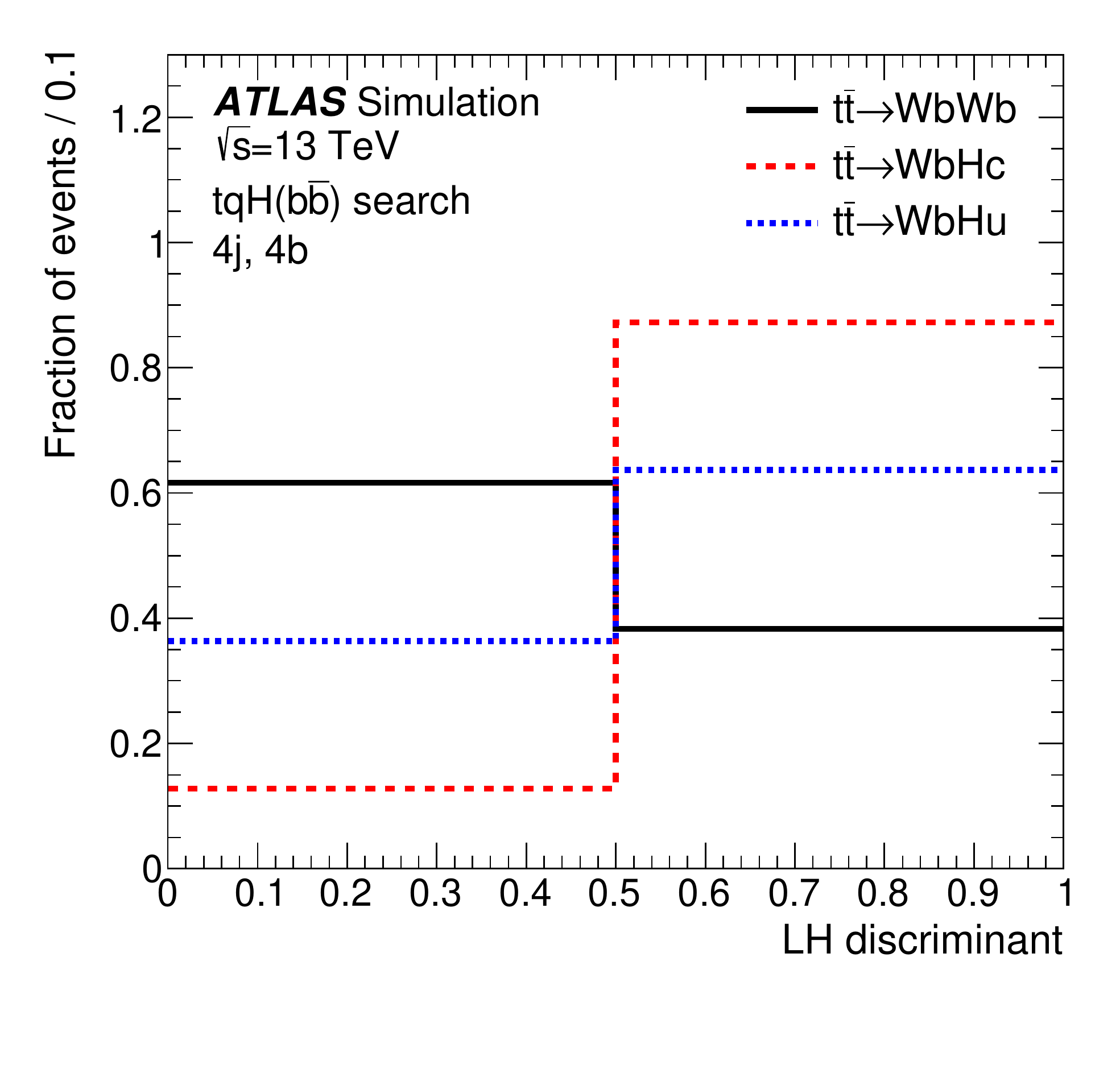}} \\
\subfloat[]{\includegraphics[width=0.33\textwidth]{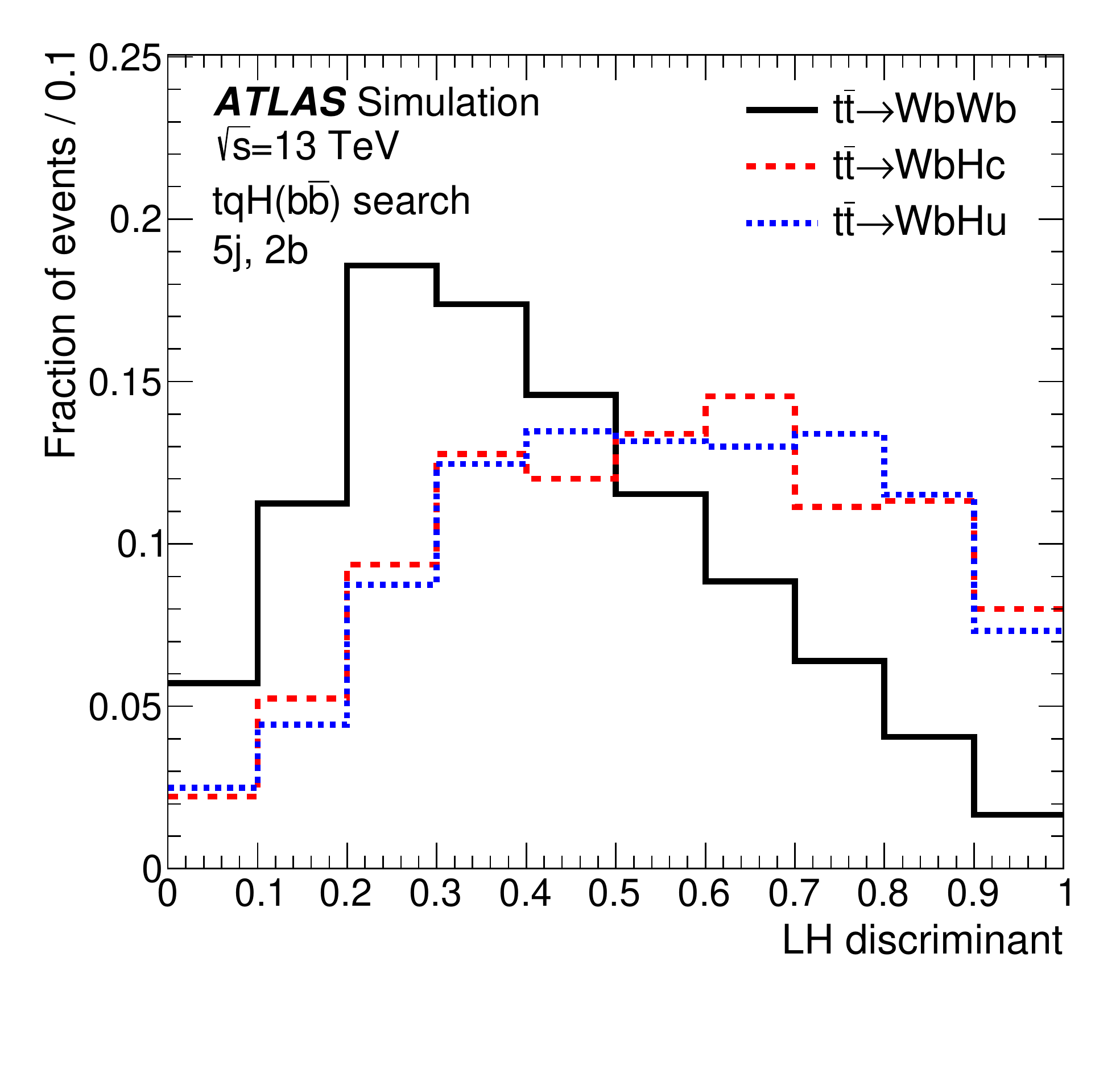}}
\subfloat[]{\includegraphics[width=0.33\textwidth]{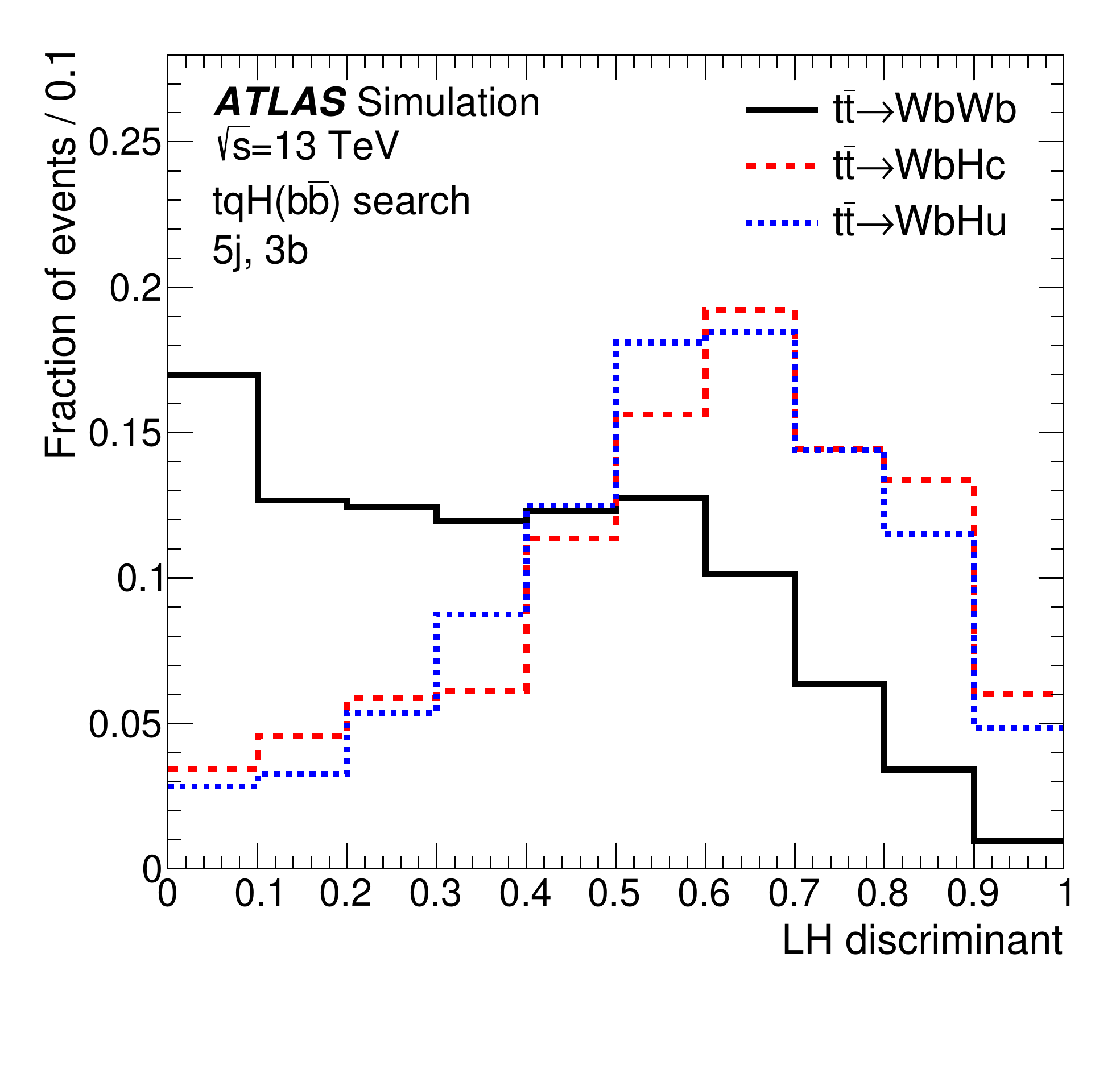}}
\subfloat[]{\includegraphics[width=0.33\textwidth]{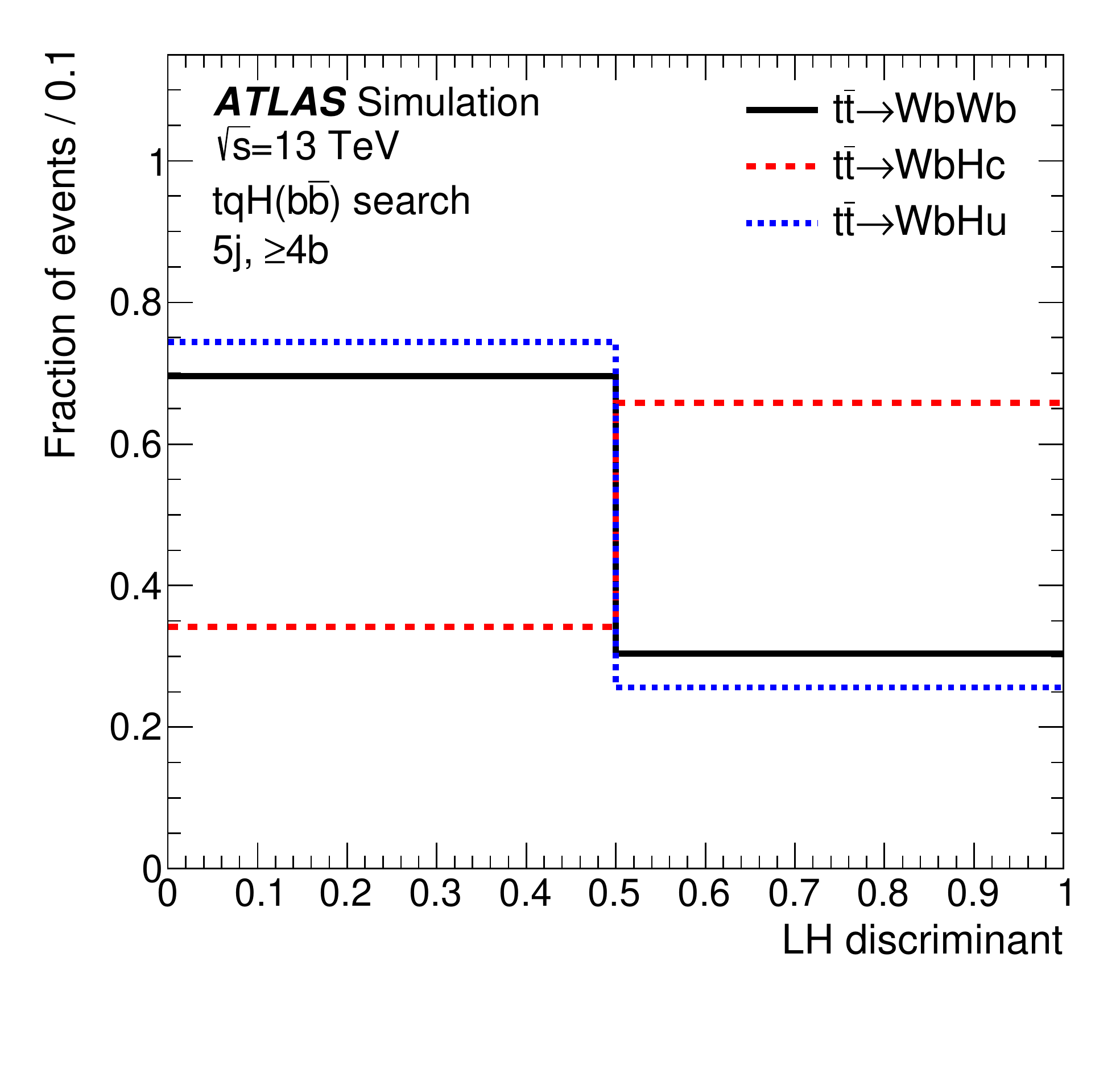}} \\
\subfloat[]{\includegraphics[width=0.33\textwidth]{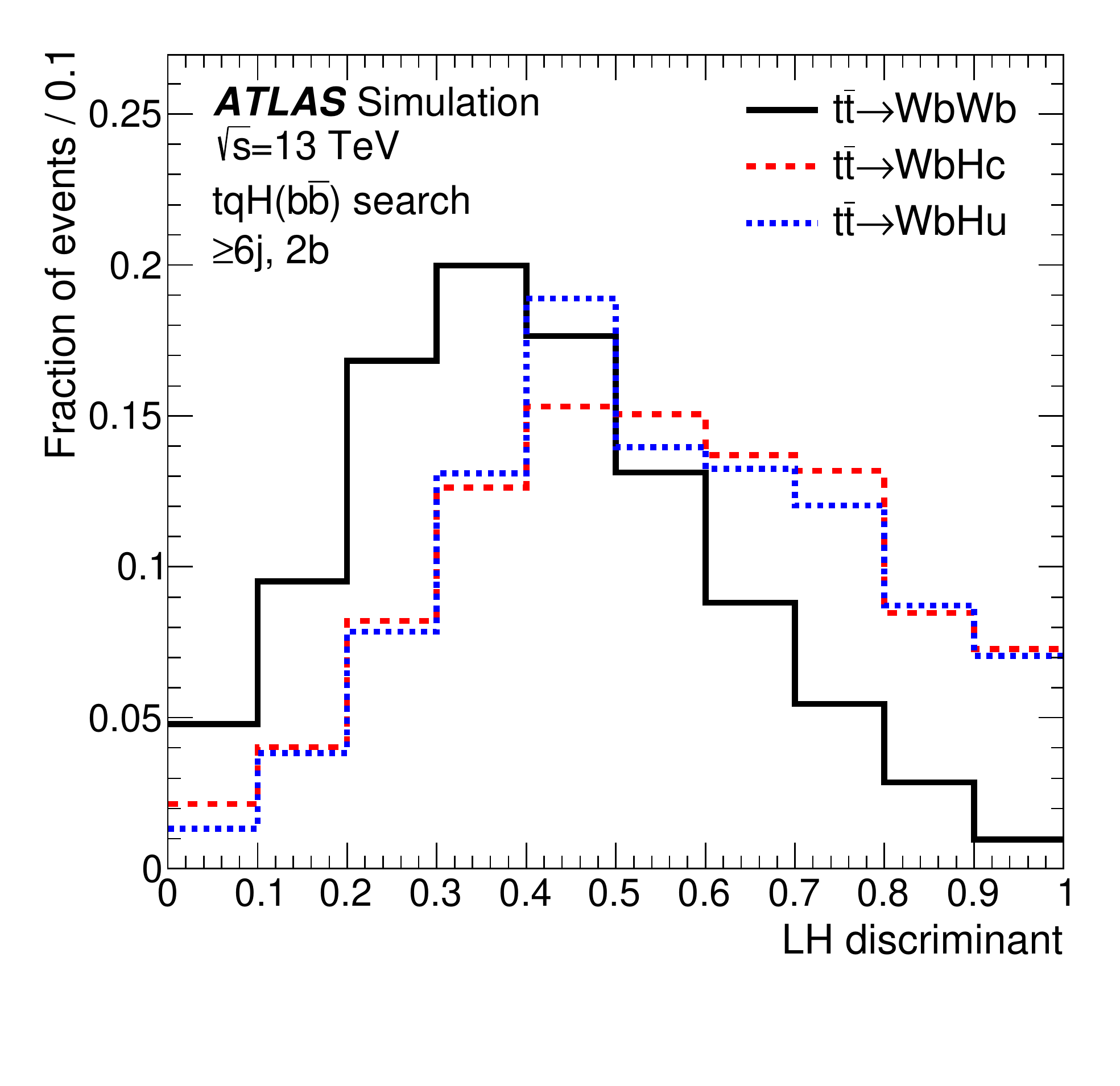}}
\subfloat[]{\includegraphics[width=0.33\textwidth]{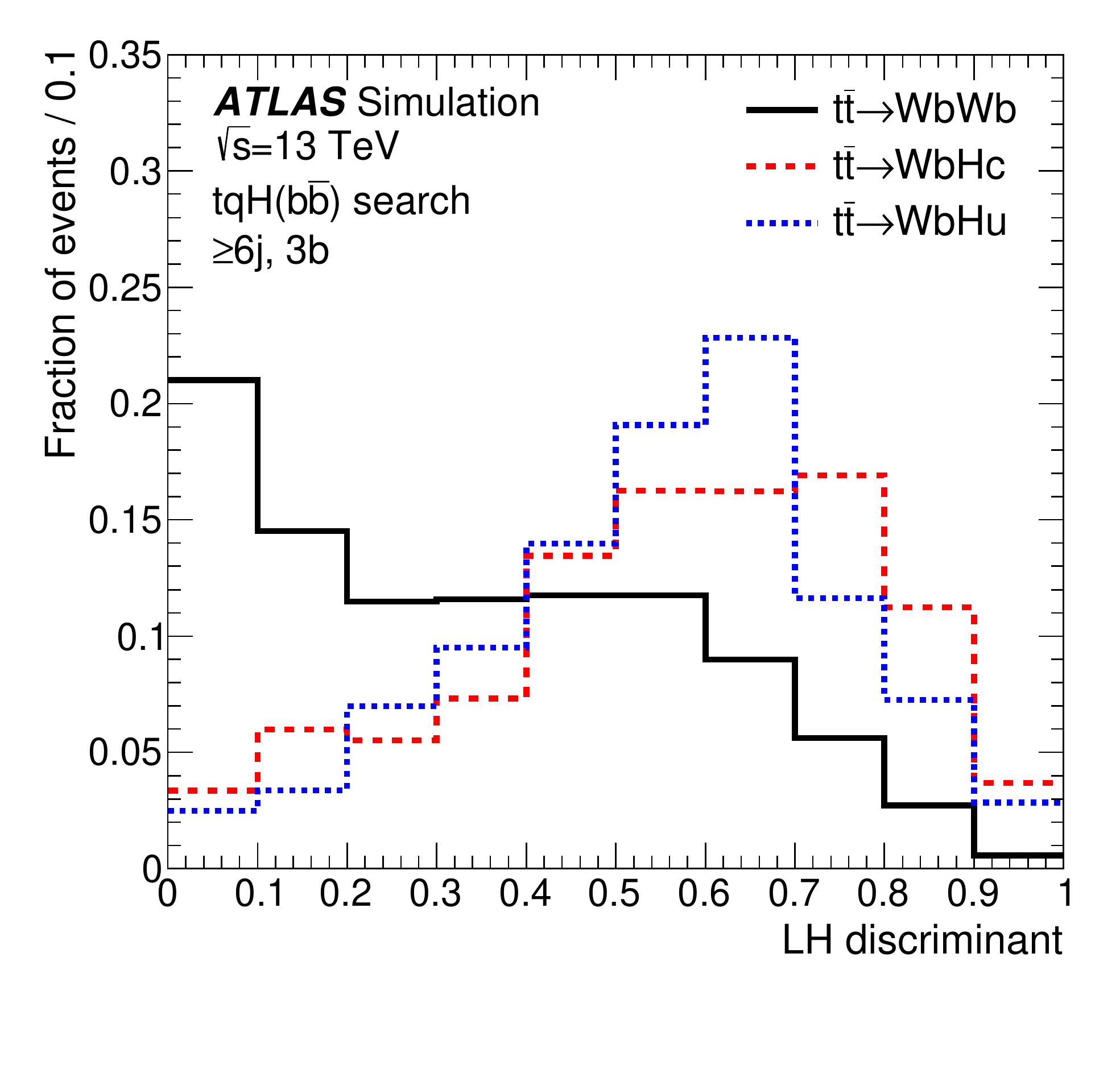}}
\subfloat[]{\includegraphics[width=0.33\textwidth]{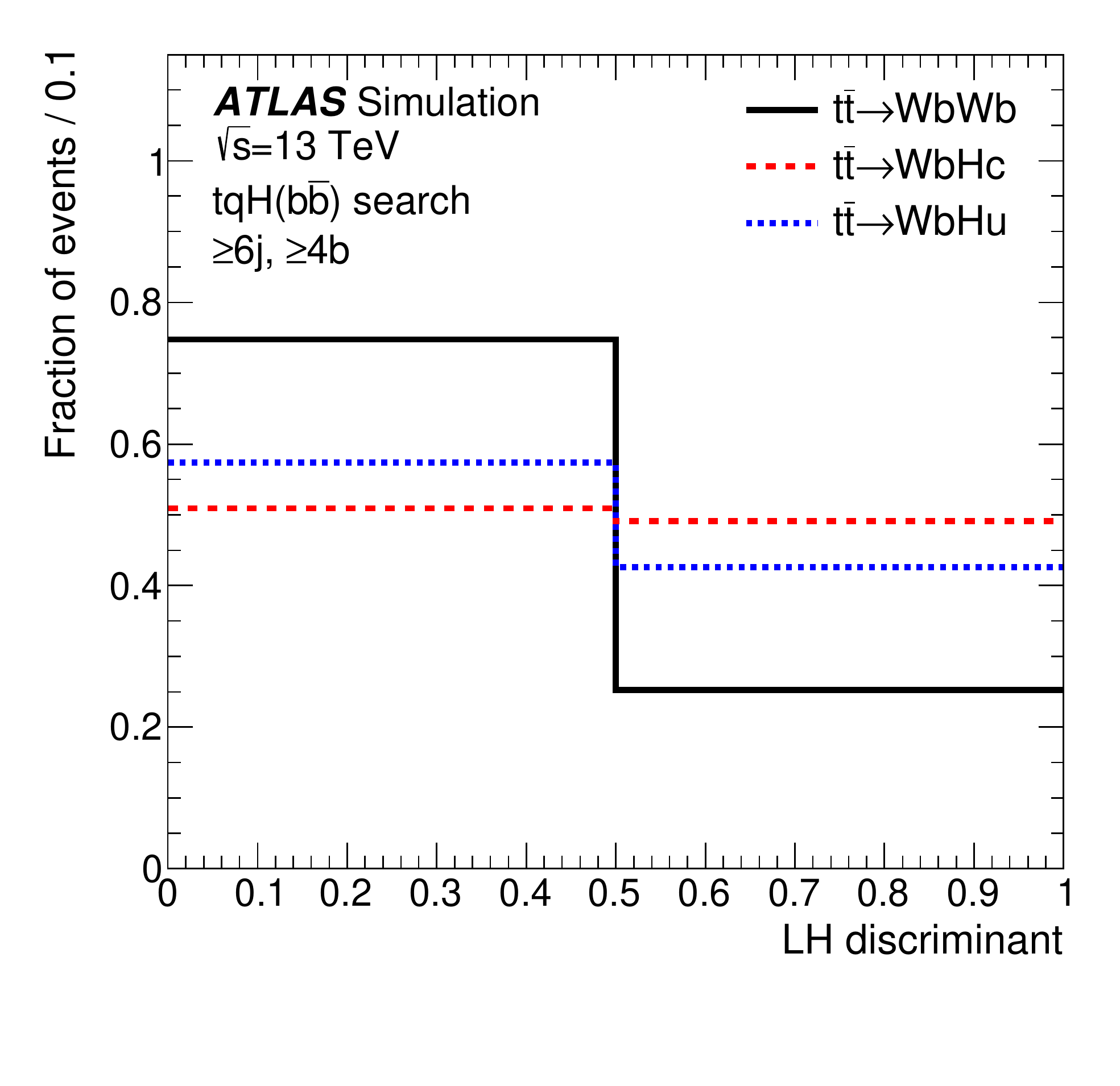}} \\
\caption{$\Hbb$ search: Comparison of the distributions of the LH discriminant after preselection
of the $\Hc$ (red dashed) and $\Hu$ (blue dotted) signals,
and the $t\bar{t}\to WbWb$ background (black solid) in different regions considered in the analysis:
(a) (4j, 2b), (b) (4j, 3b), (c) (4j, 4b), (d) (5j, 2b), (e) (5j, 3b), (f) (5j, $\geq$4b), (g) ($\geq$6j, 2b),
(h) ($\geq$6j, 3b), and (i) ($\geq$6j, $\geq$4b).
In the regions with $\geq$4 $b$-tagged jets, the signal acceptance is small, which translates
into a small number of events for the simulated samples. Therefore, only two bins are used for these distributions.}
\label{fig:LHD}
\end{center}
\end{figure*}

\FloatBarrier
 
\section{Strategy for the $\Htautau$ search}
\label{sec:strategy_Htautau}
 
The analysis strategy adopted in the $\Htautau$ search closely follows that developed in Ref.~\cite{Chen:2015nta} and is summarised in this section.
 
\subsection{Event categorisation and kinematic reconstruction}
\label{sec:htautau_reco_cat}
 
In the $\Htautau$ search, the $\ttbar \to WbHq$ signal being probed is characterised by the presence of $\tau$-leptons from the decay of
the Higgs boson and at least four jets, only one of which originates from a $b$-quark.
If one of the $\tau$-leptons decays leptonically, an isolated electron or muon and significant $\met$ is also expected.
However, in a significant fraction of the events the lowest-$\pt$ jet from the $W$ boson decay fails the minimum $\pt$ requirement of $30~\gev$,
resulting in signal events with only three jets reconstructed.
In order to optimise the sensitivity of the search, the selected events are categorised into four SRs depending
on the number of $\taulep$ and $\had$ candidates, and on the number of jets:
($\lephad$, 3j), ($\lephad$, $\geq$4j), ($\hadhad$, 3j), and ($\hadhad$, $\geq$4j).
 
This event categorisation is primarily motivated by the different quality of the event kinematic reconstruction, depending on the amount
of $\met$ in the event (larger in $\lephad$ events compared with $\hadhad$ events), and whether a jet from the hadronic top-quark decay
is missing or not (events with exactly three jets or at least four jets).
The event kinematic reconstruction is based on the strategy used in Ref.~\cite{Chen:2015nta}, and is summarised below.
 
Events with exactly three jets that are compatible with having a fully reconstructed hadronically decaying
top quark ($t \to Wb \to qqb$) are rejected, as the $t \to Hq$ decay cannot be reconstructed due to the missing light-quark jet.
This compatibility is assessed via a likelihood function that depends on the reconstructed mass of the three-jet
system and the two non-$b$-tagged jets.
For the remaining events, the selected jets are assigned to the different top-quark decay products via a criterion based on
minimising a sum of angular distances between objects. Finally, the four-momenta of the invisible decay products for each $\tau$-lepton decay
are estimated by minimising a $\chi^2$ function based on the probability density functions for the angular distance of the visible and invisible
products of the $\tau$-lepton decay, and including Gaussian constraints on the $\tau$-lepton mass, the Higgs boson mass and the
measured $\met$ within their expected resolutions. The resolution on the $\tau$-lepton mass and the Higgs boson mass are taken to be
$1.8~\gev$ and $20~\gev$, respectively, while the resolution on the measured $\met$ is parameterised as a linear function of
$\sqrt{\sum E_T}$, with $\sum E_T$ denoting the scalar sum of the $\pt$ of all physics objects contributing to the $\met$ reconstruction~\cite{Aaboud:2018tkc}.
After the $\chi^2$ minimisation, the Higgs boson four-momentum, and hence its invariant mass, as well as the
four-momentum of the parent top quark, are determined with better resolution. Following the event kinematic reconstruction, several kinematic variables
that discriminate between signal and background are defined. These variables are used in the multivariate analysis discussed in the next section.

\subsection{Multivariate discriminant}
 
\begin{table*}[t!]
\caption{\small{$\Htautau$ search: Discriminating variables used in the training of the BDT for each search region (denoted by $\times$).
The description of each variable is provided in the text.}}
\begin{center}
\begin{tabular}{ccccc}
\toprule\toprule
& \multicolumn{2}{c}{$\lephad$} & \multicolumn{2}{c}{$\hadhad$} \\
Variable & 3j & $\geq$4j & 3j & $\geq$4j  \\
\midrule
$m_{\tau\tau}^{\text{fit}}$                      	& $\times$  & $\times$  & $\times$  & $\times$ \\
$m_{Hq}$                                	& $\times$  & $\times$  & $\times$  & $\times$ \\
$m_{\text{T,lep}}$                              	& $\times$  & $\times$  &             & \\
$p_{\text{T,1}}$                             	& $\times$  & $\times$ & $\times$  & $\times$ \\
$p_{\text{T,2}}$                             	& $\times$  & $\times$  & $\times$  & $\times$ \\
$\met$ $\phi$ centrality                             	& $\times$  & $\times$  & $\times$  & $\times$ \\
$E_{\text{T},\parallel}^{\text{miss}}$          	& $\times$  & $\times$  & $\times$  & $\times$ \\
$E_{\text{T},\perp}^{\text{miss}}$          	& $\times$  & $\times$  &             & \\
$m_{b j_1}$                   		        & $\times$  & $\times$  & $\times$  & $\times$ \\
$m_{\text{lep}j}$      			        & $\times$  & $\times$  &   	  & \\
$m_{\tau j}$      			        & $\times$  & $\times$  &   	  & \\
$x_1^{\text{fit}}$				& $\times$  &	$\times$  & $\times$  & $\times$ \\
$x_2^{\text{fit}}$				& $\times$  & $\times$  & $\times$  & $\times$ \\
$m_{b j_1 j_2 }$             	   		&             & $\times$  &             & $\times$ \\
\bottomrule\bottomrule
\end{tabular}
\label{tab:mva_var}
\end{center}
\end{table*}
 
Boosted decision trees are used in each SR to improve the separation between signal and background.
In the training, only $\ttbar \to W(qq)bH(\tau\tau)q$ signal events are used against the total SM background (including both real and fake $\had$ contributions),
whereas to obtain the result the contributions from $\ttbar \to W(\ell\nu)bHq$ signal events are also taken into account.
 
A large set of potential variables were investigated in each SR separately, and only those variables that led to better discrimination
by the BDT were kept. The discrimination of a given variable was quantified by the ``separation" and ``importance" measures provided
by the TMVA package~\cite{Hocker:2007ht}.
The BDT input variables in each SR are listed in Table~\ref{tab:mva_var} and defined in the following:
 
\begin{itemize}
\item $m_{\tau\tau}^{\text{fit}}$: the invariant mass of the two $\tau$-lepton candidates after the reconstruction of the neutrinos, indicating the reconstructed Higgs boson mass.
\item $m_{Hq}$: the invariant mass of the reconstructed Higgs boson and the associated light-quark jet in the $t \to Hq$ decay, corresponding to the reconstructed mass of the parent top quark.
\item $m_{\text{T,lep}}$: the transverse mass calculated from the lepton and $\mpt$ in the $\lephad$ channel.
\item $p_{\text{T,1}}$ and $p_{\text{T,2}}$:  the transverse momenta of the lepton and $\had$ candidate (referred to as particles 1 and 2 respectively) in the $\lephad$ channel, or the transverse momenta of the leading and trailing $\had$ candidates (referred to as particles 1 and 2 respectively) in the $\hadhad$ channel.
\item $\met$ $\phi$ centrality: a variable that quantifies the angular position of $\mpt$ relative to the visible $\tau$-lepton decay products in the transverse plane. It is defined as:
\begin{equation*}
\met\; \phi\; \mathrm{centrality} = \frac{\sin(\phi_\mathrm{miss}-\phi_1)+\sin(\phi_\mathrm{miss}-\phi_2)}{\sqrt{\sin^2(\phi_\mathrm{miss}-\phi_1)+\sin^2(\phi_\mathrm{miss}-\phi_2)}}
\end{equation*}
\noindent where $\phi_\mathrm{miss}$ denotes the azimuthal angle of $\mpt$, and $\phi_1$ and $\phi_2$ denote the azimuthal angles the two $\tau$-lepton candidates
(the lepton and $\had$ candidate in the $\lephad$ channel, or the leading and trailing $\had$ candidates in the $\hadhad$ channel), referred to as particles 1 and 2 respectively.
\item $E_{\text{T},\parallel}^{\text{miss}}$: the magnitude of the projection of the original $\mpt$ vector parallel to the fitted $\mpt$ vector, minus
the magnitude of the fitted $\mpt$ vector.
\item $E_{\text{T},\perp}^{\text{miss}}$: the magnitude of the projection of the original $\mpt$ vector perpendicular to the fitted $\mpt$ vector.
\item $m_{b j_1}$: the invariant mass of the $b$-jet and the leading jet candidate from the hadronically decaying $W$ boson.
\item $m_{\text{lep}j}$: the invariant mass of the lepton and the jet that has the smallest angular distance to the $\lep$ candidate.
\item $m_{\tau j}$: the invariant mass of the  $\had$ candidate and the jet that has the smallest angular distance to the $\had$ candidate.
\item $x_{1}^{\text{fit}}$ and $x_{2}^{\text{fit}}$: the momentum fractions carried by the visible decay products from the two $\tau$-lepton candidates
(whether $\taulep$ or $\had$) per event. It is based on the best-fit four-momentum of the neutrino(s) according to the event reconstruction procedure outlined in Section~\ref{sec:htautau_reco_cat}.
\item $m_{bj_1j_2}$: the invariant mass of the $b$-jet and the two jets originating from the $W$ boson in the $t\to Wb \to j_1j_2b$ decay, corresponding to the reconstructed mass of the parent top quark. This variable is only defined for events with at least four jets.
\end{itemize}
 
Among these variables, the most discriminating are $m_{\tau\tau}^{\text{fit}}$, $p_{\text{T},2}$, $x_{1}^{\text{fit}}$ and $x_{2}^{\text{fit}}$. A comparison between data and the predicted background for some of these variables in each of the SRs considered is shown in Figures~\ref{fig:BDT_inputs_lephad} and~\ref{fig:BDT_inputs_hadhad}.
A good description of the data by the background model is observed in all cases.
The level of discrimination between signal and background achieved by the BDTs is illustrated in Figure~\ref{fig:BDT}.
 
\begin{figure*}[t]
\begin{center}
\subfloat[]{\includegraphics[width=0.40\textwidth]{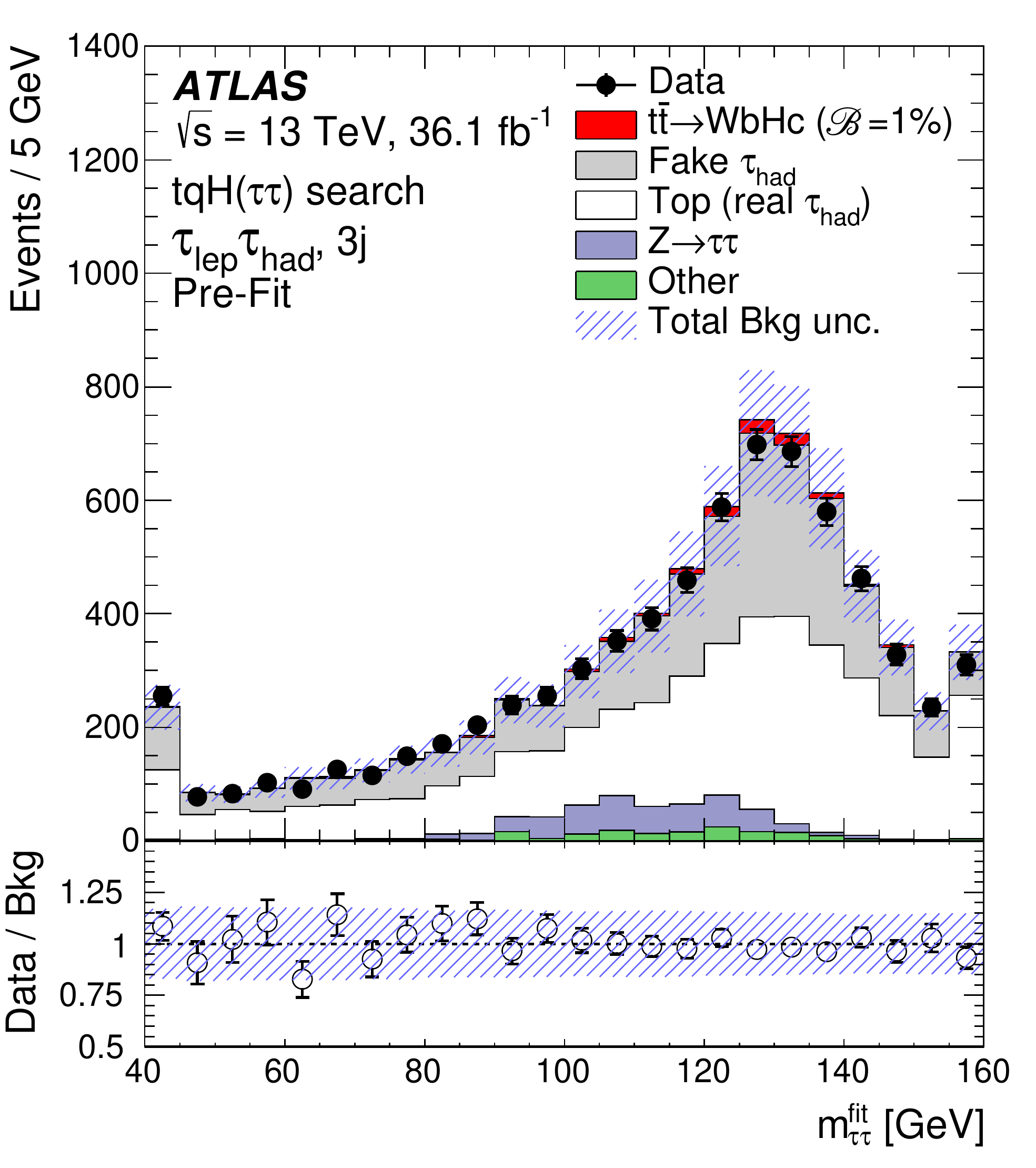}}
\subfloat[]{\includegraphics[width=0.40\textwidth]{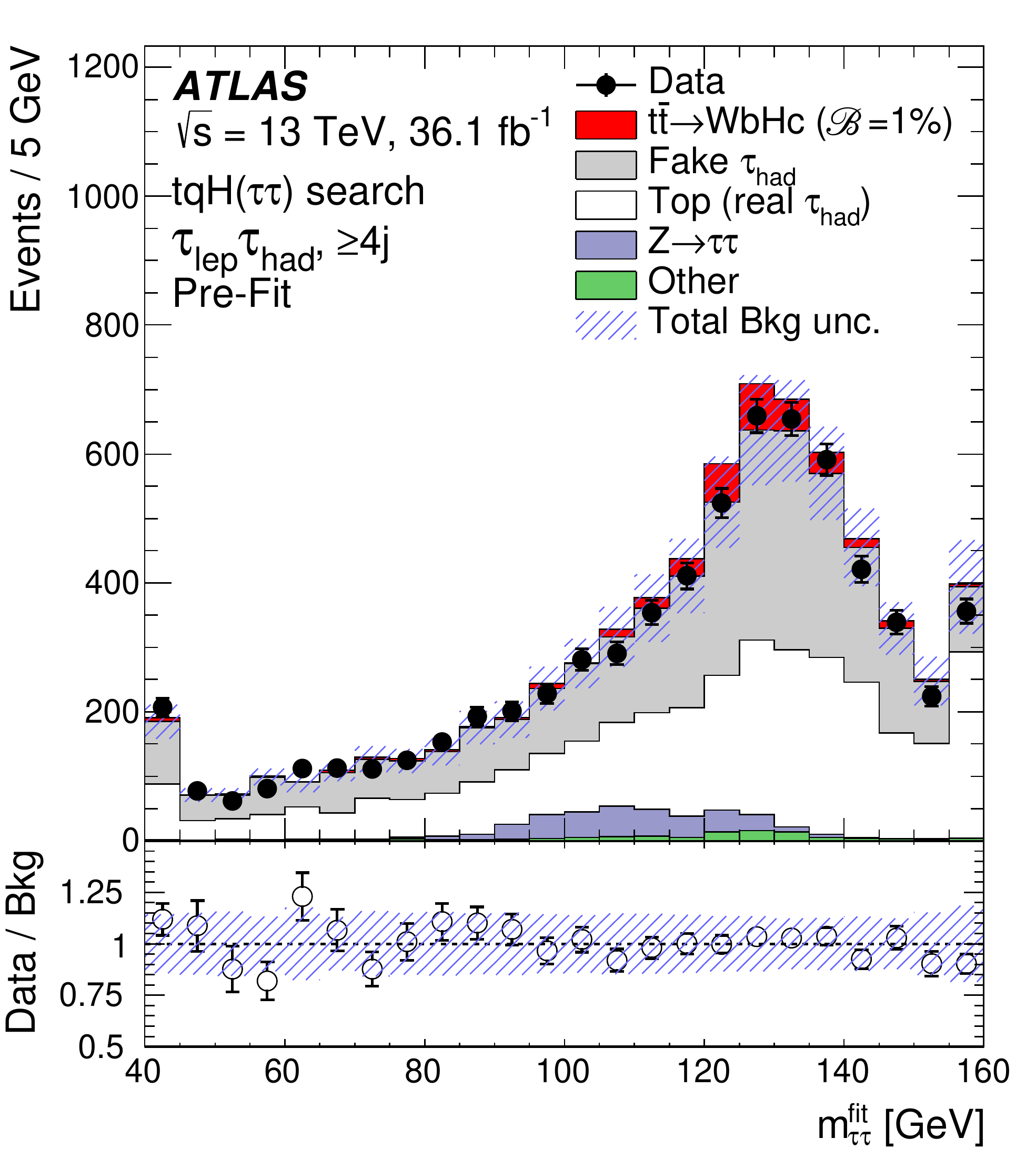}} \\
\subfloat[]{\includegraphics[width=0.40\textwidth]{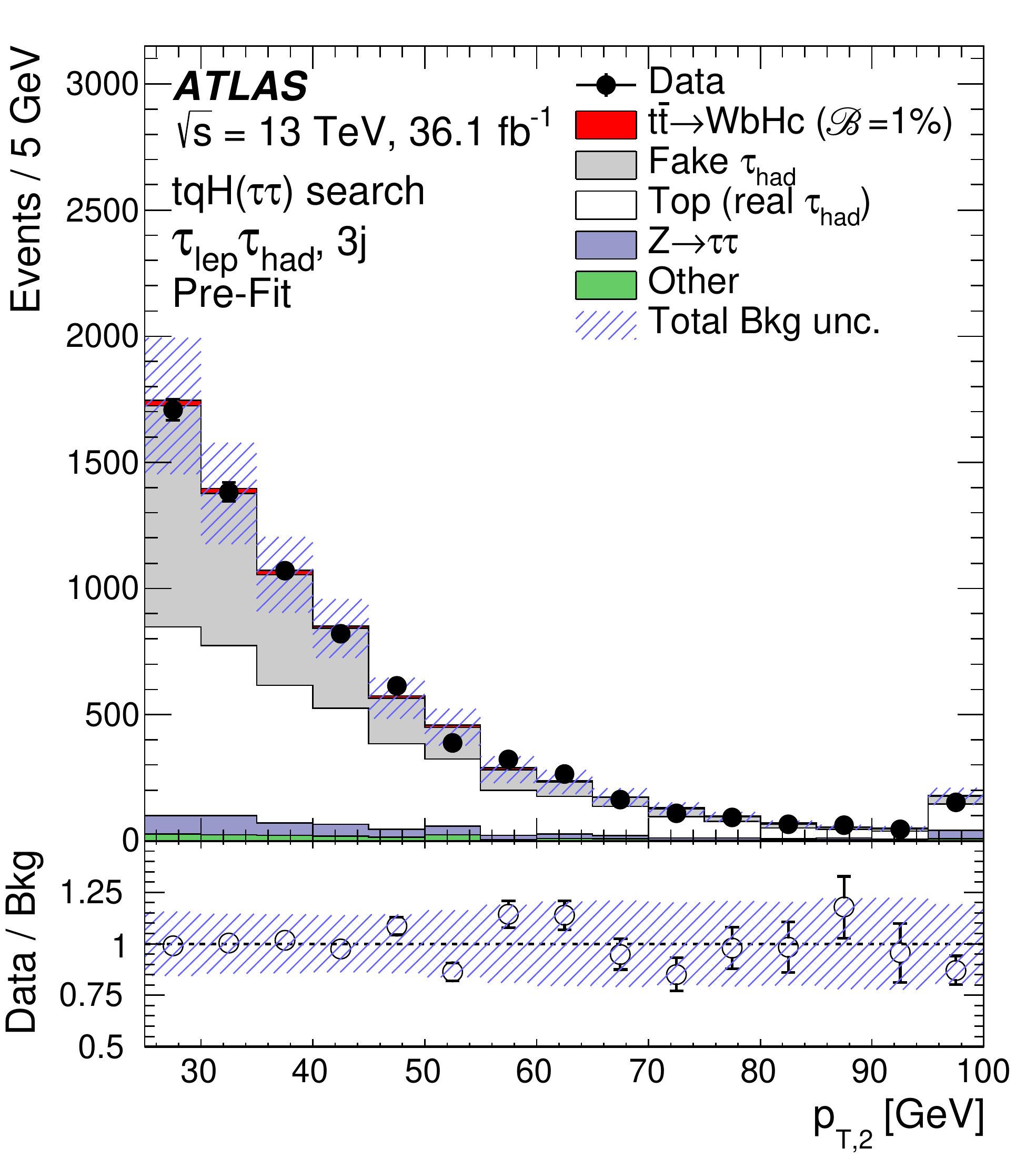}}
\subfloat[]{\includegraphics[width=0.40\textwidth]{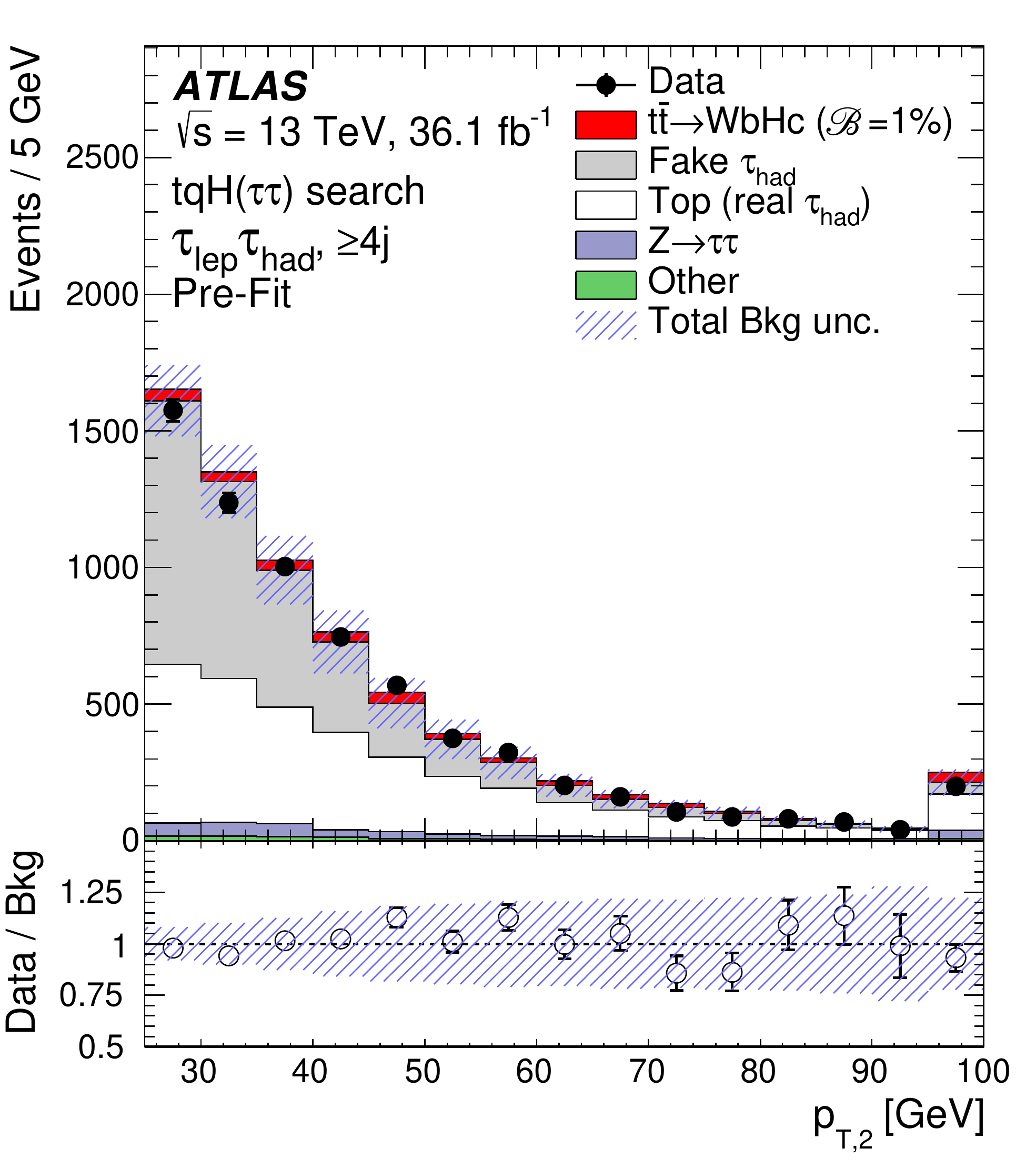}} \\
\caption{$\Htautau$ search: Comparison between the data and predicted background after preselection for the distributions of two of the most
discriminating BDT input variables in the $\lephad$ channel before the fit to data (``Pre-Fit''). The distributions are shown for
$m_{\tau\tau}^{\text{fit}}$ in (a) the ($\lephad$, 3j) region and (b) the ($\lephad$, $\geq$4j) region, and for
$p_{\text{T},2}$ in (c) the ($\lephad$, 3j)  region and (d) the ($\lephad$, $\geq$4j) region.
The contributions with real $\had$ candidates from $\ttbar$,  $\ttbar V$, $\ttbar H$, and single-top-quark backgrounds are combined into
a single background source referred to as ``Top (real $\had$)'', whereas the small contributions from
$Z\to \ell^+\ell^-$ ($\ell = e, \mu$) and diboson backgrounds are combined into ``Other''.
The expected $\Hc$ signal (solid red) corresponding to $\BR(t\to Hc)=1\%$ is also shown,
added to the background prediction.
The first and the last bins in all figures contain the underflow and overflow respectively.
The bottom panel displays the ratio of data to the SM background (``Bkg'') prediction.
The hashed area represents the total uncertainty of the background, excluding the normalisation uncertainty of the fake $\had$ background,
which is determined via a likelihood fit to data.}
\label{fig:BDT_inputs_lephad}
\end{center}
\end{figure*}
 
\begin{figure*}[t]
\begin{center}
\subfloat[]{\includegraphics[width=0.40\textwidth]{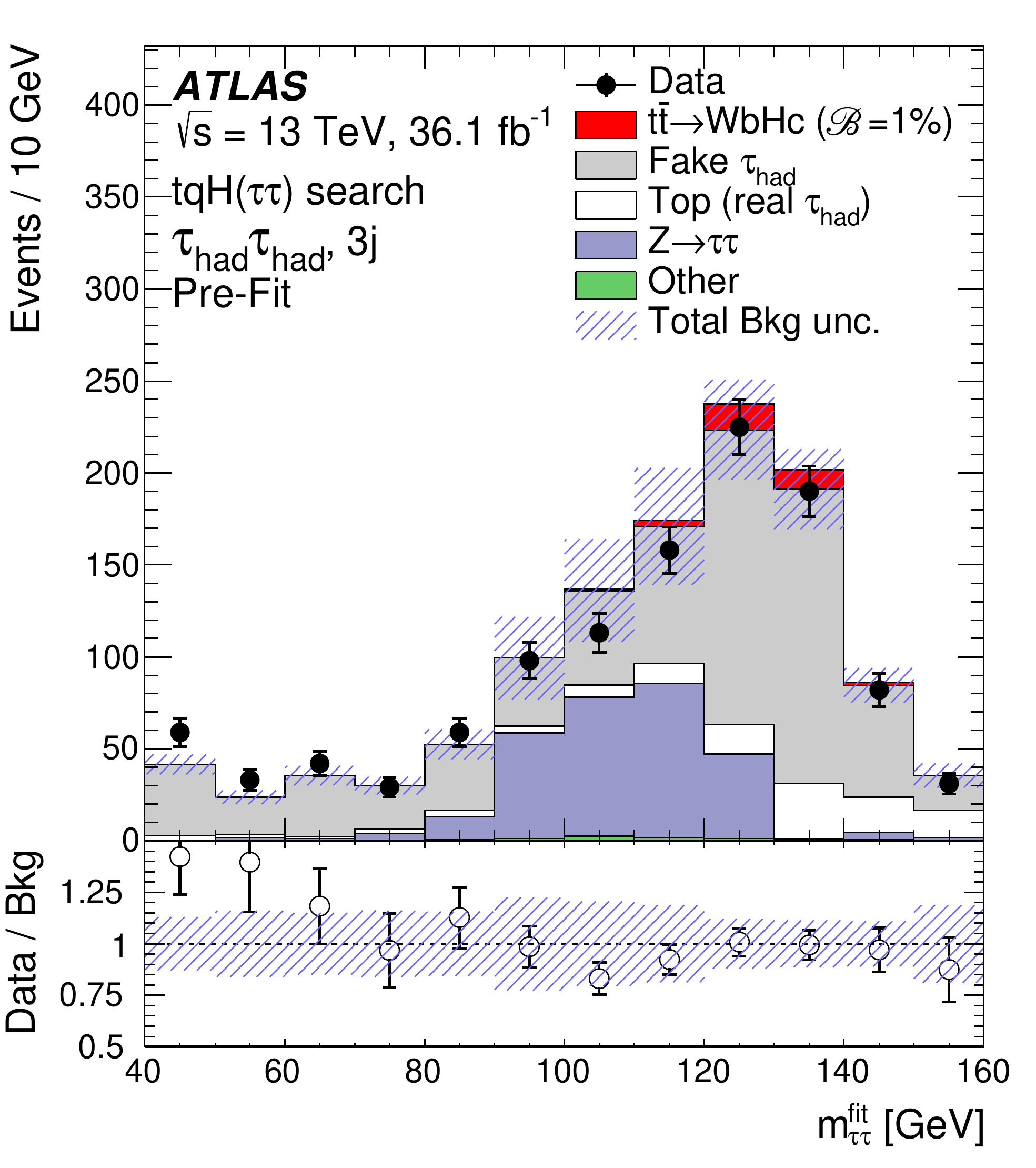}}
\subfloat[]{\includegraphics[width=0.40\textwidth]{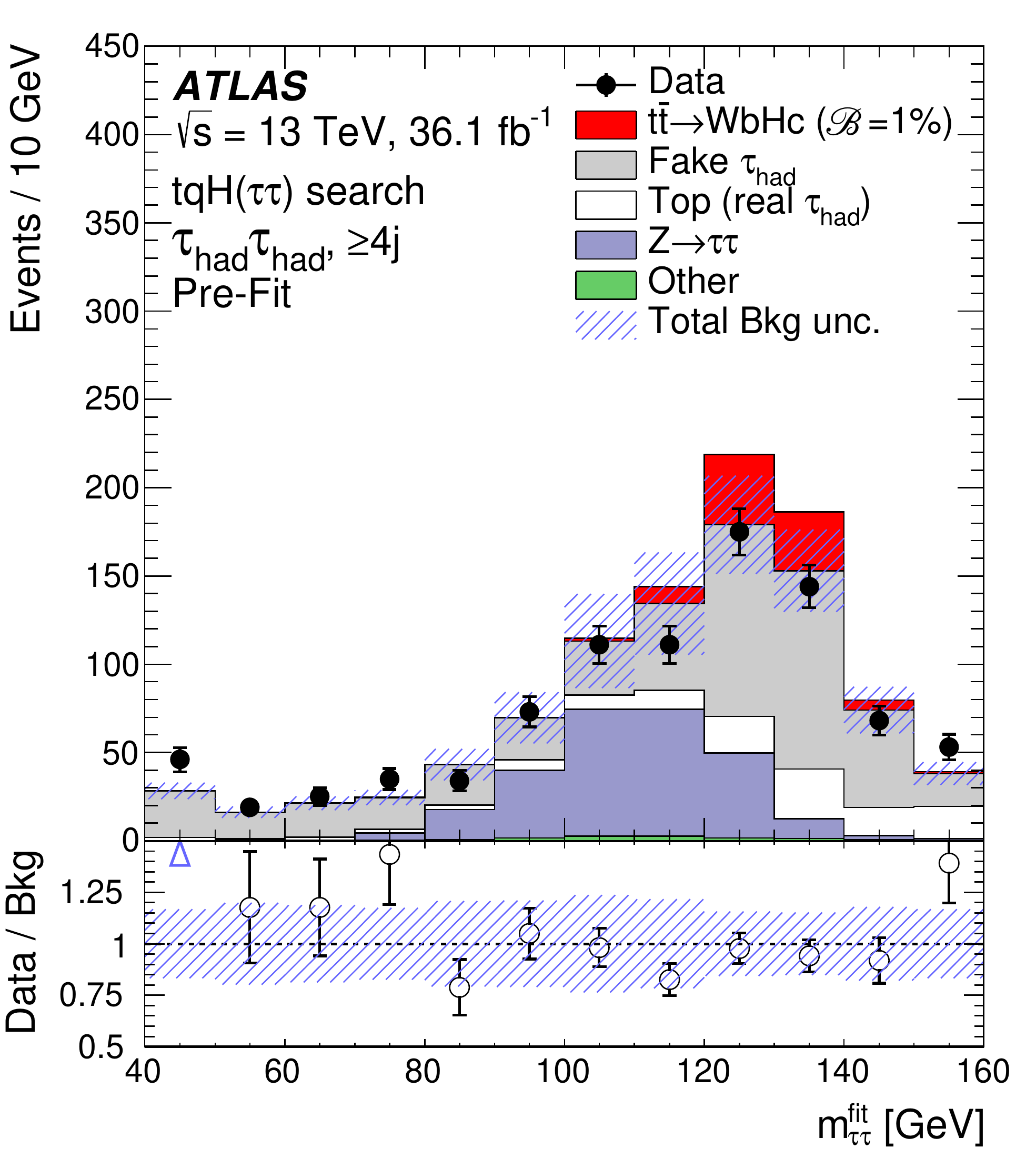}} \\
\subfloat[]{\includegraphics[width=0.40\textwidth]{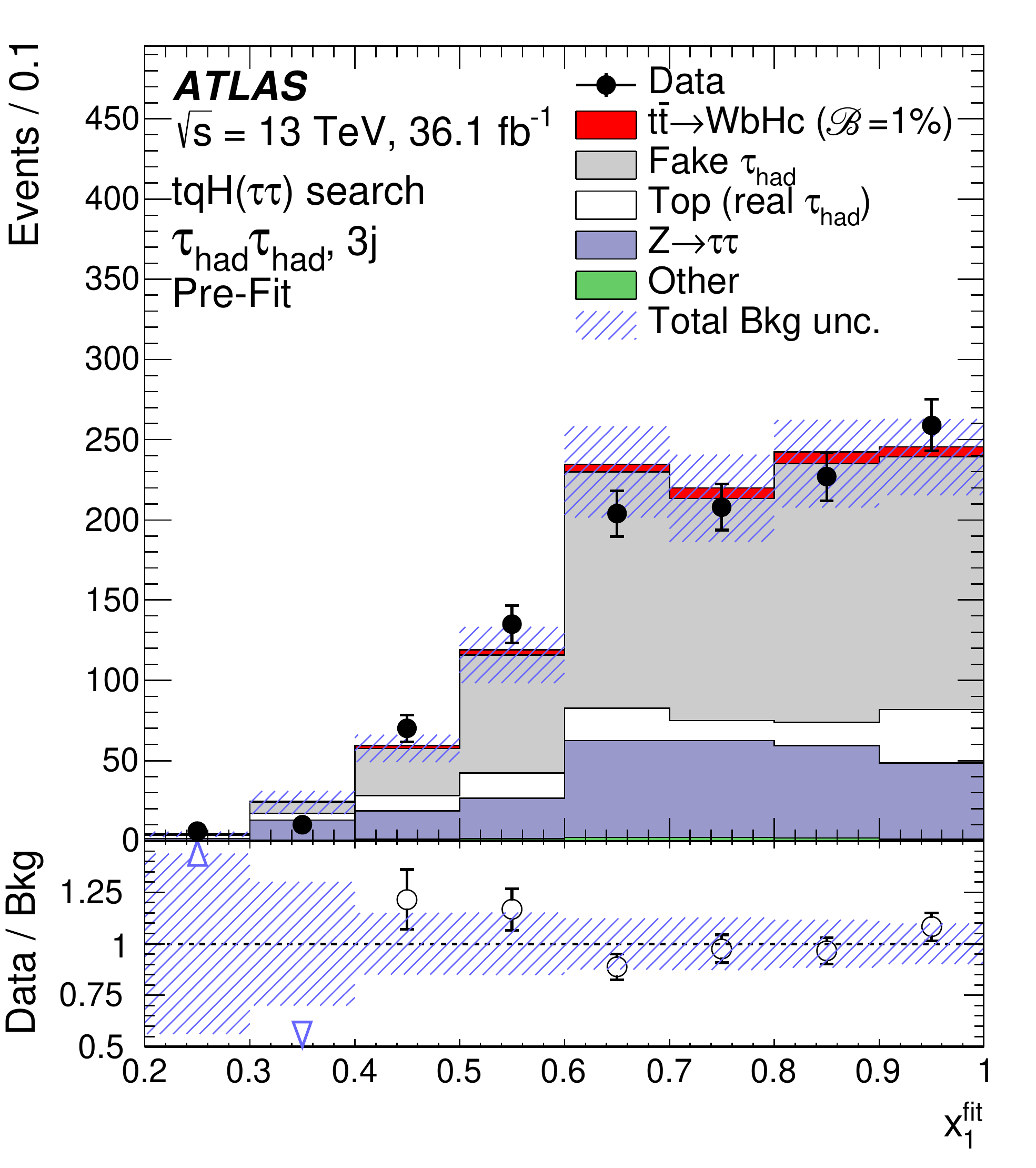}}
\subfloat[]{\includegraphics[width=0.40\textwidth]{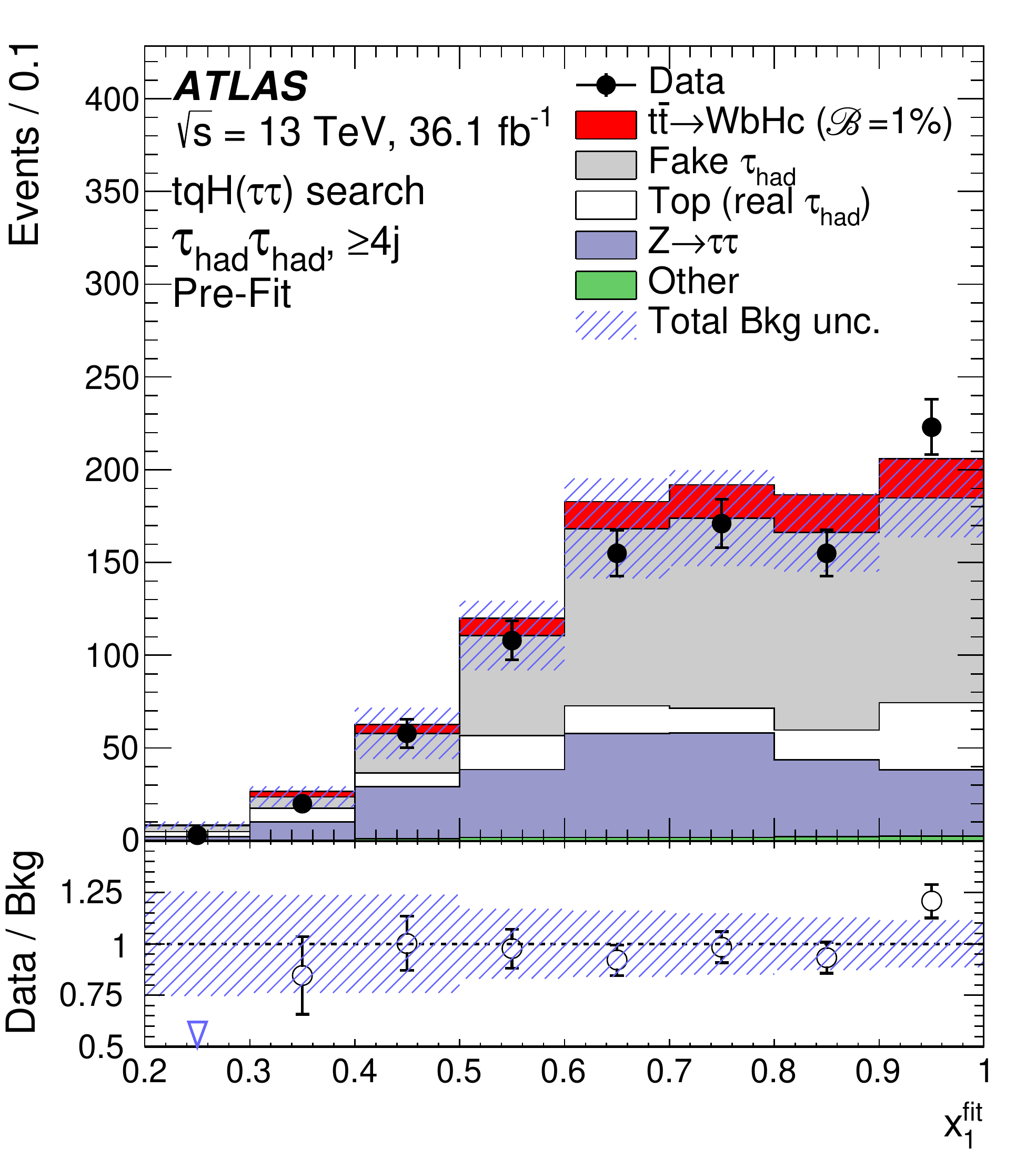}} \\
\caption{$\Htautau$ search: Comparison between the data and predicted background after preselection for the distributions of two of the most
discriminating BDT input variables in the $\hadhad$ channel before the fit to data (``Pre-Fit''). The distributions are shown for
$m_{\tau\tau}^{\text{fit}}$ in (a) the ($\hadhad$, 3j) region and (b) the ($\hadhad$, $\geq$4j) region, and for
$x_{1}^{\text{fit}}$ in (c) the ($\hadhad$, 3j)  region and (d) the ($\hadhad$, $\geq$4j) region.
The contributions with real $\had$ candidates from $\ttbar$,  $\ttbar V$, $\ttbar H$, and single-top-quark backgrounds are combined into
a single background source referred to as ``Top (real $\had$)'', whereas the small contributions from
$Z\to \ell^+\ell^-$ ($\ell = e, \mu$) and diboson backgrounds are combined into ``Other''.
The expected $\Hc$ signal (solid red) corresponding to $\BR(t\to Hc)=1\%$ is also shown,
added to the background prediction.
The first and the last bins in the figures in (a) and (b) contain the underflow and overflow respectively.
The bottom panel displays the ratio of data to the SM background (``Bkg'') prediction.
The hashed area represents the total uncertainty of the background, excluding the normalisation uncertainty of the fake $\had$ background,
which is determined via a likelihood fit to data.}
\label{fig:BDT_inputs_hadhad}
\end{center}
\end{figure*}
 
\begin{figure*}[t]
\begin{center}
\subfloat[]{\includegraphics[width=0.40\textwidth]{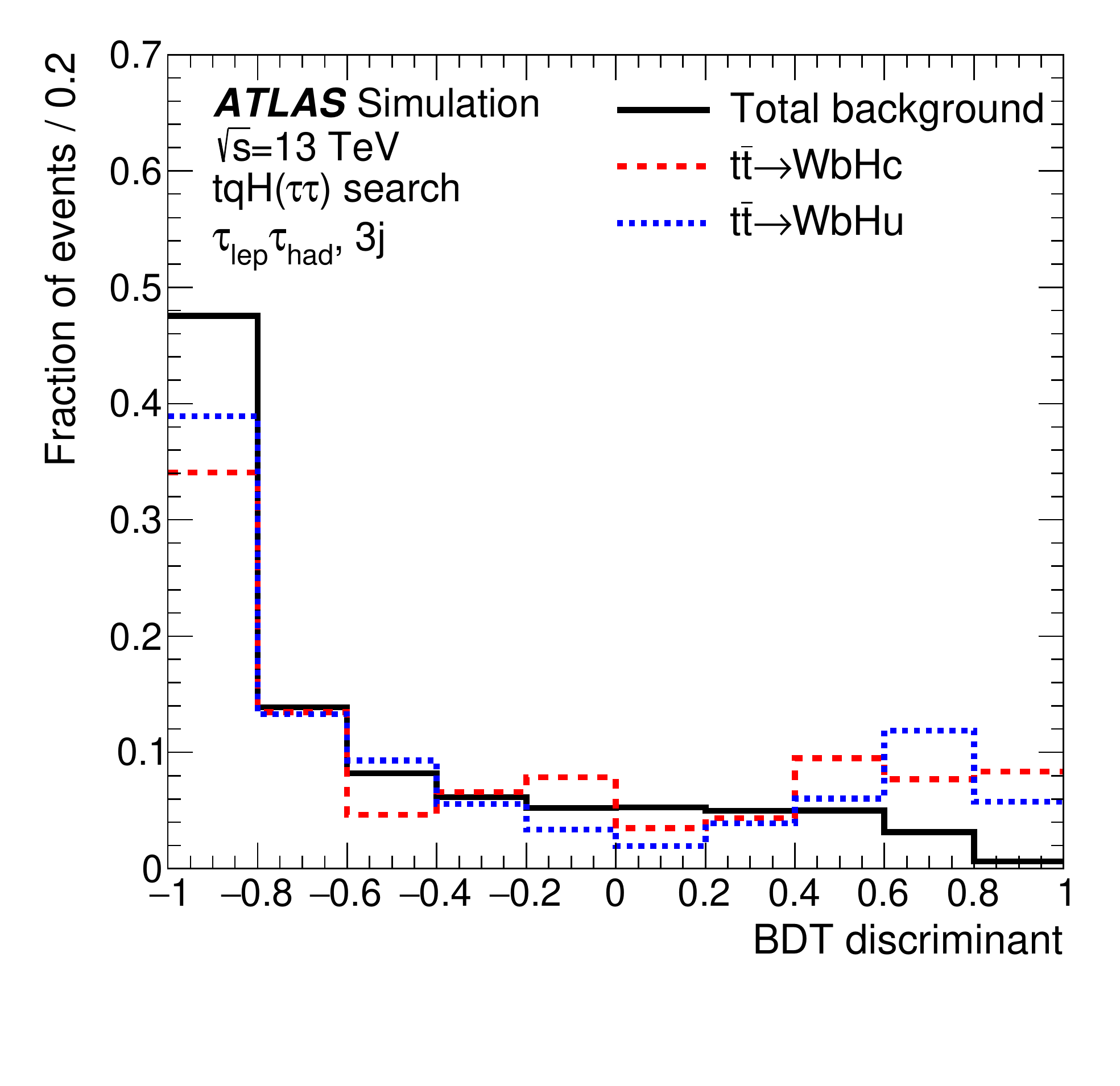}}
\subfloat[]{\includegraphics[width=0.40\textwidth]{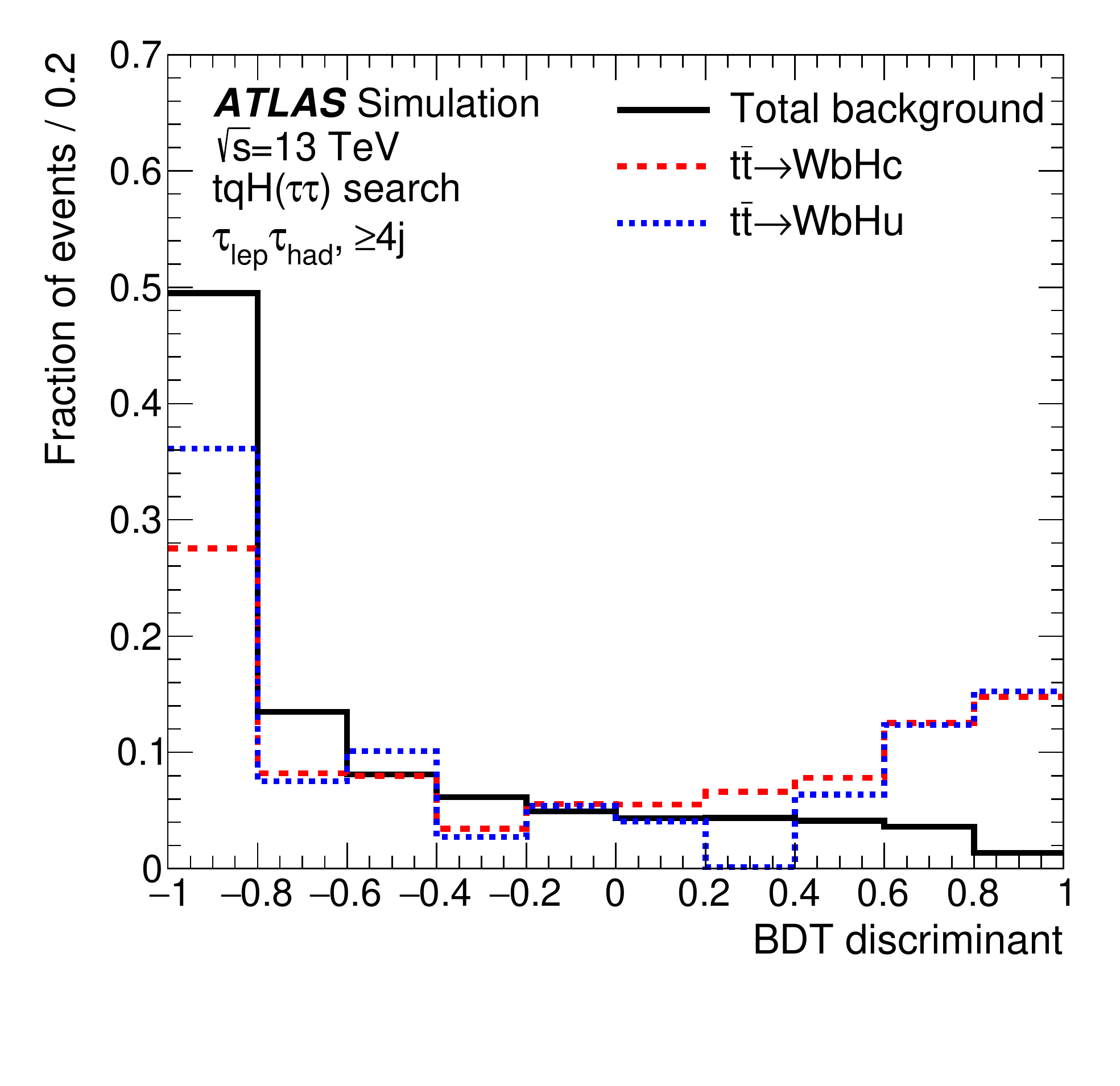}} \\
\subfloat[]{\includegraphics[width=0.40\textwidth]{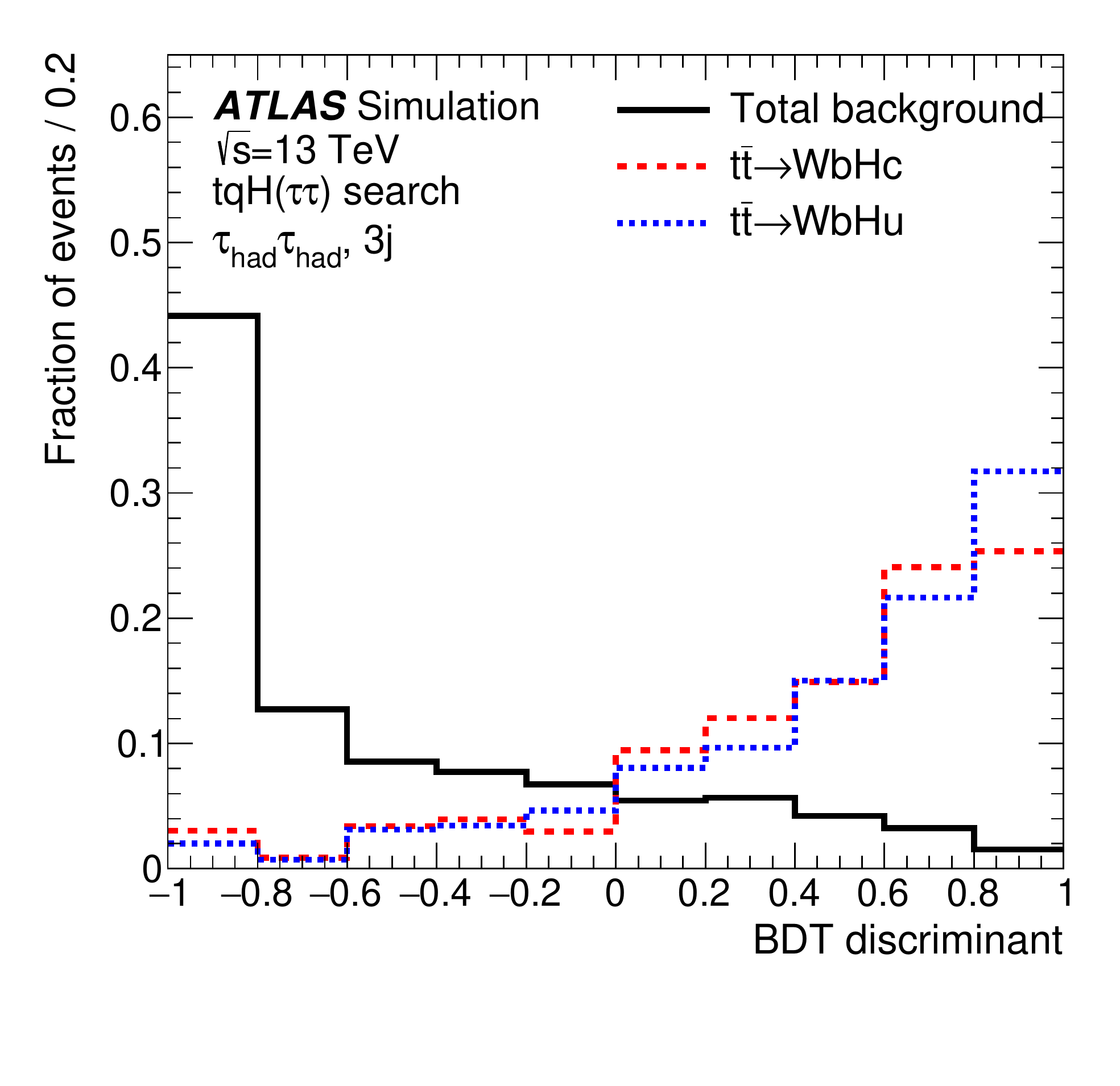}}
\subfloat[]{\includegraphics[width=0.40\textwidth]{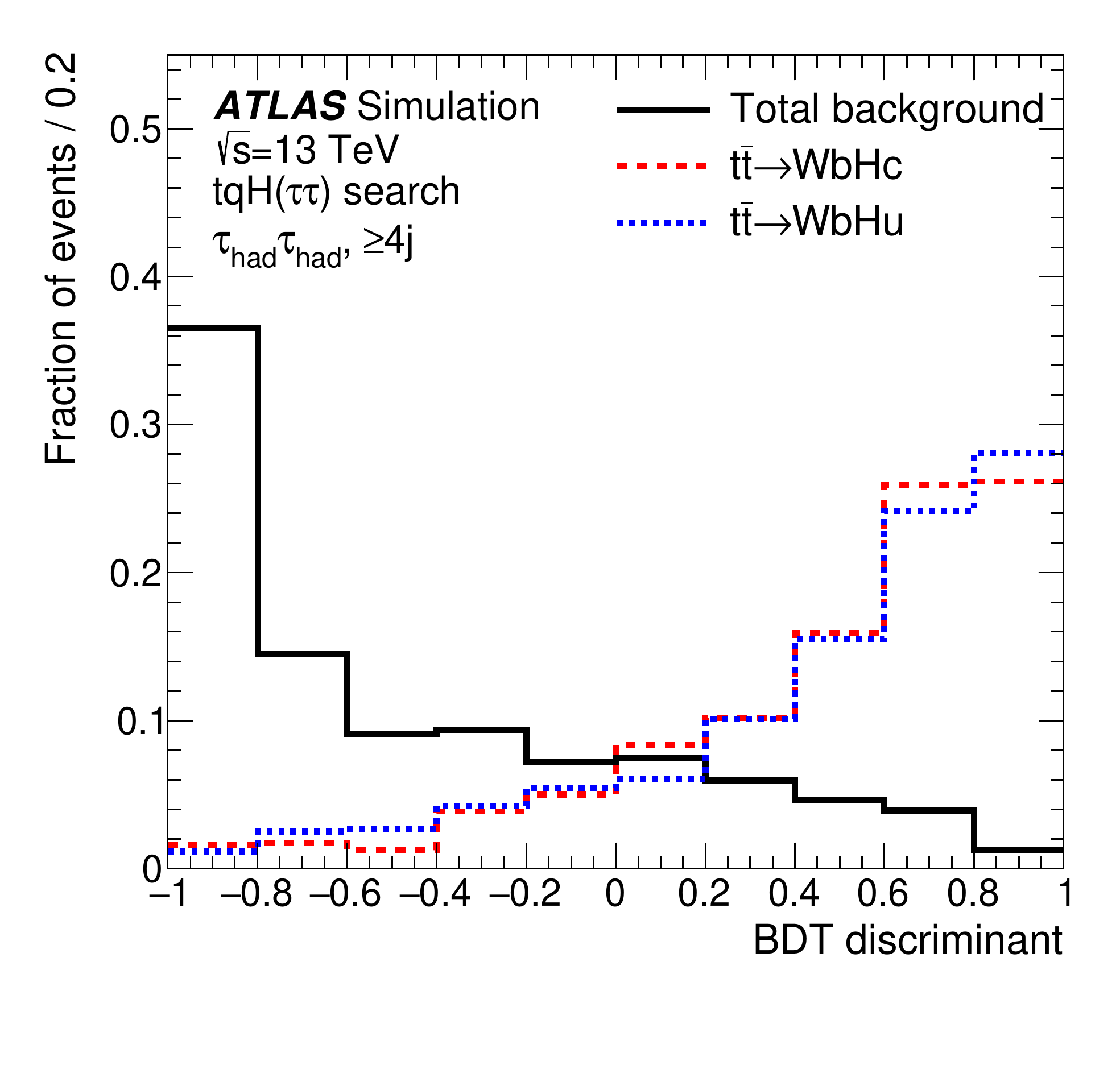}} \\
\caption{$\Htautau$ search: Comparison of the distributions of the BDT discriminant after preselection of the $\Hc$ (red dashed) and $\Hu$ (blue dotted) signals,
and the total background (black solid) in the different search regions considered:
(a) ($\lephad$, 3j), (b) ($\lephad$, $\geq$4j), (c) ($\hadhad$, 3j), and (d) ($\hadhad$, $\geq$4j). }
\label{fig:BDT}
\end{center}
\end{figure*}
\FloatBarrier
 
\section{Systematic uncertainties}
\label{sec:systematics}
 
Several sources of systematic uncertainty that can affect the normalisation of signal
and background and/or the shape of their corresponding discriminant distributions are considered.
Each source is considered to be uncorrelated with the other sources.
Correlations of a given systematic uncertainty are maintained across processes and channels
as appropriate.
The following sections describe the systematic uncertainties considered.

\subsection{Luminosity}
\label{sec:syst_lumi}
 
The uncertainty in the integrated luminosity is 2.1\%, affecting the overall normalisation of all processes estimated from the simulation.
It is derived, following a methodology similar to that detailed in Ref.~\cite{Aaboud:2016hhf}, and using the LUCID-2 detector
for the baseline luminosity measurements \cite{Avoni:2018iuv}, from a calibration of the luminosity scale using $x$--$y$ beam-separation scans.
 
\subsection{Reconstructed objects}
\label{sec:syst_objects}

Uncertainties associated with electrons, muons, and $\had$ candidates arise from the trigger, reconstruction,
identification and isolation (in the case of electrons and muons) efficiencies, as well as the momentum scale and resolution.
These are measured using $Z\to \ell^+\ell^-$ and $J/\psi\to \ell^+\ell^-$ events ($\ell =e, \mu$)~\cite{ATLAS-CONF-2016-024,Aad:2016jkr}
in the case of electrons and muons, and using $Z\to \tau^+\tau^-$ events in the case of $\had$ candidates~\cite{ATLAS-CONF-2017-029}.
 
Uncertainties associated with jets arise from the jet energy scale
and resolution, and the efficiency to pass the JVT requirements.
The largest contribution results from the jet energy scale, whose uncertainty dependence on jet $\pt$ and $\eta$, jet flavour, and pile-up treatment,
is split into 21 uncorrelated components that are treated independently~\cite{Aaboud:2017jcu}.
 
Uncertainties associated with energy scales and resolutions of leptons and jets
are propagated to $\met$. Additional uncertainties originating from the modelling
of the underlying event, in particular its impact on the $\pt$ scale and resolution
of unclustered energy, are negligible.
 
Efficiencies to tag $b$-jets and $c$-jets in the simulation are corrected to match the efficiencies in data by $\pt$-dependent factors,
whereas the light-jet efficiency is scaled by $\pt$- and $\eta$-dependent factors.
The $b$-jet efficiency is measured in a data sample enriched in $\ttbar$ events~\cite{Aaboud:2018xwy}, while the $c$-jet efficiency is measured
using $\ttbar$ events~\cite{ATLAS-CONF-2018-001} or $W$+$c$-jet events~\cite{Aad:2015ydr}.
The light-jet efficiency is measured in a multijet data sample enriched in light-flavour jets~\cite{ATLAS-CONF-2018-006}.
Since the $\ttbar$ sample used to measure the $c$-jet tagging efficiency overlaps with the analysis sample, the $\Hbb$ search uses
instead the $W$+$c$-jet scale factors.
In the case of the $\Hbb$ ($\Htautau$) search, the uncertainties in these scale factors include
a total of 6 independent sources affecting $b$-jets, 1 (2) source(s) affecting $c$-jets, and 17 sources affecting light-jets.
These systematic uncertainties are taken as uncorrelated between $b$-jets, $c$-jets, and light-jets.
An additional uncertainty is included due to the extrapolation of these corrections to jets
with $\pt$ beyond the kinematic reach of the data calibration samples used ($\pt>300~\gev$ for $b$- and $c$-jets,
and $\pt>750~\gev$ for light-jets); it is taken to be correlated among the three jet flavours.
Since the fraction of signal and background in this kinematic regime is very small, these uncertainties have a negligible
impact in the analyses.
Finally, an uncertainty related to the application of $c$-jet scale factors to $\tau$-jets is considered,
which also has a negligible impact.
 
\subsection{Background modelling}
\label{sec:syst_bkgmodeling}
 
A number of sources of systematic uncertainty affecting the modelling of $t\bar{t}$+jets are considered.
An uncertainty of  6\% is assigned to the inclusive $\ttbar$ production
cross section~\cite{Czakon:2011xx}, including contributions from varying the factorisation and renormalisation
scales, as well as from the top-quark mass, the PDF and $\alpha_{\textrm{S}}$. The latter two represent the largest contribution
to the overall theoretical uncertainty in the cross section and were calculated using the PDF4LHC prescription~\cite{Botje:2011sn}
with the MSTW 2008 68\% CL NNLO, CT10 NNLO~\cite{Lai:2010vv,Gao:2013xoa} and NNPDF2.3 5F FFN~\cite{Ball:2012cx} PDF sets.
The uncertainty associated with the choice of NLO generator is derived by comparing the nominal prediction from
{\powheg}+{\pythiaeight} with a prediction from \textsc{Sherpa}~2.2.1. For the latter, the matrix-element calculation is performed
for up to two partons at NLO and up to four partons at LO using \textsc{Comix} and \textsc{OpenLoops}, and
merged with the {\sherpa} parton shower using the ME+PS@NLO prescription.
The uncertainty due to the choice of parton shower and hadronisation (PS \& Had) model is derived
by comparing the predictions from {\powheg} interfaced either to {\pythiaeight} or {\herwig7}.
The latter uses the MMHT2014 LO~\cite{Harland-Lang:2014zoa} PDF set in combination with the H7UE tune~\cite{Bellm:2015jjp}.
The uncertainty in the modelling of additional radiation is assessed with two alternative {\powheg}+{\pythiaeight} samples:
a sample with increased radiation (referred to as radHi) is obtained by decreasing the renormalisation and factorisation scales
by a factor of two, doubling the $h_{\textrm{damp}}$ parameter, and using the Var3c upward variation of the A14 parameter set;
a sample with decreased radiation (referred to as radLow) is obtained by increasing the scales by a factor of two
and using the Var3c downward variation of the A14 set~\cite{ATL-PHYS-PUB-2016-004}.
 
In the case of the $\Hbb$ search, where the $\ttbar$+HF background plays a prominent role (see Fig.~\ref{fig:Hbb_Summary}), a more
detailed treatment of its associated systematic uncertainties is used. In particular, since several analysis
regions have a sufficiently large number of \ttbin\ background events, its normalisation is
determined in the fit to data.
In the case of the \ttcin\ normalisation, an uncertainty of 50\% is assumed, as the fit to the data is unable
to precisely determine it, and the analysis has very limited sensitivity to this uncertainty.
Since the diagrams that contribute to $\ttbar$+light-jets, \ttcin, and \ttbin\
production are different, all above uncertainties in $\ttbar$+jets
background modelling (NLO generator, PS \& Had, and radHi/radLow), except the uncertainty of the inclusive cross section, are
considered to be uncorrelated among these processes.
Additional uncertainties of the \ttbin\ background are considered associated with the NLO prediction from {\ShOL},
which is used for reweighting the nominal {\powheg}+{\pythiaeight} prediction.
These include three different scale variations, a different shower-recoil model scheme, and
two alternative PDF sets (MSTW 2008 NLO and NNPDF2.3 NLO). Additional uncertainties are assessed for
the contributions to the \ttbin\ background originating from multiple parton interactions.
Finally, an additional uncertainty is assigned to the \ttbin\ background by comparing
the predictions from {\powheg}+{\pythiaeight} and {\ShOL} 4F (5F vs 4F).
In the derivation of the above uncertainties, the overall normalisations of the \ttcin\ and \ttbin\ backgrounds
at the particle level are fixed to the nominal prediction. In order to maintain the inclusive $\ttbar$ cross section,
the normalisation of the $\ttbar$+light-jets background at the particle level is adjusted accordingly.
 
Uncertainties affecting the normalisation of the $V$+jets background are estimated for the sum
of $W$+jets and $Z$+jets, and separately for $V$+light-jets, $V$+$\geq$1$c$+jets, and $V$+$\geq$1$b$+jets subprocesses.
The total normalisation uncertainty of $V$+jets processes is estimated by comparing the data and total background prediction in
the different analysis regions considered, but requiring exactly zero $b$-tagged jets. Agreement between data and predicted background
in these modified regions, which are dominated by $V$+light-jets, is found to be within approximately 30\%. This bound is taken to
be the normalisation uncertainty, correlated across all $V$+jets subprocesses.
Since {\sherpa}~2.2 has been found to underestimate $V$+heavy-flavour production by about a factor
of 1.3~\cite{Aaboud:2017xsd}, additional 30\% normalisation uncertainties are assumed for $V$+$\geq$1$c$+jets and $V$+$\geq$1$b$+jets
subprocesses, considered uncorrelated between them.
 
Uncertainties affecting the modelling of the single-top-quark background include a
$+5\%$/$-4\%$ uncertainty of the total cross section estimated as a weighted average
of the theoretical uncertainties in $t$-, $tW$- and $s$-channel production~\cite{Kidonakis:2011wy,Kidonakis:2010ux,Kidonakis:2010tc}.
Additional uncertainties associated with the modelling of additional radiation are assessed by comparing the nominal
samples with alternative samples where generator parameters are varied.
For the $t$- and $tW$-channel processes, an uncertainty due to the choice of parton shower and hadronisation model is derived
by comparing events produced by {\powheg} interfaced to {\pythia}~6 or {\herwigpp}.
These uncertainties are treated as fully correlated among single-top-quark production processes, but uncorrelated with the
corresponding uncertainty of the $\ttbar$+jets background.
An additional systematic uncertainty in $tW$-channel production concerning the separation
between $t\bar{t}$ and $tW$ at NLO is assessed by comparing
the nominal sample, which uses the diagram removal scheme~\cite{Frixione:2008yi}, with an alternative sample
using the diagram subtraction scheme~\cite{Frixione:2008yi}.
 
Uncertainties of the diboson background normalisation include 5\% from the NLO theory cross sections~\cite{Campbell:1999ah,Campbell:2011bn},
as well as an additional 24\% normalisation uncertainty added in quadrature for each additional inclusive jet-multiplicity bin, based on a
comparison among different algorithms for merging LO matrix elements and parton showers~\cite{Alwall:2007fs}
(it is assumed that two jets originate from the $W/Z$ decay, as in $WW/WZ \to \ell \nu jj$).
Therefore, the total normalisation uncertainty is $5\% \oplus \sqrt{N-2}\times 24\%$, where $N$ is the selected jet multiplicity,
resulting in 34\%, 42\%, and 48\%, for events with exactly 4 jets, exactly 5 jets, and $\geq$6 jets, respectively.
Recent comparisons between data and {\sherpa}~2.1.1 for $WZ(\to \ell\nu\ell\ell) + \geq$4 jets show
agreement within the experimental uncertainty of approximately 40\%~\cite{Aaboud:2016yus}, which further justifies the above uncertainty.
Given the very small contribution of this background to the total prediction, the final result is not affected by the assumed modelling
uncertainties.
 
Uncertainties of the $\ttbar V$ and $\ttbar H$ cross sections are 15\% and $+10\%$/$-13\%$, respectively,
from the uncertainties of their respective NLO theoretical cross sections~\cite{Campbell:2012dh,Garzelli:2012bn,deFlorian:2016spz}.
 
Uncertainties of the data-driven multijet background in the $\Hbb$ search include
contributions from the limited size of the data sample, particularly at high jet and $b$-tag multiplicities, as
well as from the uncertainty in the rate of fake leptons, estimated in
different control regions (e.g.\ selected with an upper requirement on either $\met$ or $\mtw$).
A combined normalisation uncertainty of 50\% due
to all these effects is assigned, which is taken as correlated across jet
and $b$-tag multiplicity bins, but uncorrelated between electron and muon channels.
No explicit shape uncertainty is assigned since the large statistical uncertainties associated with
the multijet background prediction, which are uncorrelated
between bins in the final discriminant distribution, effectively cover all possible shape uncertainties.
 
Uncertainties of the data-driven fake $\had$ background in the $\Htautau$ search are obtained by using additional signal-depleted regions. The construction is similar to that of the SRs and corresponding CRs discussed in Section~\ref{sec:fakeleptons}, but employing further loosened $\had$ identification criteria, and thus referred to as ``loose SR''  and ``loose CR''.
In each loose SR, after subtracting the small simulation-predicted contribution from real $\had$ candidates,
the relative difference in the shape of the distribution between the remaining data and the fake $\had$ background estimate based on its associated loose CR
is assigned as an uncertainty of the prediction in the nominal SR.
In addition, a 30\% uncertainty is applied to the fraction of $\ttbar$ events with a fake $\had$ candidate from the simulation that are added to the fake $\had$ template in the $\hadhad$ channel as part of the fake $\had$ background estimation procedure. This uncertainty, associated with the modelling of the fake $\had$ rate by the simulation, is estimated by comparing data and simulation in a sample enriched in $\ttbar$ dilepton events plus a fake $\had$ candidate.
The same uncertainty is assigned to the selected signal events with fake $\had$ candidates.
In addition, a systematic uncertainty is assigned to account for the different fractional composition of particles (various types of leptons and partons) producing the fake $\had$ candidates
between each SR and its corresponding CR in the $\ttbar$ simulation.
Finally, the normalisation of the fake $\had$ background in each SR is determined in the fit to data.
 
\subsection{Signal modelling}
\label{sec:syst_sigmodeling}
 
Several normalisation and shape uncertainties are taken into account for the $\Hq$ signal.
The uncertainty of the $\ttbar$ cross section also applies to the $\Hq$ signal and is taken to be the same as,
and fully correlated with, the uncertainty assigned to the $\ttbar \to WbWb$ background.
Uncertainties of the Higgs boson branching ratios are taken into account
following the recommendation in Ref.~\cite{deFlorian:2016spz}.
Additional uncertainties associated with the modelling of additional radiation, with the choice of NLO generator, and
with the choice of parton shower and hadronisation model, are estimated from the comparison of the nominal
and alternative $\ttbar \to WbWb$ background samples (discussed in Section~\ref{sec:syst_bkgmodeling}) and applied to $\Hq$ signal.
These modelling uncertainties are taken to be uncorrelated with those affecting the $\ttbar \to WbWb$ background.
 
\section{Statistical analysis}
\label{sec:stat_analysis}
 
For each search, the final discriminant distributions across all analysis regions considered are jointly analysed to test for the
presence of a signal. The statistical analysis uses a binned likelihood function ${\cal L}(\mu,\theta)$ constructed as
a product of Poisson probability terms over all bins considered in the search. This function depends
on the signal-strength parameter $\mu$, defined as a factor multiplying the expected yield of $\Hq$ signal events
normalised to a reference branching ratio $\BR_{\mathrm{ref}}(t\to Hq)=1\%$,
and $\theta$, a set of nuisance parameters that encode the effect of systematic uncertainties on the signal and background expectations.
Therefore, the expected total number of events in a given bin depends on $\mu$ and $\theta$.
All nuisance parameters are subject to Gaussian or log-normal constraints in the likelihood, with the exception of a few parameters
that control the normalisation of some background components (e.g.\ the \ttbin\ background in the case of the $\Hbb$ search), which
are treated as free parameters in the fit.
 
For a given value of $\mu$, the nuisance parameters $\theta$ allow variations of the expectations for signal and background
according to the corresponding systematic uncertainties, and their fitted values result in the deviations from
the nominal expectations that globally provide the best fit to the data.
This procedure allows a reduction of the impact of systematic uncertainties on
the search sensitivity by taking advantage of the highly populated background-dominated bins included in the likelihood fit.
Statistical uncertainties in each bin of the predicted final discriminant distributions are taken into account by dedicated parameters in the fit.
The best-fit $\BR(t\to Hq)$ is obtained by performing a binned likelihood fit to the data under the signal-plus-background
hypothesis, maximising the likelihood function ${\cal L}(\mu,\theta)$ over $\mu$ and $\theta$.
 
The fitting procedure was initially validated through extensive studies using mock data, defined as the sum of all predicted backgrounds
plus an injected signal of variable strength, as well as by performing fits to real data where bins of the final discriminant variable with
a signal contamination above 5\% are excluded (referred to as blinding requirements).
In both cases, the robustness of the model for systematic uncertainties is established by verifying the stability of the fitted background
when varying assumptions about some of the leading sources of uncertainty.
After this, the blinding requirements
are removed in the data and a fit under the signal-plus-background hypothesis is performed. Further checks involve the comparison of the fitted
nuisance parameters before and after removal of the blinding requirements, and their values are found to be consistent. In addition, it is verified that the
fit is able to correctly determine the strength of a simulated signal injected into the real data.
 
The test statistic $q_\mu$ is defined as the profile likelihood ratio,
$q_\mu = -2\ln({\cal L}(\mu,{\hat{\theta}}_\mu)/{\cal L}(\hat{\mu},\hat{\theta}))$,
where $\hat{\mu}$ and $\hat{\theta}$ are the values of the parameters that
maximise the likelihood function (subject to the constraint $0\leq \hat{\mu} \leq \mu$), and ${\hat{\theta}}_\mu$ are the values of the
nuisance parameters that maximise the likelihood function for a given value of $\mu$.
The test statistic $q_\mu$ is evaluated with the {\textsc RooFit} package~\cite{Verkerke:2003ir,RooFitManual}.
A related statistic is used to determine whether the observed data is compatible with the background-only hypothesis (the so-called discovery test)
by setting $\mu=0$ in the profile likelihood ratio and leaving $\hat{\mu}$ unconstrained: $q_0 = -2\ln({\cal L}(0,{\hat{\theta}}_0)/{\cal L}(\hat{\mu},\hat{\theta}))$.
The $p$-value (referred to as $p_0$), representing the level of agreement between the data and the background-only hypothesis, is estimated by integrating
the distribution of $q_0$ based on the asymptotic formulae in Ref.~\cite{Cowan:2010js},
above the observed value of $q_0$ in the data.
Upper limits on $\mu$, and thus on
$\BR(t\to Hq)$, are derived by using $q_\mu$ in the CL$_{\textrm{s}}$ method~\cite{Junk:1999kv,Read:2002hq}.
For a given signal scenario, values of the $\BR(t\to Hq)$ yielding CL$_{\textrm{s}} < 0.05$,
where CL$_{\textrm{s}}$ is computed using the asymptotic approximation~\cite{Cowan:2010js}, are excluded at $\geq 95\%$ CL.

 
\section{Results}
\label{sec:result}
 
This section presents the results obtained from the individual searches for $\Hq$, as well as their combination,
following the statistical analysis discussed in Section~\ref{sec:stat_analysis}.
 
\subsection{$\Hbb$ search}
\label{sec:results_Hbb}
 
A binned likelihood fit under the signal-plus-background hypothesis
is performed on the LH discriminant distributions in the nine analysis regions considered.
In the regions with exactly three $b$-tagged jets, which have the highest sensitivity, the full LH distribution is used with ten equal-width bins.
In contrast, in the regions with at least four $b$-tagged jets,
which have a limited number of data events and a small signal fraction, only two equal-width bins are used. Finally, in the regions with exactly two $b$-tagged jets
the total event yield after requiring the LH discriminant to be above 0.6, is used.
The unconstrained parameters of the fit are the signal strength and a global normalisation factor applied to the \ttbin\ background
common to all analysis regions.
Figures~\ref{fig:prepostfit_unblinded_WbHc_3btagex} and~\ref{fig:prepostfit_unblinded_WbHc_4btagin} show a comparison of the LH discriminant for
data and prediction in the regions with exactly three and at least four $b$-tagged jets,
both before and after performing the fit to data, in the case of the $\Hc$ search.
Tables summarising the pre-fit and post-fit yields can be found in Appendix~\ref{sec:prepostfit_yields_Hbb_appendix}.
 
The best-fit branching ratio obtained is $\BR(t\to Hc)=[-0.2^{+0.7}_{-0.7}\,(\mathrm{stat})^{+2.2}_{-2.3}\,(\mathrm{syst})] \times 10^{-3}$,
assuming $\BR(t\to Hu)=0$.
A similar fit is performed for the $\Hu$ search, yielding $\BR(t\to Hu)=[0.2^{+0.8}_{-0.7}\,(\mathrm{stat})^{+2.5}_{-2.9}\,(\mathrm{syst})] \times 10^{-3}$,
assuming $\BR(t\to Hc)=0$.
The total uncertainties of the measured branching ratios are dominated by systematic uncertainties.
 
The large number of events in the analysis regions considered, together with their different background compositions, allows
the fit to place constraints on the combined effect of several sources of systematic uncertainty.
As a result, an improved background prediction is obtained with a significantly reduced uncertainty, not only in the
signal-depleted regions, but also in the most sensitive analysis regions for this search, (4j, 3b) and (5j, 3b).
The regions with two $b$-tagged jets are used to constrain the leading uncertainties affecting the $\ttbar$+light-jets background prediction,
while the channels with at least four $b$-tagged jets are sensitive to the uncertainties affecting the $\ttbar$+HF background prediction.
In particular, one of the main corrections applied by the fit is an increase of the $\ttbin$ normalisation by a factor of $1.17 \pm  0.15$
relative to the nominal prediction by adjusting the corresponding nuisance parameter.  The $\ttcin$ normalisation is also increased, by a factor of $1.34 \pm 0.40$.
These corrections are in agreement with those found in Ref.~\cite{Aaboud:2017rss}.
Additionally, a few nuisance parameters are adjusted by the fit, with the largest effects corresponding to the
leading nuisance parameters related to the $b$-tagging and $c$-tagging calibrations (by about 0.8 standard deviations),
and those related to $\ttbin$ and $\ttcin$ modelling, which are based on a comparison with alternative generators (by 0.5 standard deviations or less).
The leading uncertainties affecting the signal extraction by the fit are related to the $c$-tagging calibration ($\Delta\BR$ $\sim$ $1.5 \times 10^{-3}$),
followed by the $\ttbar$+light-jets PS \& Had uncertainty ($\Delta\BR$ $\sim$ $1.2 \times 10^{-3}$). Smaller contributions ($\Delta\BR$ $\sim$ 0.5--$1.0 \times 10^{-3}$ each)
result from the uncertainties associated with the $\ttbin$ 5F vs 4F comparison, the dependence of jet energy scale on the jet flavour,
the uncertainty of the $\ttcin$ normalisation, and the limited size of the simulated samples in some of the bins with the highest signal-to-background ratio.
The uncertainty most strongly constrained by the fit is that related to the $c$-tagging calibration. It is reduced by about a factor of two of its value as originally determined
in $W$+$c$-jet events~\cite{Aad:2015ydr}. This is possible because the fit exploits the large number of $\ttbar$ events with two and three $b$-tagged jets
to effectively perform a $c$-tagging calibration, whose results are found to be consistent with those of Ref.~\cite{ATLAS-CONF-2018-001}.
Beyond the constraints on a few individual uncertainties, the significant reduction of the total background uncertainty achieved by the fit primarily derives
from the anti-correlations found among systematic uncertainties from different sources.
 
In the absence of a significant excess of data events above the background expectation, 95\% CL limits are set on $\BR(t\to Hc)$ and $\BR(t\to Hu)$.
The observed (expected) 95\% CL upper limits on the branching ratios
are $\BR(t\to Hc)<4.2 \times 10^{-3}\,(4.0 \times 10^{-3})$ and $\BR(t\to Hu)<5.2 \times 10^{-3}\,(4.9 \times 10^{-3})$.

\begin{figure*}[htbp]
\begin{center}
\subfloat[]{\includegraphics[width=0.33\textwidth]{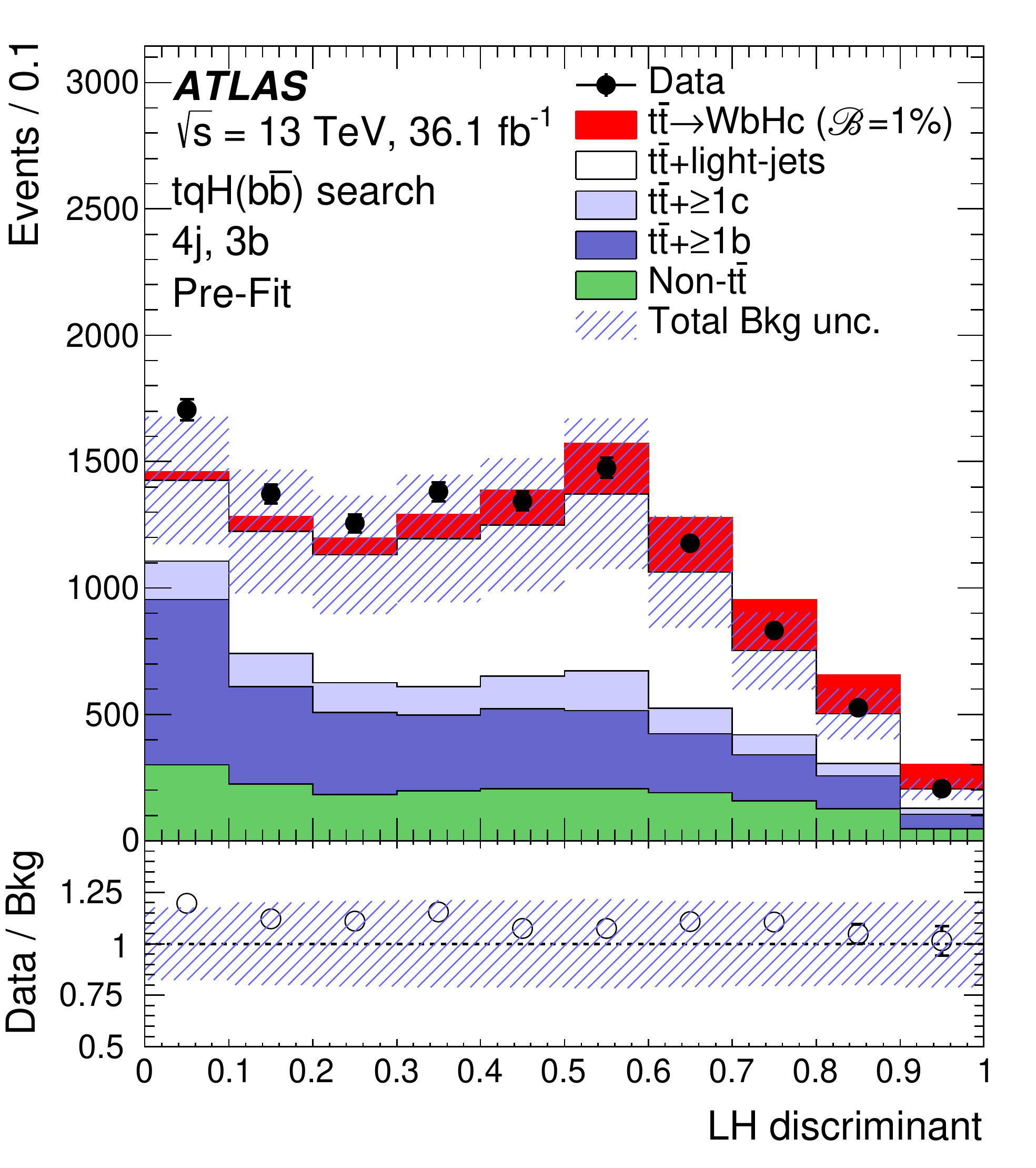}}
\subfloat[]{\includegraphics[width=0.33\textwidth]{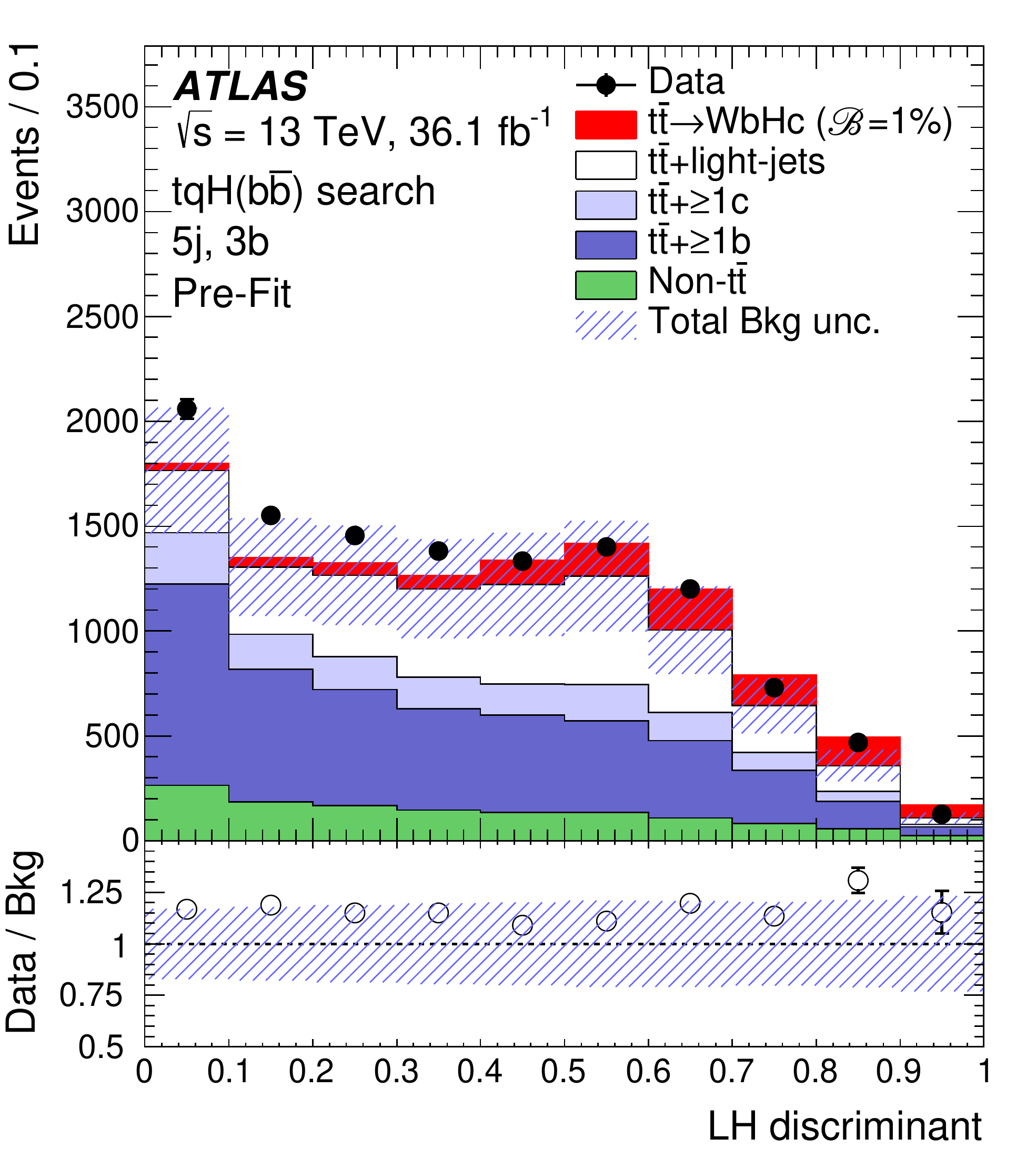}}
\subfloat[]{\includegraphics[width=0.33\textwidth]{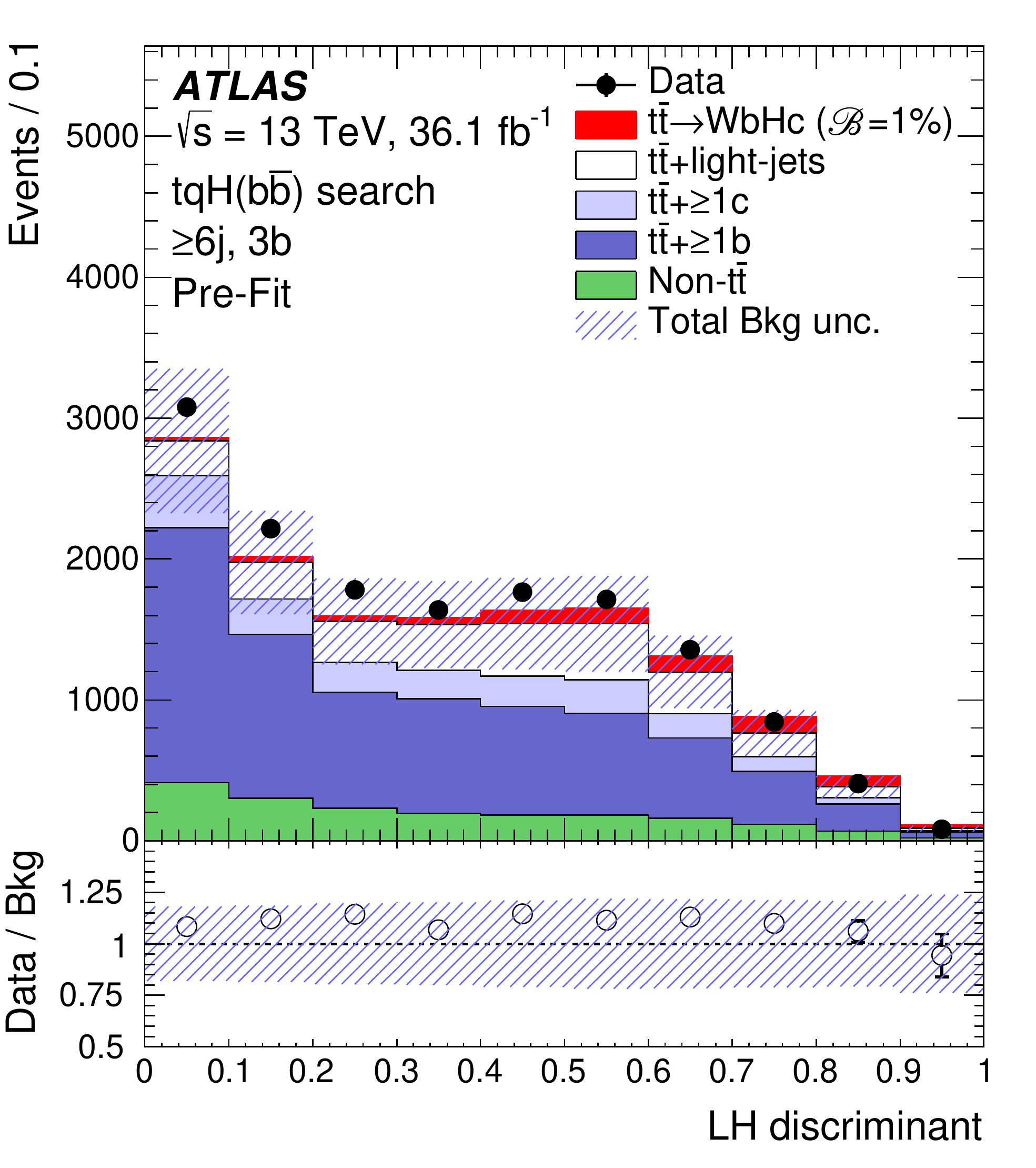}} \\
\subfloat[]{\includegraphics[width=0.33\textwidth]{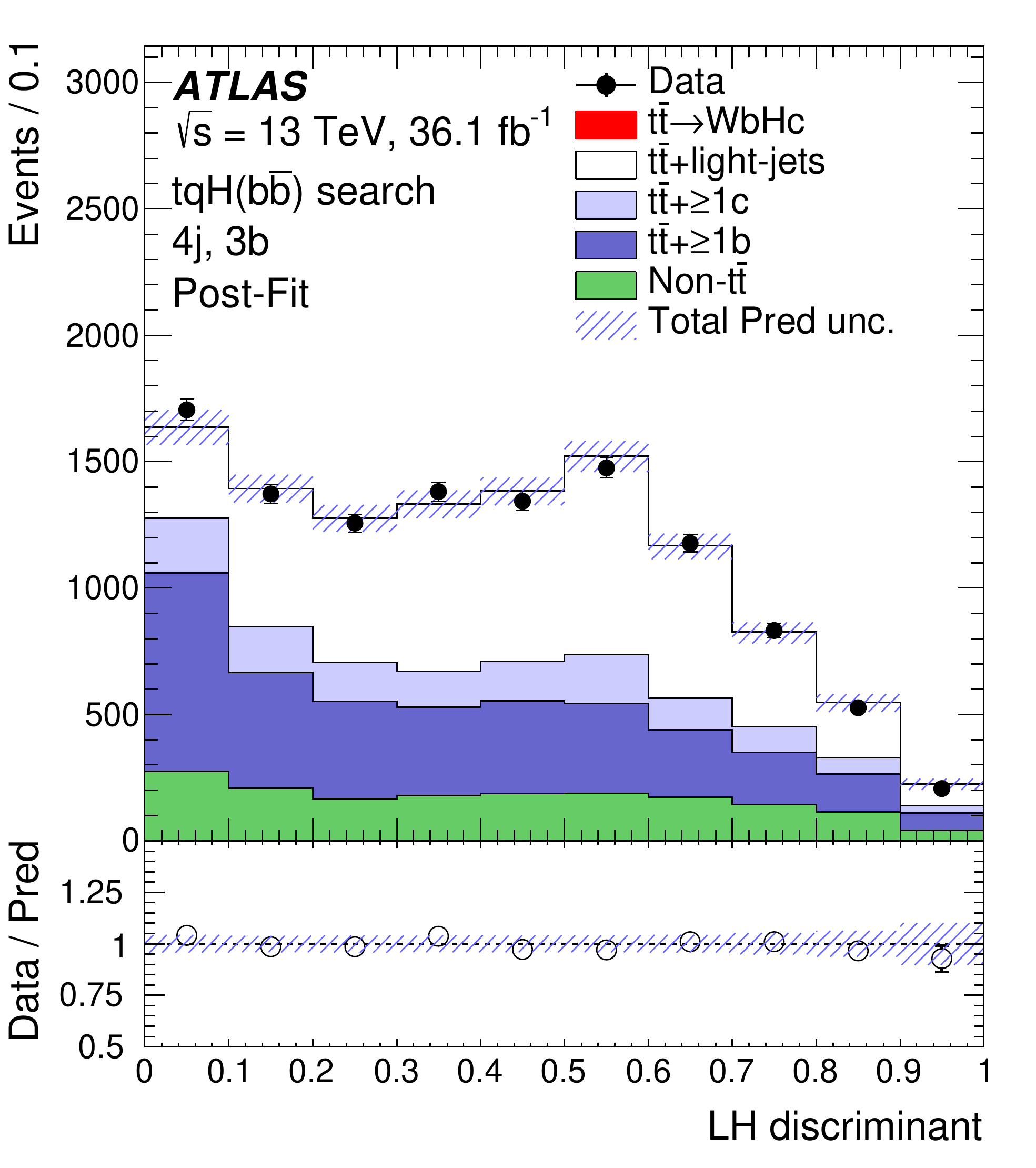}}
\subfloat[]{\includegraphics[width=0.33\textwidth]{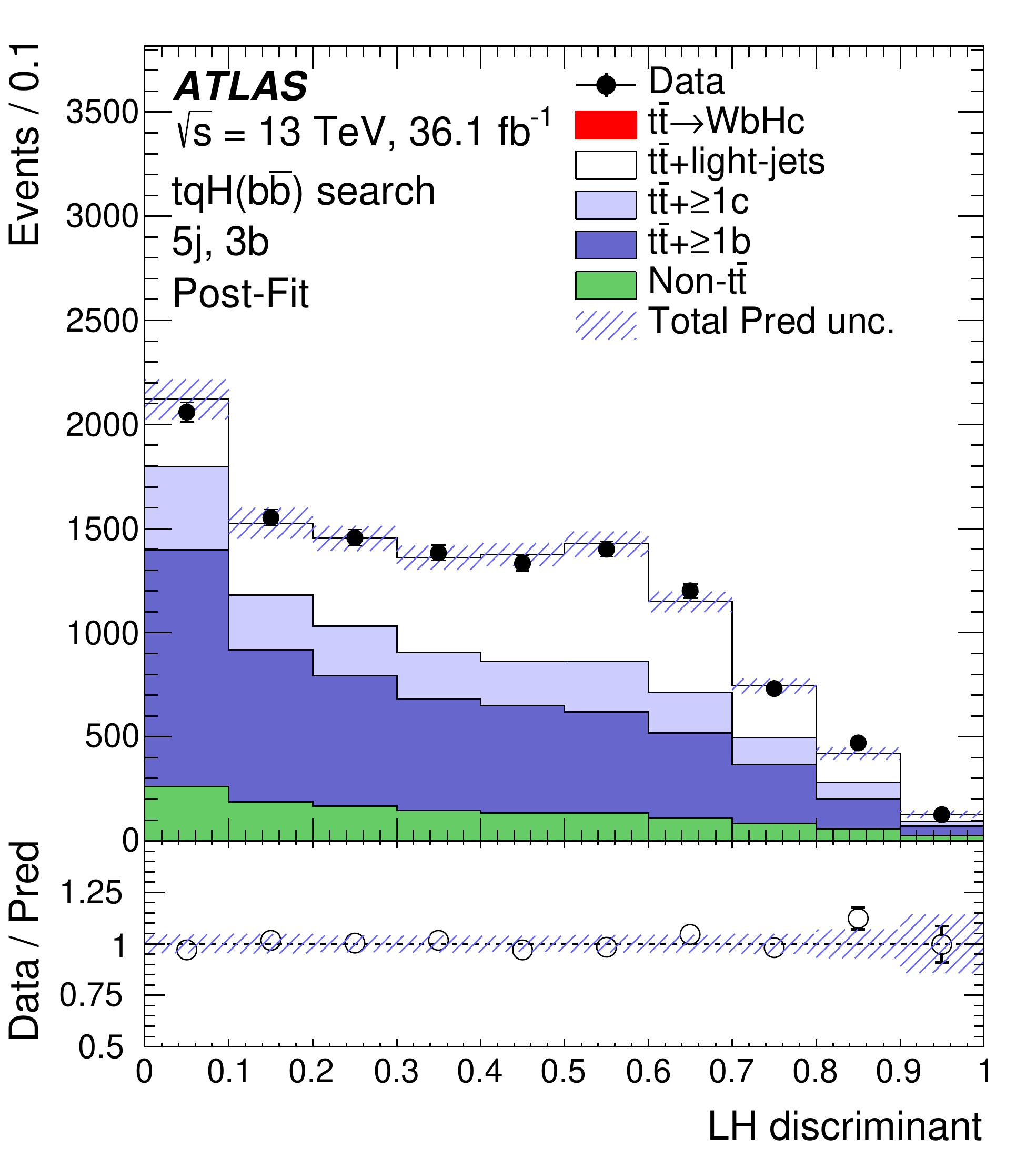}}
\subfloat[]{\includegraphics[width=0.33\textwidth]{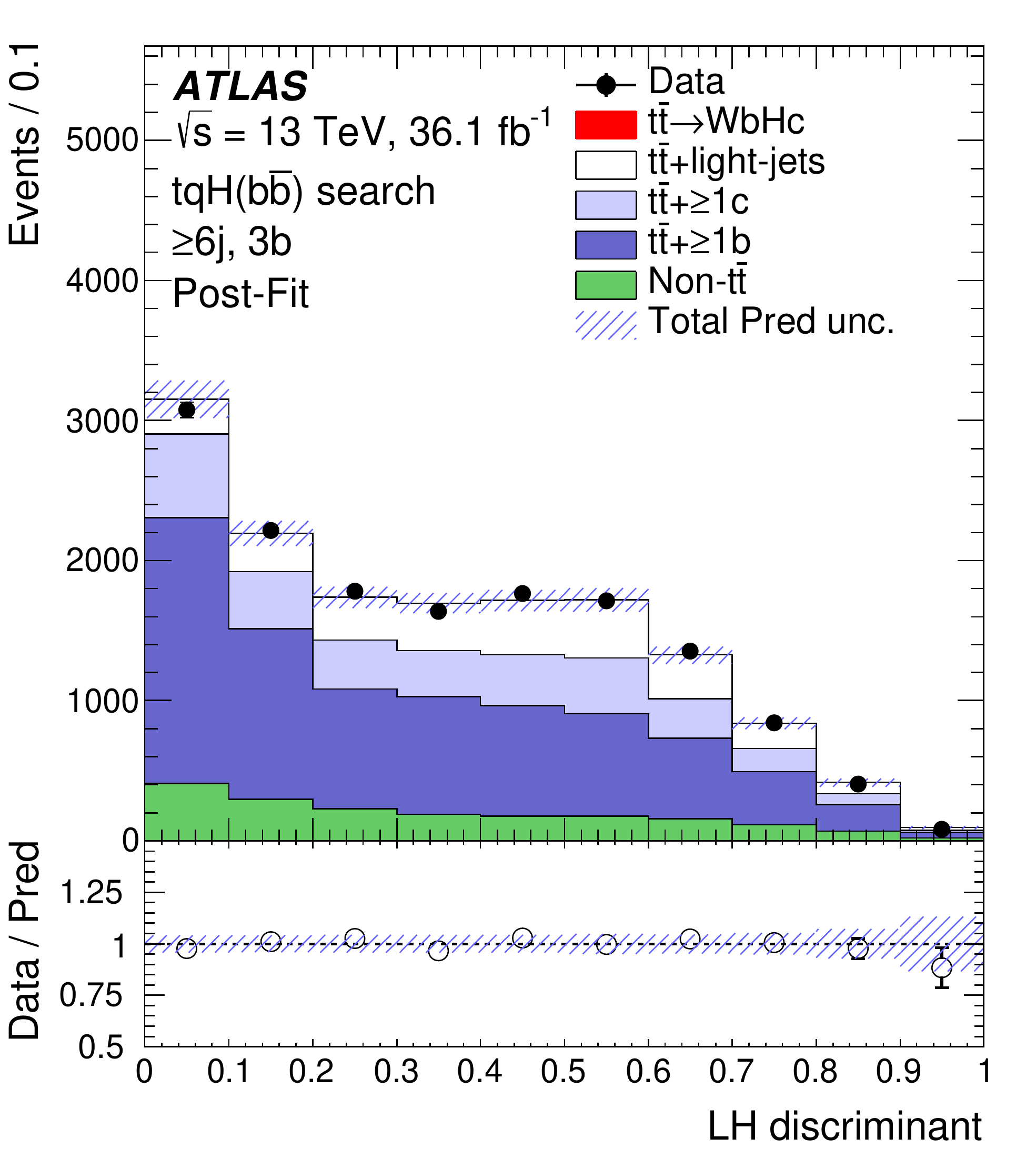}} \\
\caption{\small{$\Hbb$ search: Comparison between the data and prediction for the LH discriminant distribution in the regions with three $b$-tagged jets,
before and after the fit to data  (``Pre-Fit'' and ``Post-Fit'', respectively) under the signal-plus-background hypothesis.
Shown are the (4j, 3b) region (a) pre-fit and (d) post-fit,  the (5j, 3b) region (b) pre-fit and (e) post-fit, and
the ($\geq$6j, 3b) region (c) pre-fit and (f) post-fit.
The small contributions from $\ttbar V$, $\ttbar H$, single-top-quark, $W/Z$+jets, diboson, and multijet backgrounds are combined into a single background source
referred to as ``Non-$\ttbar$''.
In the pre-fit figures the expected $\Hc$ signal (solid red) corresponding to $\BR(t\to Hc)=1\%$ is also shown,
added to the background prediction. In the post-fit figures, the $\Hc$ signal is normalised using the best-fit branching ratio,
$\BR(t\to Hc)=(-0.2^{+2.3}_{-2.4}) \times 10^{-3}$.
The bottom panels display the ratios of data to either the SM background prediction before the fit (``Bkg'')  or the total signal-plus-background
prediction after the fit (``Pred'').
The hashed area represents the total uncertainty of the background.
In the case of the pre-fit background uncertainty, the normalisation uncertainty of the $\ttbin$ background is not included. }}
\label{fig:prepostfit_unblinded_WbHc_3btagex}
\end{center}
\end{figure*}
 
\begin{figure*}[htbp]
\begin{center}
\subfloat[]{\includegraphics[width=0.33\textwidth]{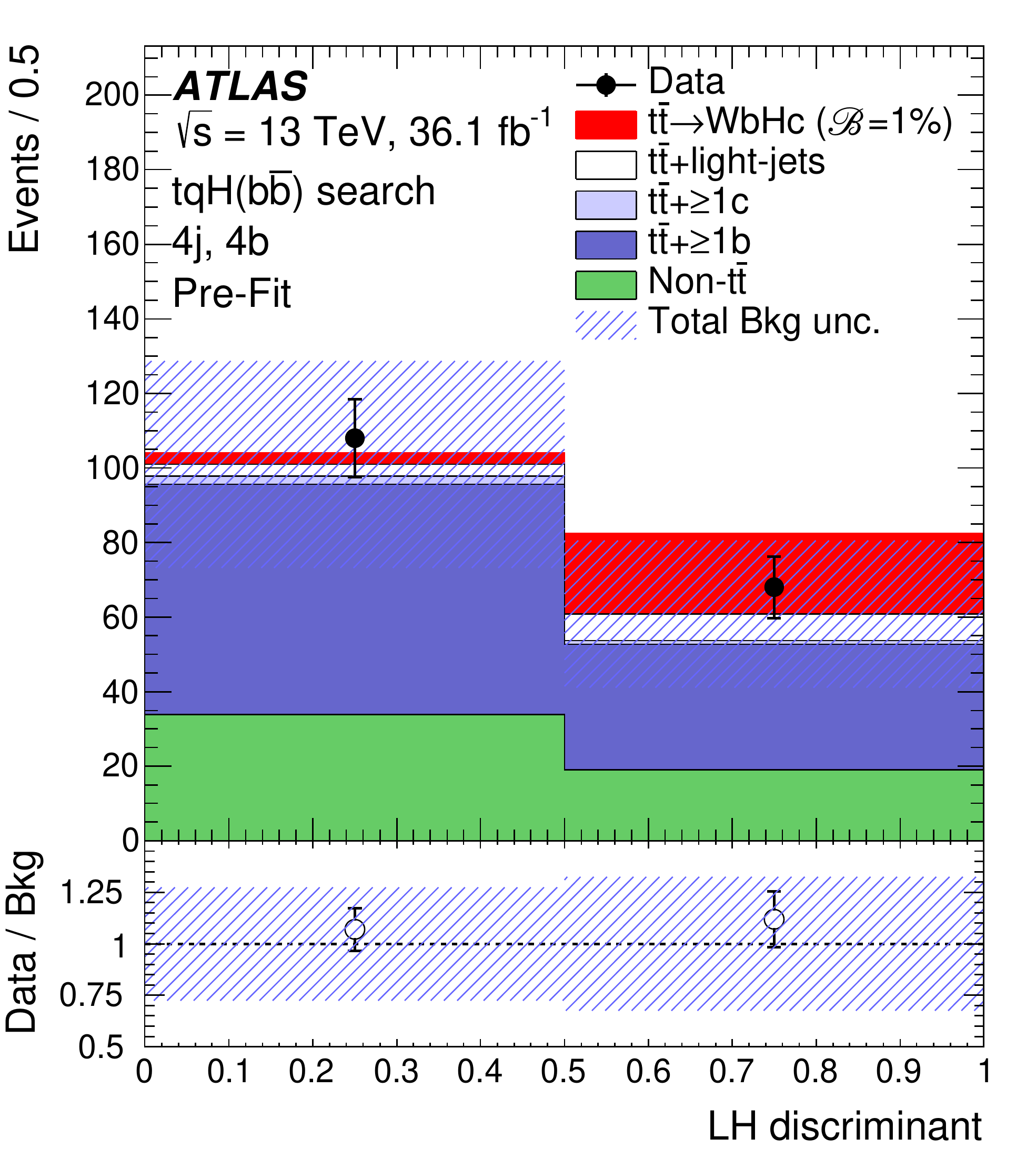}}
\subfloat[]{\includegraphics[width=0.33\textwidth]{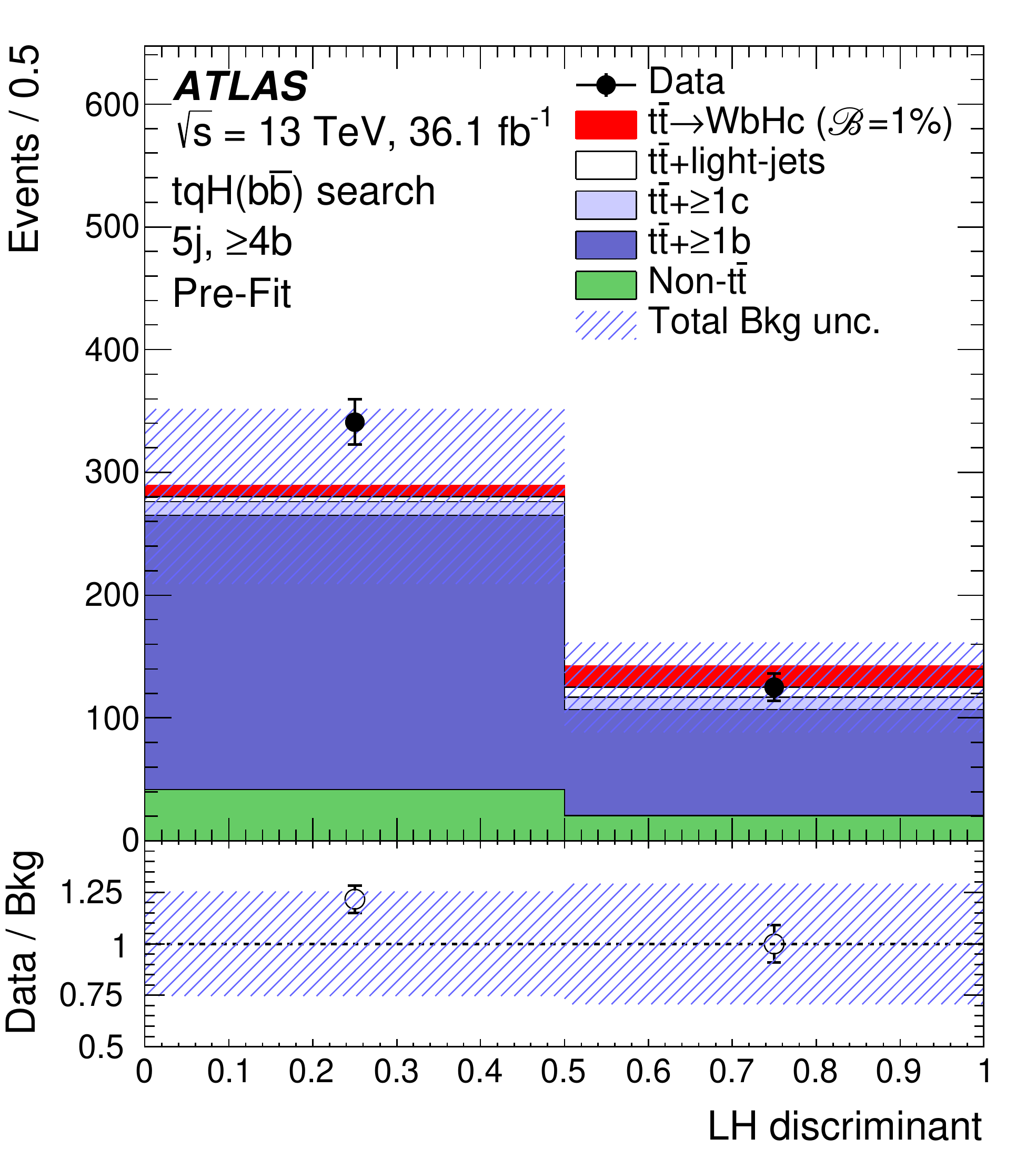}}
\subfloat[]{\includegraphics[width=0.33\textwidth]{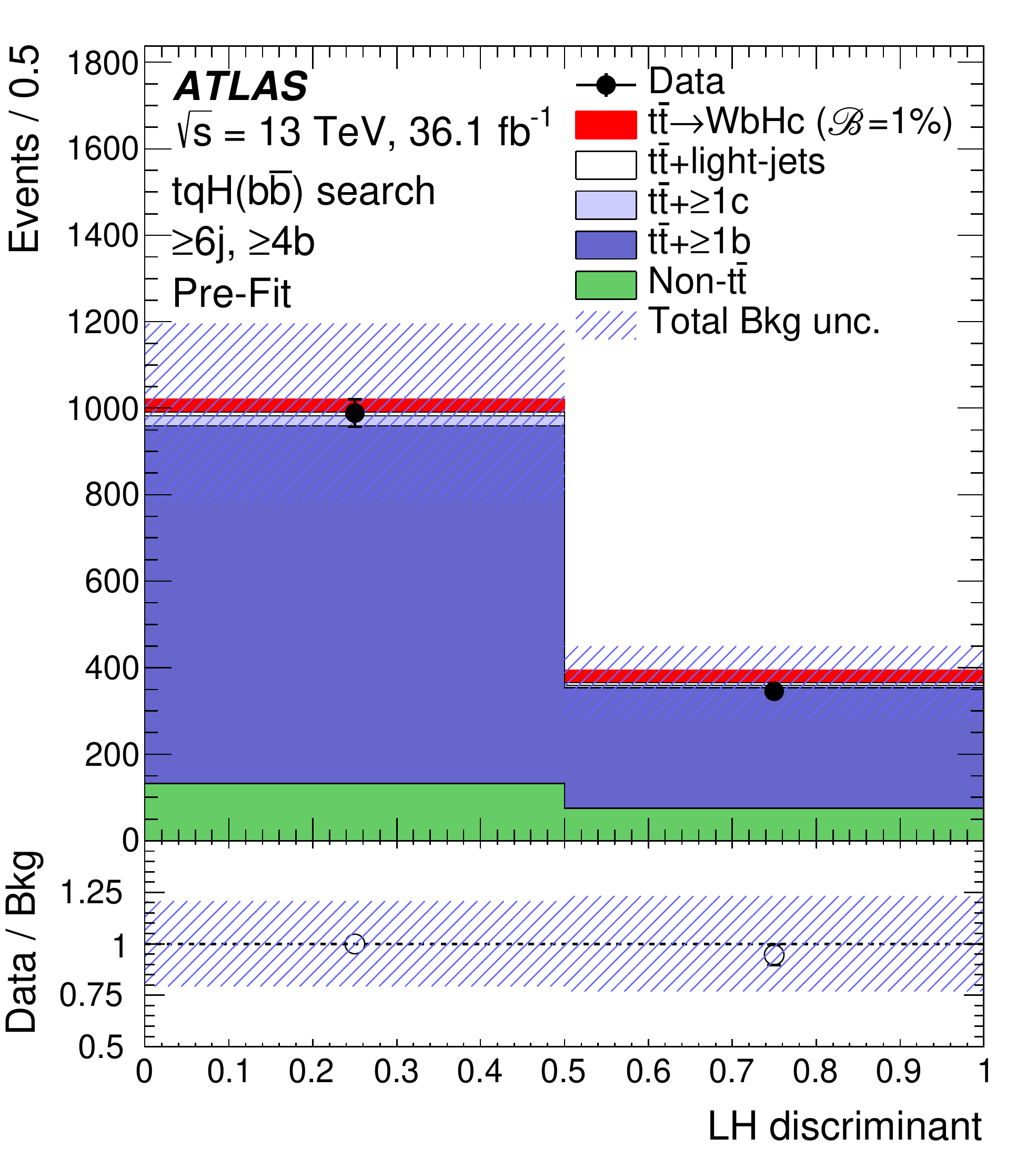}} \\
\subfloat[]{\includegraphics[width=0.33\textwidth]{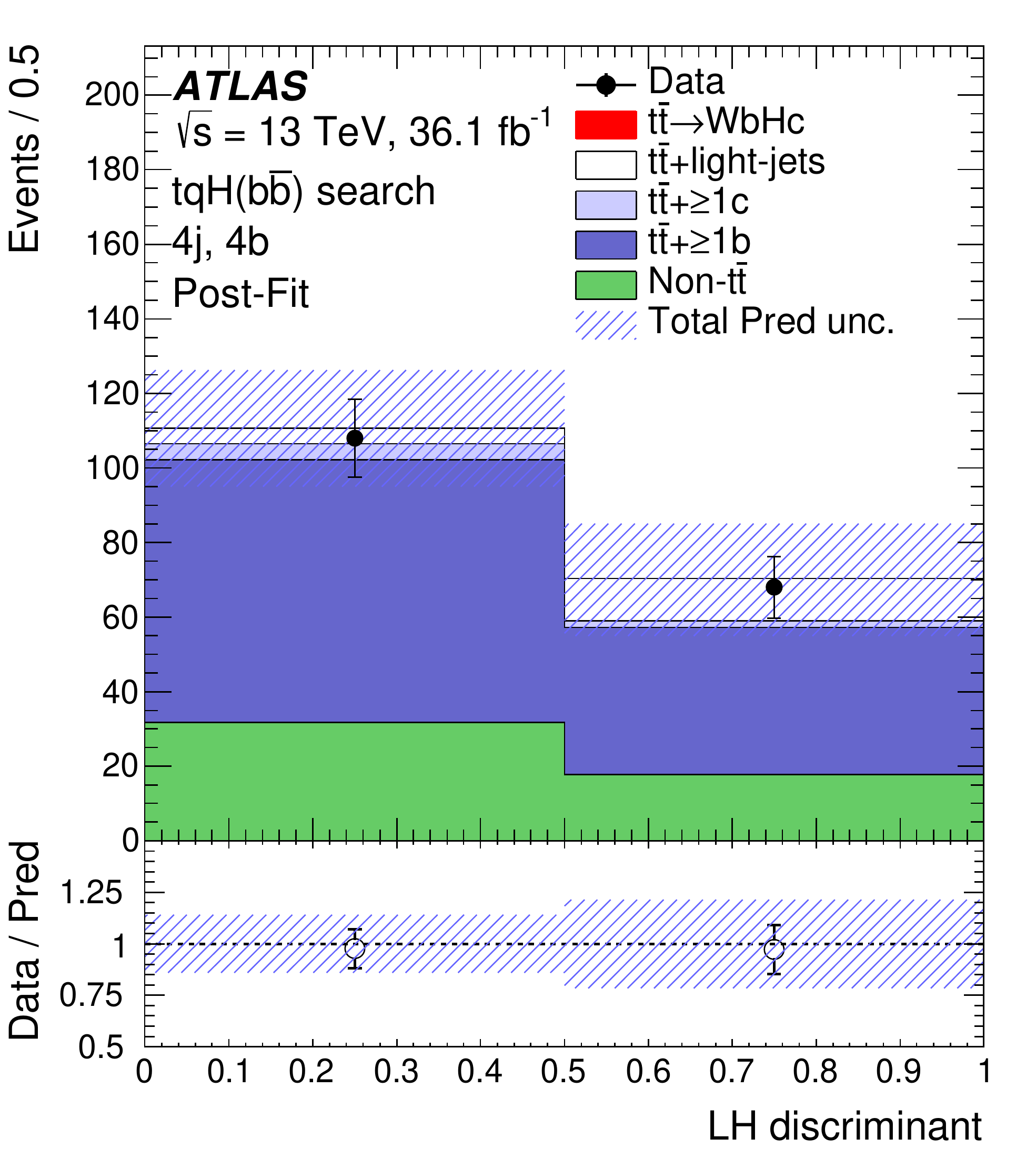}}
\subfloat[]{\includegraphics[width=0.33\textwidth]{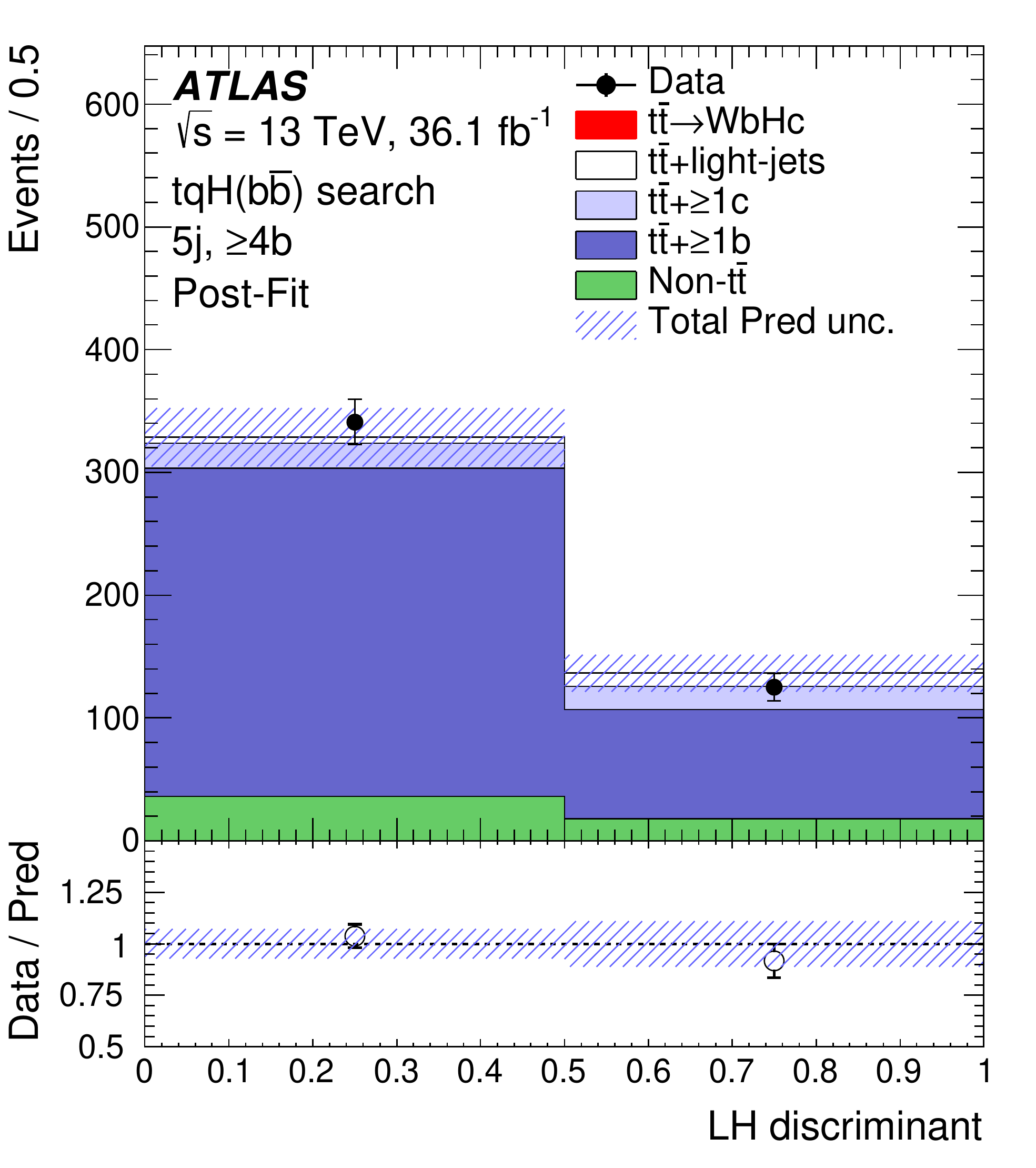}}
\subfloat[]{\includegraphics[width=0.33\textwidth]{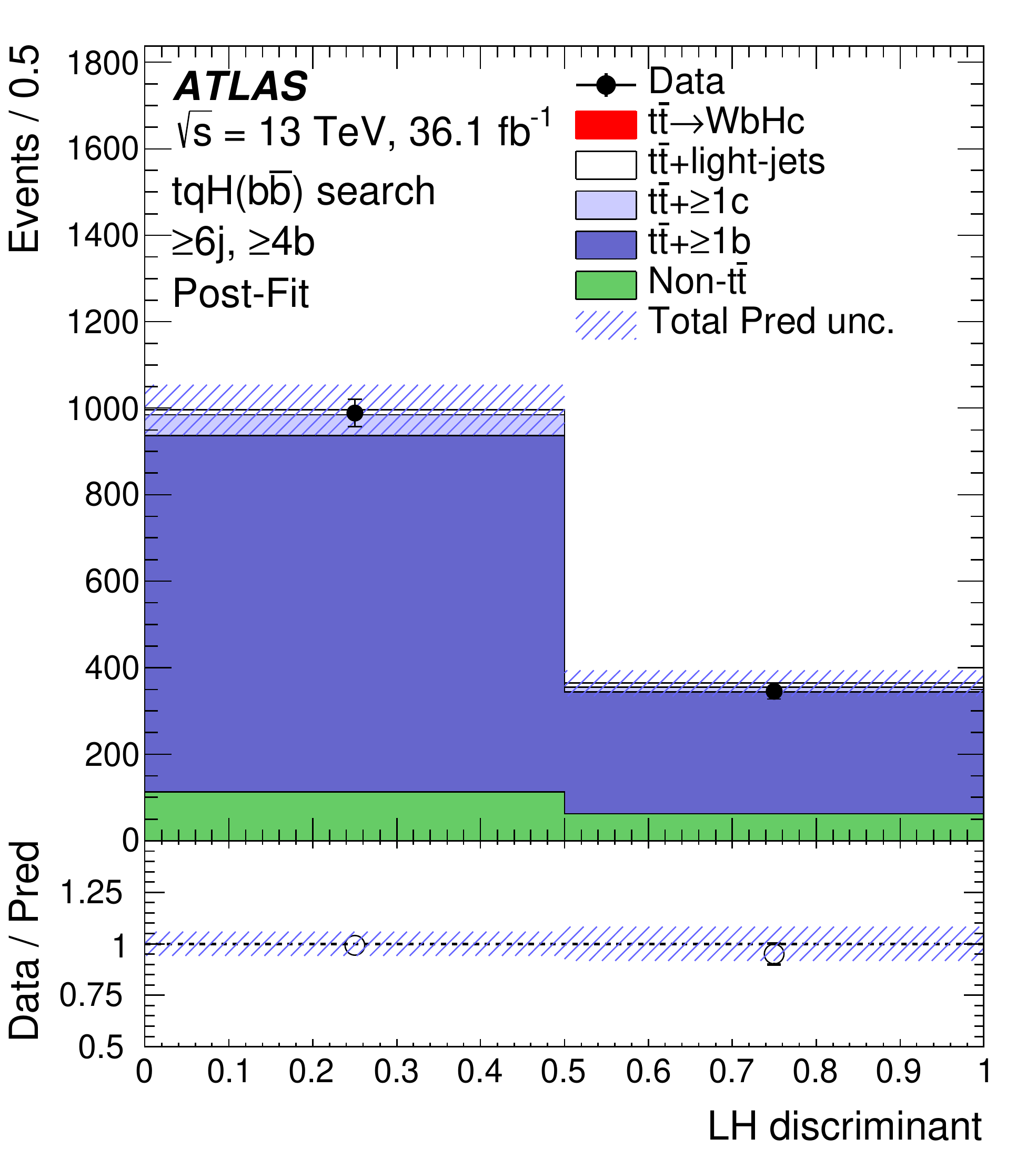}} \\
\caption{\small{$\Hbb$ search: Comparison between the data and prediction for the LH discriminant distribution in the regions with at least four $b$-tagged jets,
before and after the fit to data  (``Pre-Fit'' and ``Post-Fit'', respectively) under the signal-plus-background hypothesis.
Shown are the (4j, 4b) region (a) pre-fit and (d) post-fit,  the (5j, $\geq$4b) region (b) pre-fit and (e) post-fit, and
the ($\geq$6j, $\geq$4b) region (c) pre-fit and (f) post-fit.
The small contributions from $\ttbar V$, $\ttbar H$, single-top-quark, $W/Z$+jets, diboson, and multijet backgrounds are combined into a single background source
referred to as ``Non-$\ttbar$''.
In the pre-fit figures the expected $\Hc$ signal (solid red) corresponding to $\BR(t\to Hc)=1\%$ is also shown,
added to the background prediction. In the post-fit figures, the $\Hc$ signal is normalised using the best-fit branching ratio,
$\BR(t\to Hc)=(-0.2^{+2.3}_{-2.4}) \times 10^{-3}$.
The bottom panels display the ratios of data to either the SM background prediction before the fit (``Bkg'')  or the total signal-plus-background
prediction after the fit (``Pred'').
The hashed area represents the total uncertainty of the background.
In the case of the pre-fit background uncertainty, the normalisation uncertainty of the $\ttbin$ background is not included. }}
\label{fig:prepostfit_unblinded_WbHc_4btagin}
\end{center}
\end{figure*}

\subsection{$\Htautau$ search}
\label{sec:results_Htautau}
 
A binned likelihood fit under the signal-plus-background hypothesis is performed on the BDT discriminant distributions in the four
analysis regions considered. The unconstrained parameters of the fit are the signal
strength, and four independent parameters associated with the normalisation of the fake $\had$ background in each of the analysis regions.
No significant pulls or constraints are obtained for the fitted nuisance parameters, resulting in a post-fit background prediction in each analysis region that is
very close to the pre-fit prediction, albeit with reduced uncertainties due to the anti-correlations among sources of systematic uncertainty resulting from the fit.
Figure~\ref{fig:prepostfit_unblinded_WbHc_lh} shows a comparison of the data and prediction for the BDT discriminant distribution in
the ($\lephad$, 3j) and ($\lephad$, $\geq$4j) regions, both pre- and post-fit to data, in the case of the $\Hc$ search.
A similar comparison for the ($\hadhad$, 3j) and ($\hadhad$, $\geq$4j) regions is shown in Figure~\ref{fig:prepostfit_unblinded_WbHc_hh}.
Tables summarising the pre-fit and post-fit yields can be found in Appendix~\ref{sec:prepostfit_yields_Htautau_appendix}.
 
The best-fit branching ratio obtained is $\BR(t\to Hc)=[-4.4^{+7.7}_{-7.0}\,(\mathrm{stat})^{+6.2}_{-4.9}\,(\mathrm{syst})] \times 10^{-4}$, assuming $\BR(t\to Hu)=0$.
The best-fit normalisation factors for the fake $\had$ background are: $0.82 \pm 0.23$ in the ($\lephad$, 3j) region, $0.84^{+0.25}_{-0.28}$ in the ($\lephad$, $\geq$4j) region,
$0.94^{+0.18}_{-0.17}$ in the ($\hadhad$, 3j) region, and $0.90 \pm 0.26$ in the ($\hadhad$, $\geq$4j) region.
A similar fit is performed for the $\Hu$ search, yielding $\BR(t\to Hu)=[-5.3^{+7.3}_{-6.5}\,(\mathrm{stat})^{+5.3}_{-4.2}\,(\mathrm{syst})] \times 10^{-4}$,
assuming $\BR(t\to Hc)=0$. The obtained normalisation factors for the fake $\had$ background agree within 1\% with those obtained by the $\Hc$ search.
In both cases, the uncertainty of the measured branching ratio is dominated by the statistical uncertainty.
The main contributions to the total systematic uncertainty arise from the fake $\had$ background estimation and the uncertainty associated
with the different responses to quark-initiated and gluon-initiated jets.
No significant excess of data events above the background expectation is found,
and observed (expected) 95\% CL limits are set on $\BR(t\to Hc)$ and $\BR(t\to Hu)$:
$\BR(t\to Hc)<1.9 \times 10^{-3}\,(2.1 \times 10^{-3})$ and $\BR(t\to Hu)<1.7 \times 10^{-3}\,(2.0 \times 10^{-3})$.
These results are dominated by the $\hadhad$ channel, which has a sensitivity a factor of two better than that of the $\lephad$ channel.
 
\begin{figure*}[htbp]
\begin{center}
\subfloat[]{\includegraphics[width=0.40\textwidth]{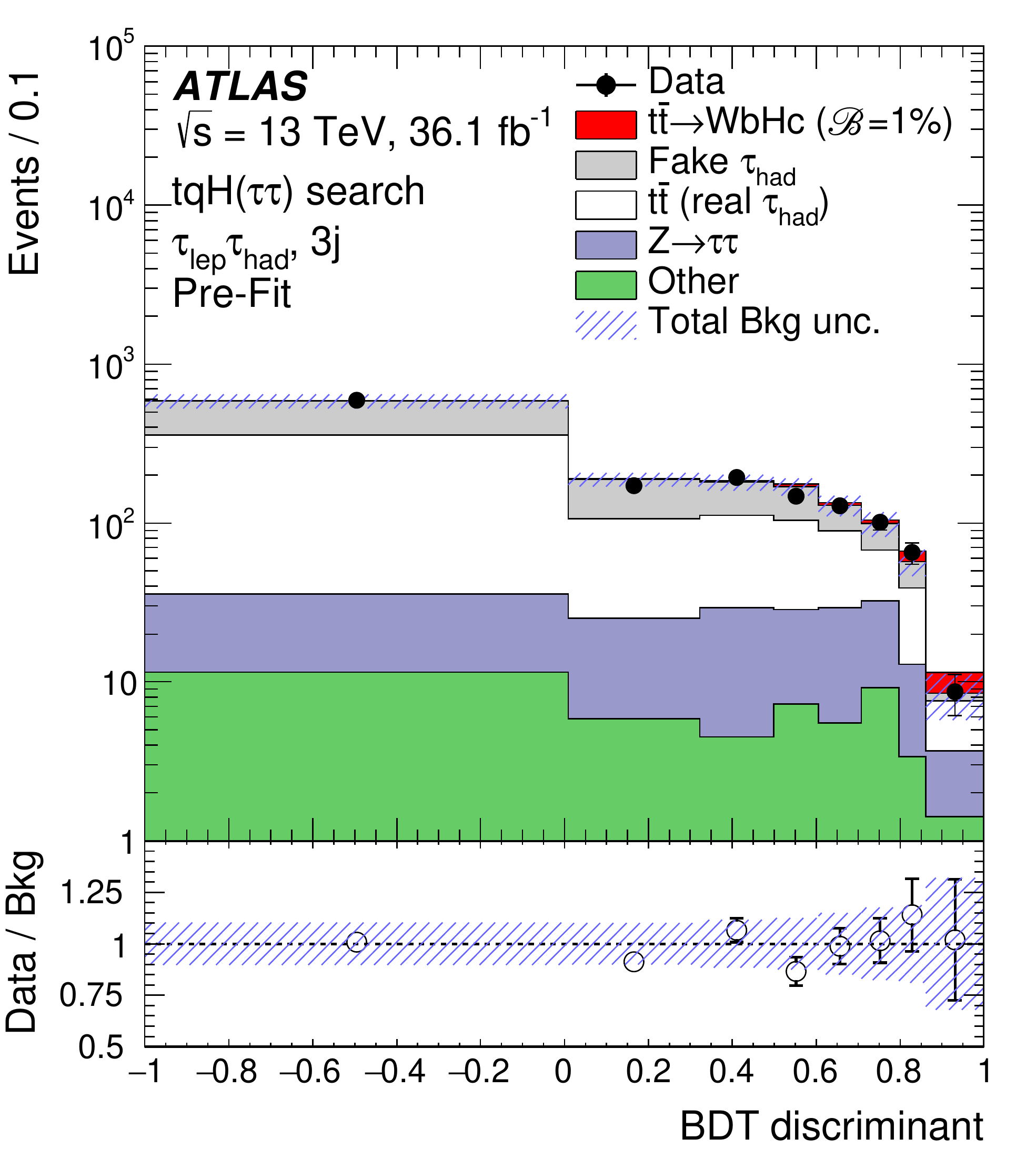}}
\subfloat[]{\includegraphics[width=0.40\textwidth]{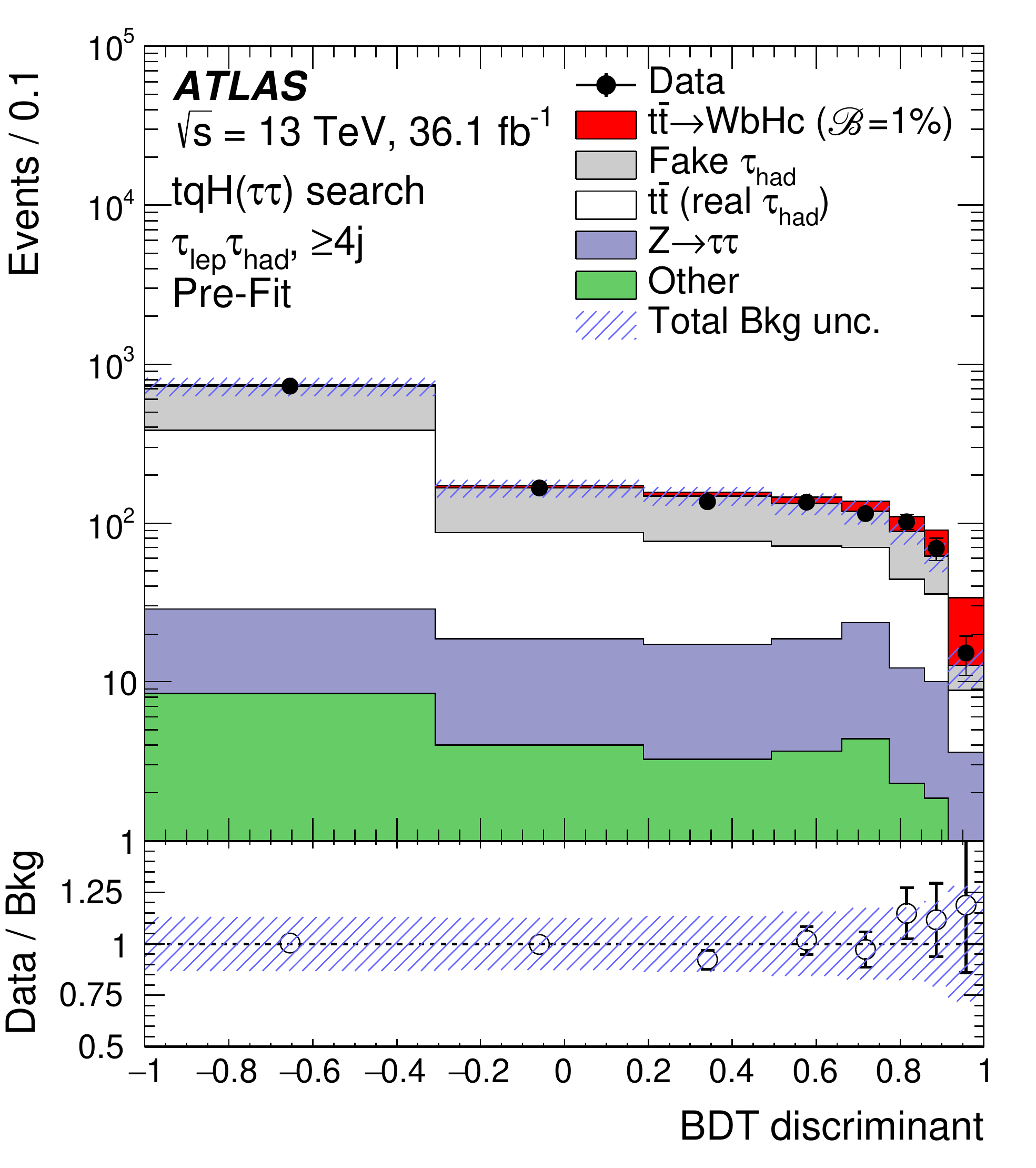}} \\
\subfloat[]{\includegraphics[width=0.40\textwidth]{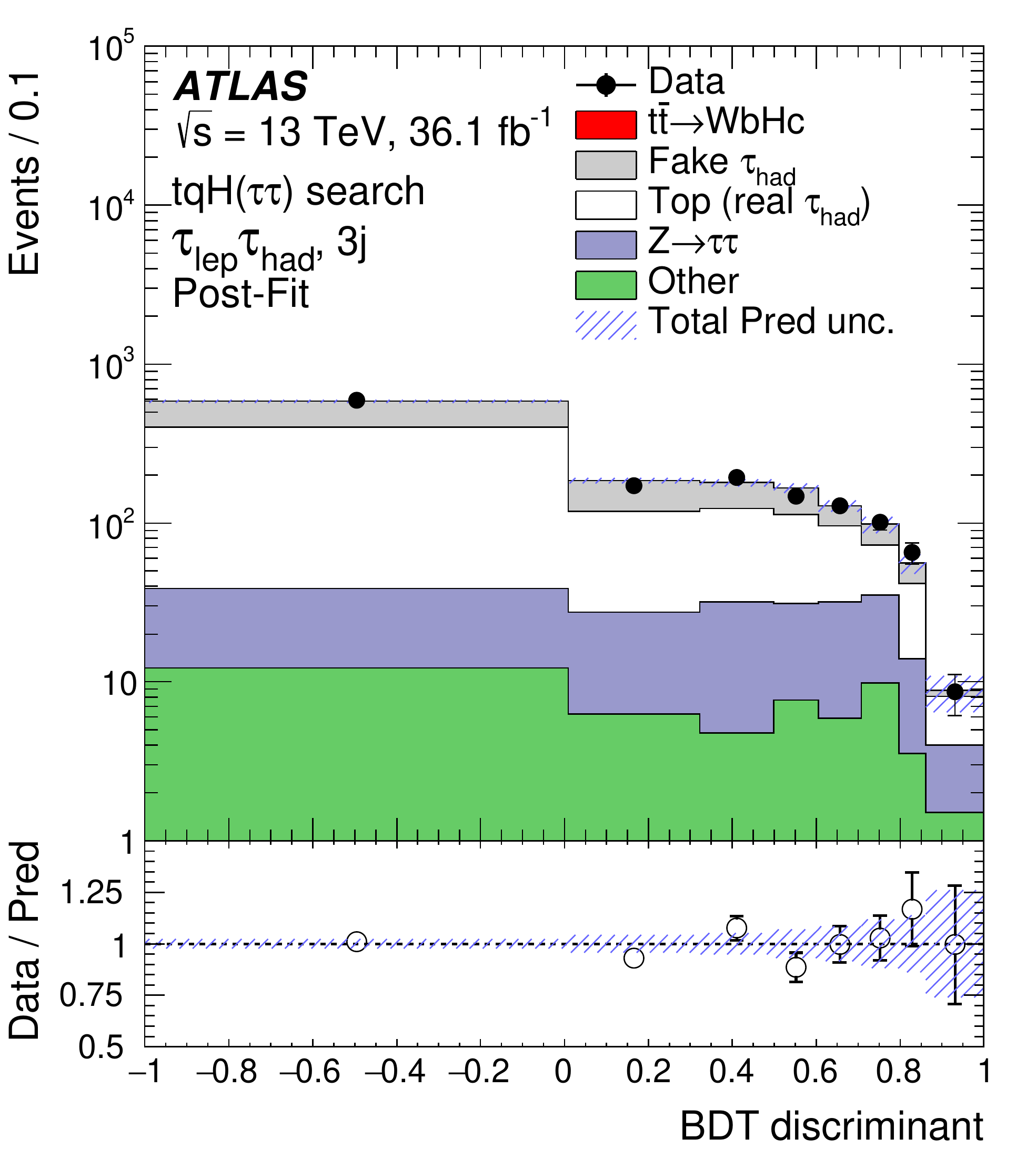}}
\subfloat[]{\includegraphics[width=0.40\textwidth]{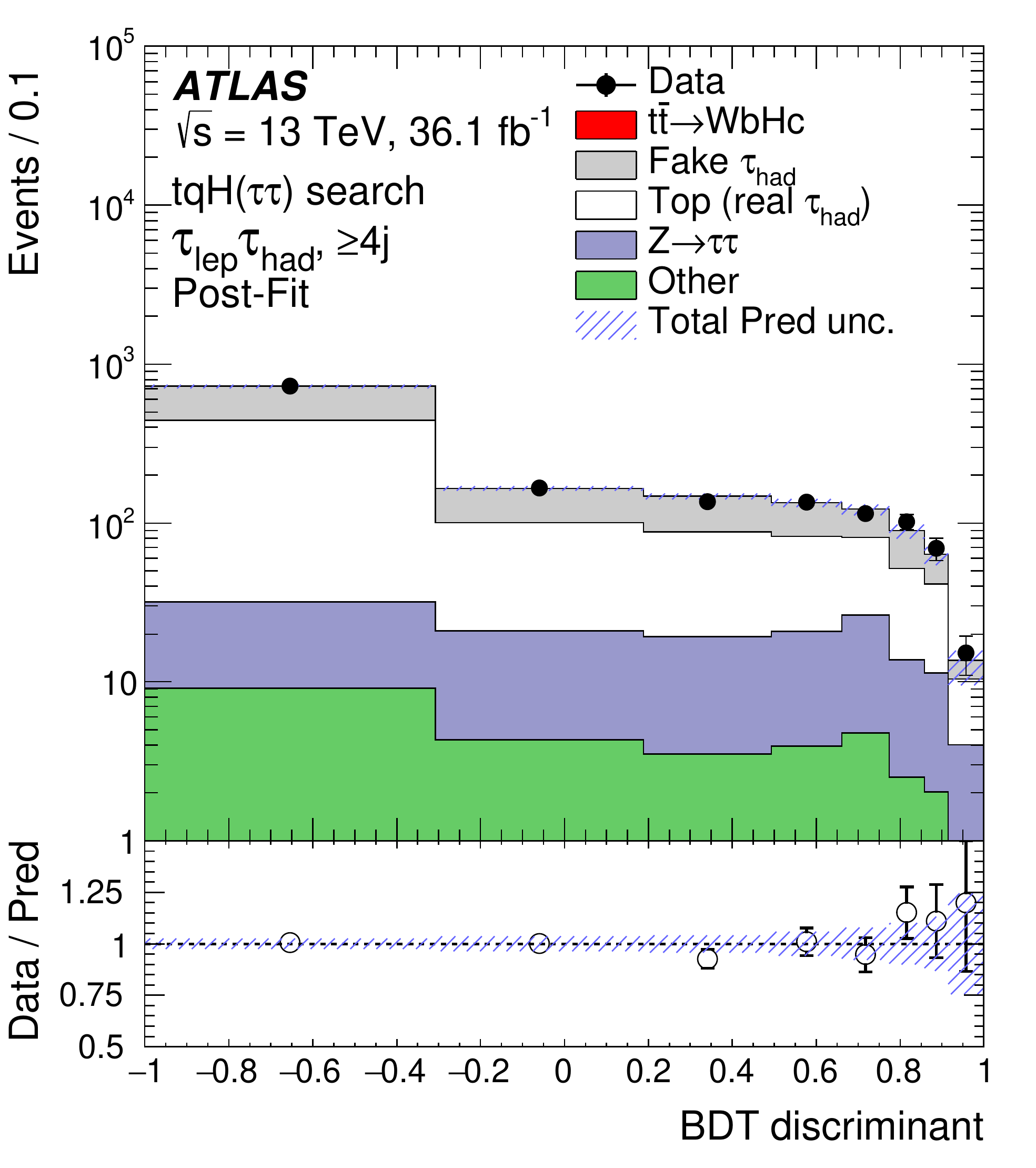}} \\
\caption{\small{$\Htautau$ search: Comparison between the data and prediction for the BDT discriminant distribution in the
$\lephad$ channel, before and after the fit to data  (``Pre-Fit'' and ``Post-Fit'', respectively) under the signal-plus-background hypothesis.
Shown are the ($\lephad$, 3j) region (a) pre-fit and (c) post-fit, and the ($\lephad$, $\geq$4j) region (b) pre-fit and (d) post-fit.
The contributions with real $\had$ candidates from $\ttbar$,  $\ttbar V$, $\ttbar H$, and single-top-quark backgrounds are combined into
a single background source referred to as ``Top (real $\had$)'', whereas the small contributions from
$Z\to \ell^+\ell^-$ ($\ell = e, \mu$) and diboson backgrounds are combined into ``Other''.
In the pre-fit figures the expected $\Hc$ signal (solid red) corresponding to $\BR(t\to Hc)=1\%$ is also shown,
added to the background prediction. In the post-fit figures, the $\Hc$ signal is normalised using the best-fit branching ratio,
$\BR(t\to Hc)=(-4.4^{+9.9}_{-8.5})\times 10^{-4}$.
The bottom panels display the ratios of data to either the SM background prediction before the fit (``Bkg'')  or the total signal-plus-background
prediction after the fit (``Pred'').
The hashed area represents the total uncertainty of the background.
In the case of the pre-fit background uncertainty, the normalisation uncertainty of the fake $\had$ background is not included.}}
\label{fig:prepostfit_unblinded_WbHc_lh}
\end{center}
\end{figure*}
 
\begin{figure*}[htbp]
\begin{center}
\subfloat[]{\includegraphics[width=0.40\textwidth]{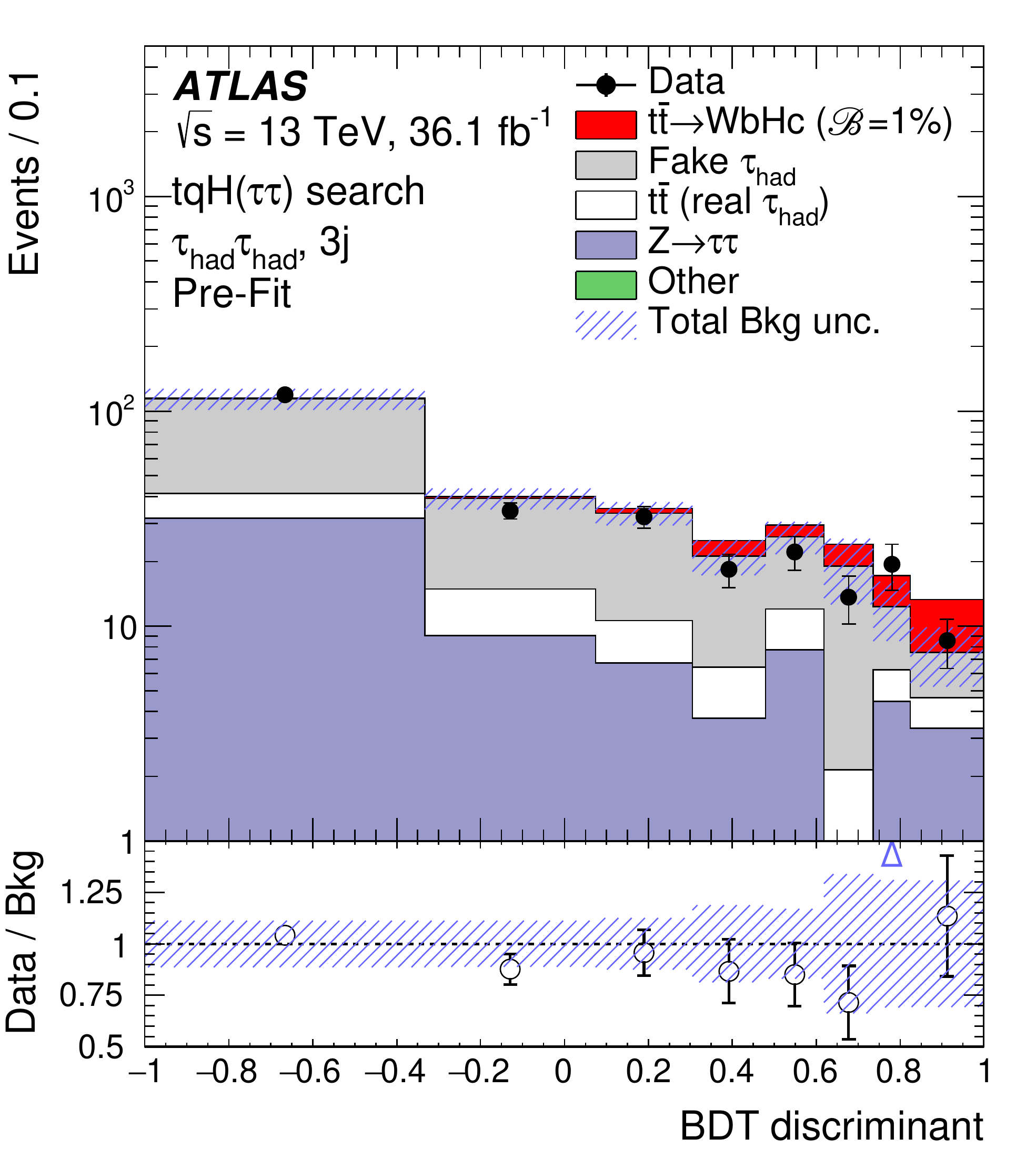}}
\subfloat[]{\includegraphics[width=0.40\textwidth]{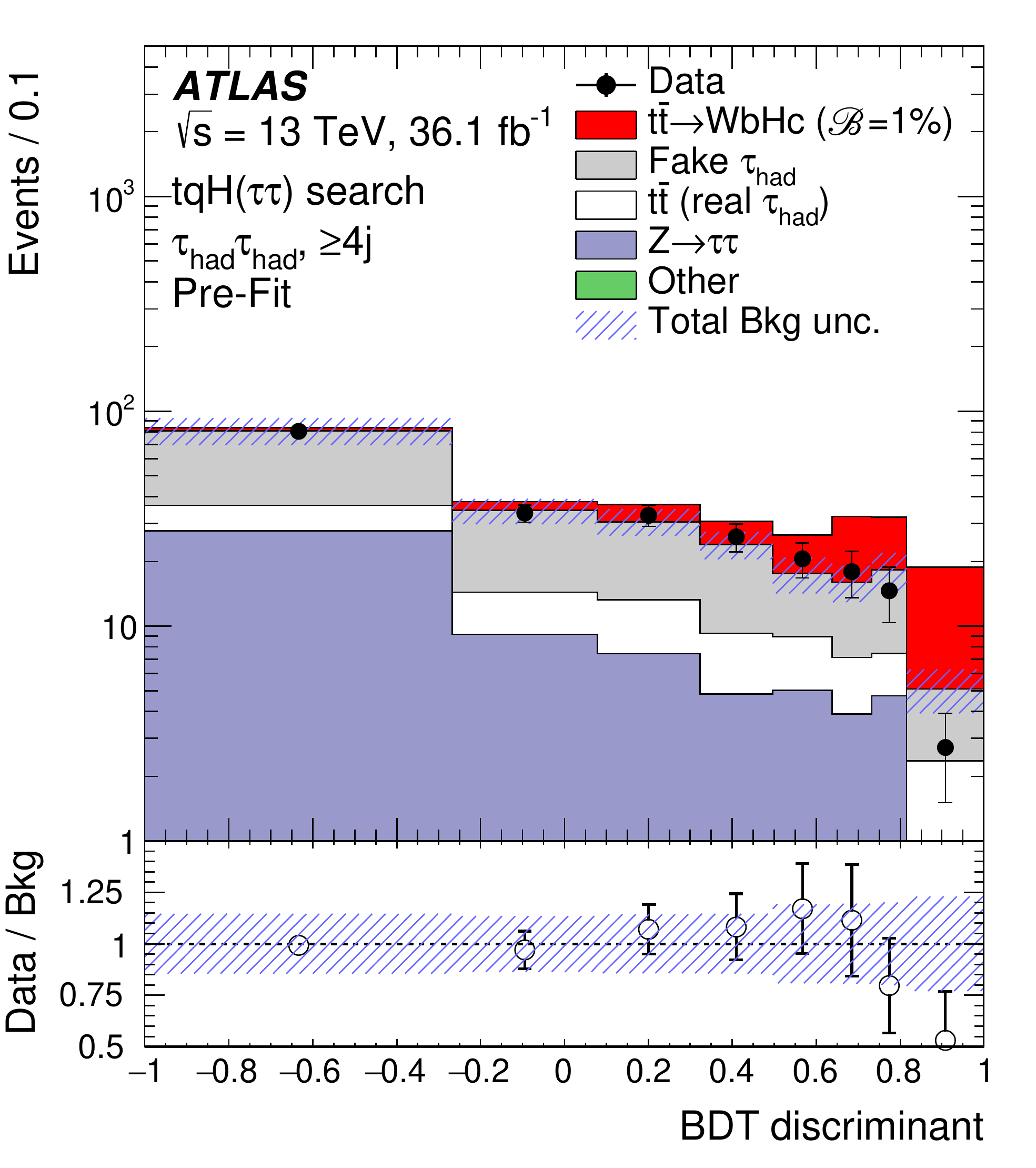}} \\
\subfloat[]{\includegraphics[width=0.40\textwidth]{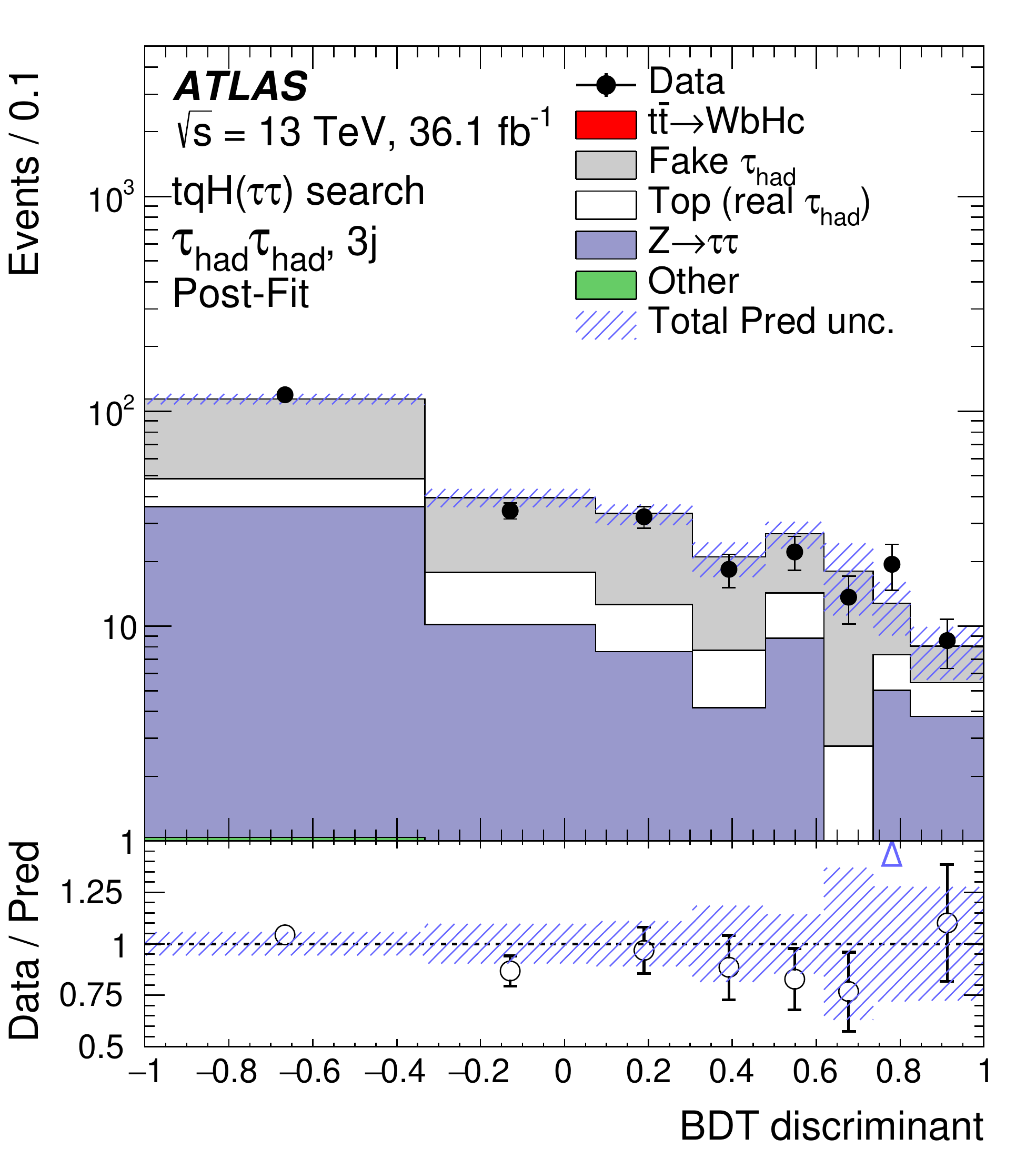}}
\subfloat[]{\includegraphics[width=0.40\textwidth]{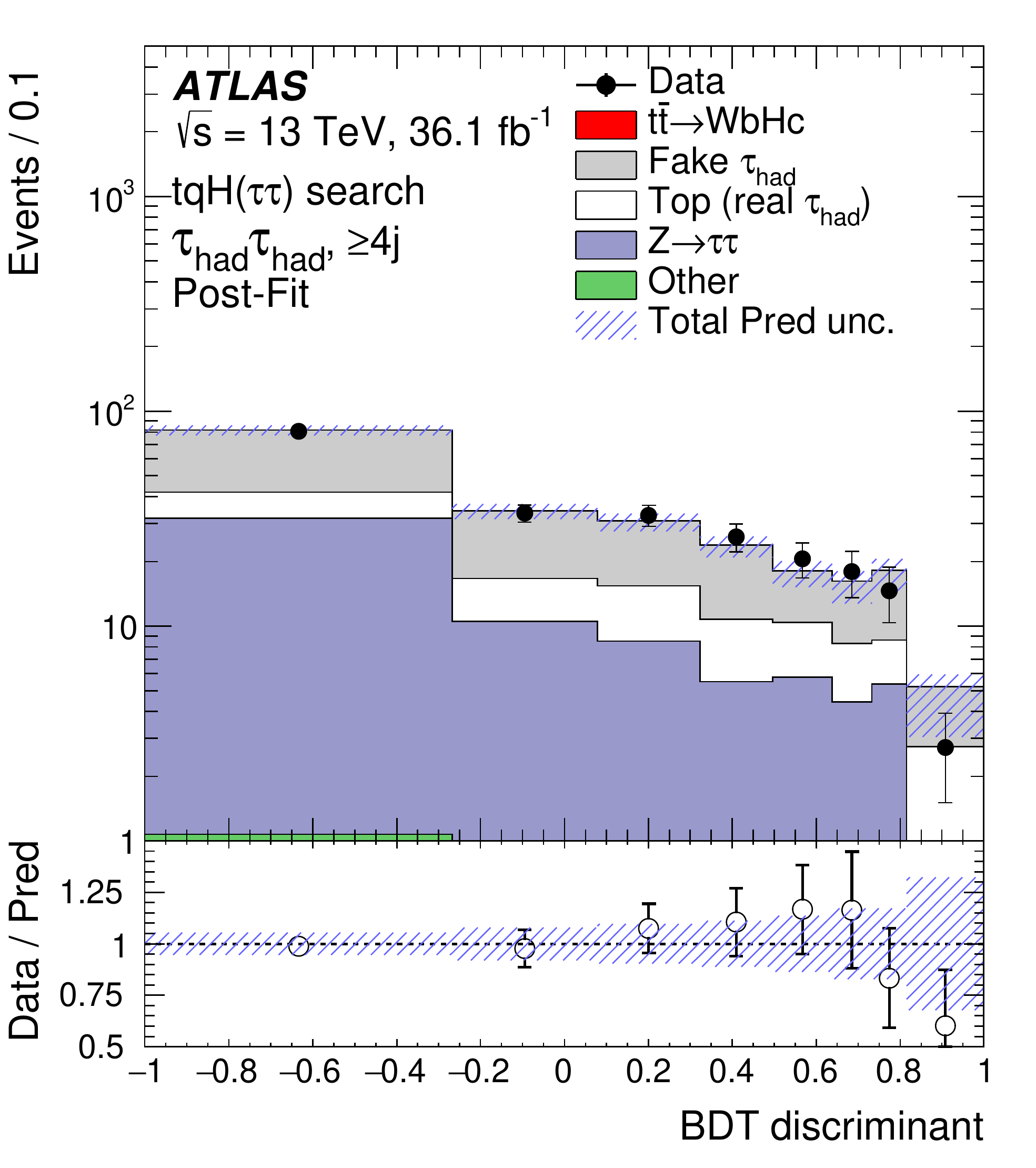}} \\
\caption{\small{$\Htautau$ search: Comparison between the data and prediction for the BDT discriminant distribution in the
$\hadhad$ channel, before and after the fit to data  (``Pre-Fit'' and ``Post-Fit'', respectively) under the signal-plus-background hypothesis.
Shown are the ($\hadhad$, 3j) region (a) pre-fit and (c) post-fit, and the ($\hadhad$, $\geq$4j) region (b) pre-fit and (d) post-fit.
The contributions with real $\had$ candidates from $\ttbar$,  $\ttbar V$, $\ttbar H$, and single-top-quark backgrounds are combined into
a single background source referred to as ``Top (real $\had$)'', whereas the small contributions from
$Z\to \ell^+\ell^-$ ($\ell = e, \mu$) and diboson backgrounds are combined into ``Other''.
In the pre-fit figures the expected $\Hc$ signal (solid red) corresponding to $\BR(t\to Hc)=1\%$ is also shown,
added to the background prediction. In the post-fit figures, the $\Hc$ signal is normalised using the best-fit branching ratio,
$\BR(t\to Hc)=(-4.4^{+9.9}_{-8.5})\times 10^{-4}$.
The bottom panels display the ratios of data to either the SM background prediction before the fit (``Bkg'')  or the total signal-plus-background
prediction after the fit (``Pred'').
The blue triangles indicate points that are outside the vertical range of the figure.
The hashed area represents the total uncertainty of the background.
In the case of the pre-fit background uncertainty, the normalisation uncertainty of the fake $\had$ background is not included.}}
\label{fig:prepostfit_unblinded_WbHc_hh}
\end{center}
\end{figure*}
 
\subsection{Combination of ATLAS searches}
\label{sec:results_combo}
 
The $\Hbb$ and $\Htautau$ searches are combined with the ATLAS searches in diphoton~\cite{Aaboud:2017mfd} and multilepton~\cite{Aaboud:2018pob} final states
of events in the same data set, referred to as ``$\Hgg$ search'' and ``$\HML$ search'', respectively.
Since all searches, with the exception of the $\Hbb$ search, are dominated by the data statistical uncertainty,
and in each search the dominant systematic uncertainties are different, the combined result is
insensitive to the assumed correlations of systematic uncertainties across searches.
Therefore, the only systematic uncertainties taken to be fully correlated among the four searches are
those affecting the integrated luminosity, the $\ttbar$ cross section, signal modelling, a subset of the uncertainties
on the Higgs boson branching ratios (those associated with uncertainties in $\alpha_\mathrm{S}$ and $m_b$),
and a subset of jet-related uncertainties (jet energy resolution and JVT requirement).
The rest of the jet-related uncertainties (jet energy scale and $b$-tagging) are taken as fully correlated among
the $\Hbb$, $\Htautau$, and $\HML$ searches, but uncorrelated with the $\Hgg$ search. The rest of the uncertainties,
e.g.\ those related to leptons and to background modelling, are taken as uncorrelated among the four searches.

\begin{figure*}[t!]
\begin{center}
\includegraphics[width=0.7\textwidth]{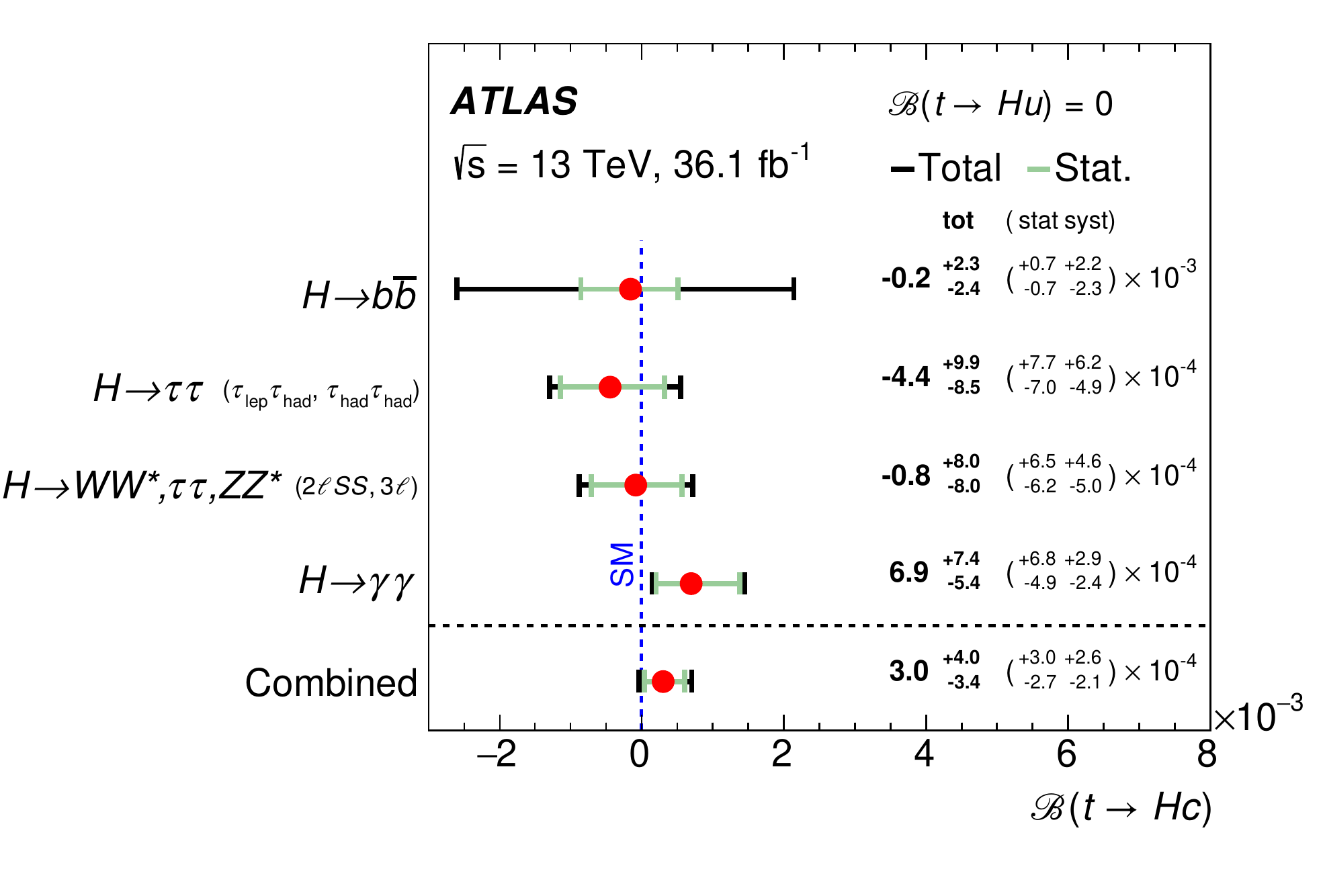}
\caption{\small {Summary of the best-fit $\BR(t\to Hc)$ for the individual searches as well as their combination,
assuming $\BR(t\to Hu)=0$. }}
\label{fig:summary_printnum_hc}
\end{center}
\end{figure*}
\begin{figure*}[h!]
\begin{center}
\includegraphics[width=0.7\textwidth]{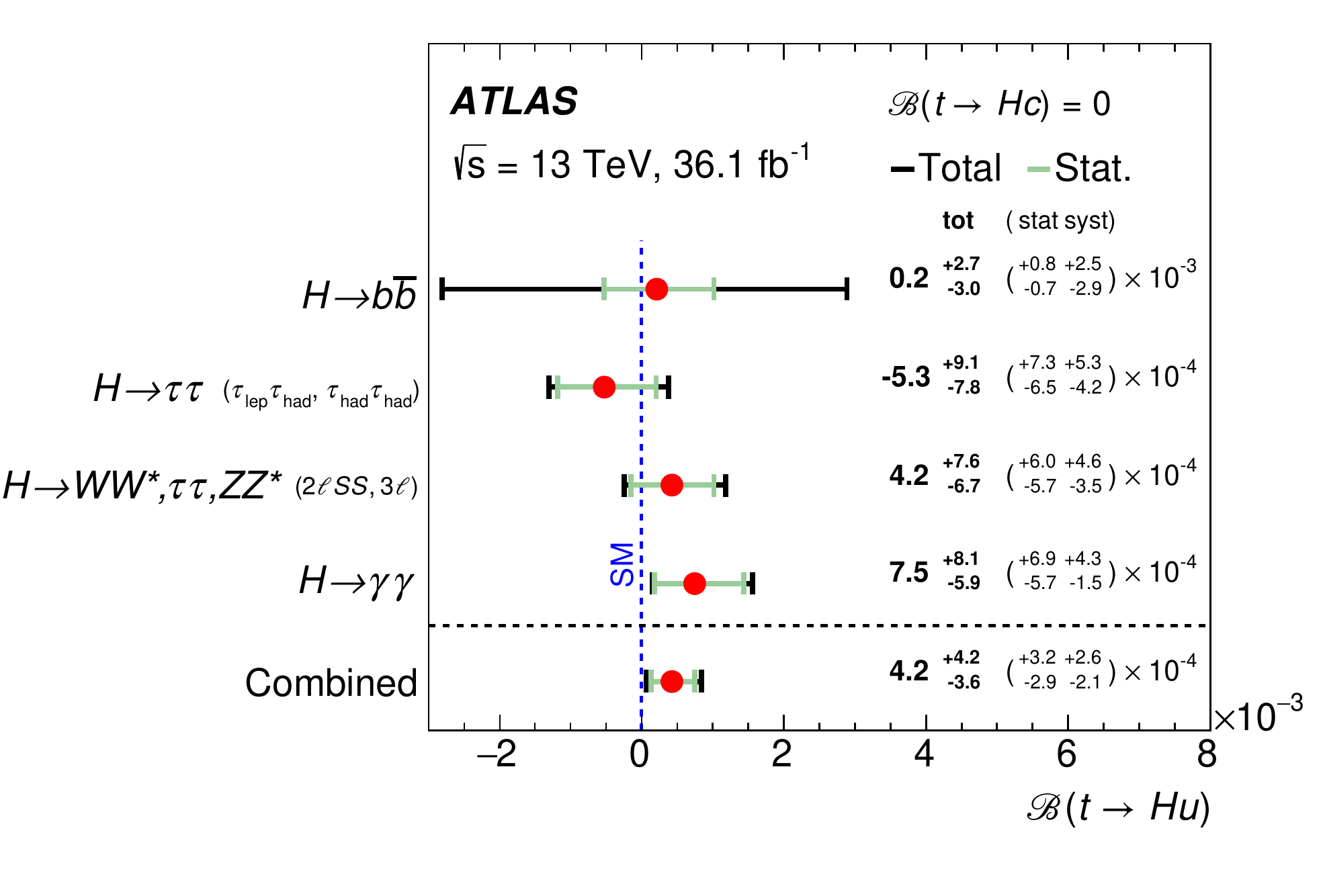}
\caption{\small {Summary of the best-fit $\BR(t\to Hu)$ for the individual searches as well as their combination,
assuming $\BR(t\to Hc)=0$. }}
\label{fig:summary_printnum_hu}
\end{center}
\end{figure*}
 
The first set of combined results is obtained for each branching ratio separately, setting the other branching ratio to zero.
The best-fit combined branching ratios are $\BR(t\to Hc)=[3.0^{+3.0}_{-2.7}\,(\mathrm{stat})^{+2.6}_{-2.1}\,(\mathrm{syst})] \times 10^{-4}$ and
$\BR(t\to Hu)=[4.2^{+3.2}_{-2.9}\,(\mathrm{stat})^{+2.6}_{-2.1}\,(\mathrm{syst})] \times 10^{-4}$.
A comparison of the best-fit branching ratios for the individual searches and their combination is shown in Figure~\ref{fig:summary_printnum_hc}
for $\BR(t\to Hc)$ and Figure~\ref{fig:summary_printnum_hu} for $\BR(t\to Hu)$.
The observed (expected) 95\% CL combined upper limits on the branching ratios are
$\BR(t\to Hc)<1.1 \times 10^{-3}\,(8.3 \times 10^{-4})$ and $\BR(t\to Hu)<1.2 \times 10^{-3}\,(8.3 \times 10^{-4})$.
A summary of the upper limits on the branching ratios obtained by the individual searches, as well as their combination, is given
in Table~\ref{tab:limits_summary} and in Figures~\ref{fig:limits_combo_1D_hc} and~\ref{fig:limits_combo_1D_hu}.

Upper limits on the branching ratios $\BR(t\to Hq)$ ($q=u,c$) can be translated into upper limits on the non-flavour-diagonal Yukawa couplings $\lamHq$
appearing in the Lagrangian~\cite{Harnik:2012pb}:
\begin{equation*}
{\cal L}_\mathrm{FCNC} = -\lambda_{t_\mathrm{L} q_\mathrm{R}} \bar{t}_\mathrm{L} q_\mathrm{R} H - \lambda_{q_\mathrm{L} t_\mathrm{R}} \bar{q}_\mathrm{L} t_\mathrm{R} H  + \mathrm{h.c.}
\end{equation*}
The branching ratio $\BR(t\to Hq)$ is estimated as the ratio of its partial width~\cite{Zhang:2013xya} to the SM $t \to Wb$ partial width~\cite{Denner:1990ns},
which is assumed to be dominant. Both predicted partial widths include next-to-leading-order QCD corrections.
Using the expression derived in Ref.~\cite{Aad:2014dya}, the coupling $|\lamHq|$ can be extracted as $| \lamHq | = (1.92 \pm 0.02) \sqrt{\BR(t\to Hq)}$.
The $\lamHq$ coupling corresponds to the sum in quadrature of the couplings relative to the two possible chirality combinations of the quark fields,
$\lamHq \equiv \sqrt{ |\lambda_{t_\mathrm{L} q_\mathrm{R}}|^2 +   |\lambda_{q_\mathrm{L} t_\mathrm{R}}|^2 }$~\cite{Harnik:2012pb}.
The observed (expected) upper limits on the couplings from the combination of the searches are $|\lamHc|<0.064\,(0.055)$ and $|\lamHu|<0.066\,(0.055)$.
 
\begin{table}[t!]
\caption{\small{Summary of 95\% CL upper limits on $\BR(t \to Hc)$ and $\BR(t \to Hu)$, in each case neglecting the other decay mode. Signatures with two same-charge (three) leptons and no $\had$ candidates are denoted by $2\ell$SS ($3\ell$). }}
\begin{center}
\begin{tabular}{lcc}
\toprule\toprule
& \multicolumn{1}{c}{95\% CL upper limits} & \multicolumn{1}{c}{95\% CL upper limits}  \\
& \multicolumn{1}{c}{on $\BR(t \to Hc)$} & \multicolumn{1}{c}{on $\BR(t \to Hu)$} \\
&  Observed (Expected) & Observed (Expected)  \\
\midrule\midrule
$H \to b\bar{b}$ & $4.2 \times 10^{-3}$ ($4.0 \times 10^{-3}$) & $5.2 \times 10^{-3}$ ($4.9 \times 10^{-3}$) \\
$H \to \tau\tau$ ($\lephad$, $\hadhad$) & $1.9 \times 10^{-3}$ ($2.1 \times 10^{-3}$) & $1.7 \times 10^{-3}$ ($2.0 \times 10^{-3}$) \\
$H \to WW^*, \tau\tau, ZZ^*$ ($2\ell$SS, $3\ell$)~\cite{Aaboud:2018pob}  & $1.6 \times 10^{-3}$ ($1.5 \times 10^{-3}$) & $1.9 \times 10^{-3}$ ($1.5 \times 10^{-3}$) \\
$H \to \gamma\gamma$~\cite{Aaboud:2017mfd} & $2.2 \times 10^{-3}$ ($1.6 \times 10^{-3}$) & $2.4 \times 10^{-3}$ ($1.7 \times 10^{-3}$) \\
\midrule
Combination  & $1.1 \times 10^{-3}$ ($8.3 \times 10^{-4}$) & $1.2 \times 10^{-3}$ ($8.3 \times 10^{-4}$) \\
\bottomrule\bottomrule
\end{tabular}
\label{tab:limits_summary}
\end{center}
\end{table}
 
\begin{figure*}[h!]
\begin{center}
\subfloat[]{\includegraphics[width=0.7\textwidth]{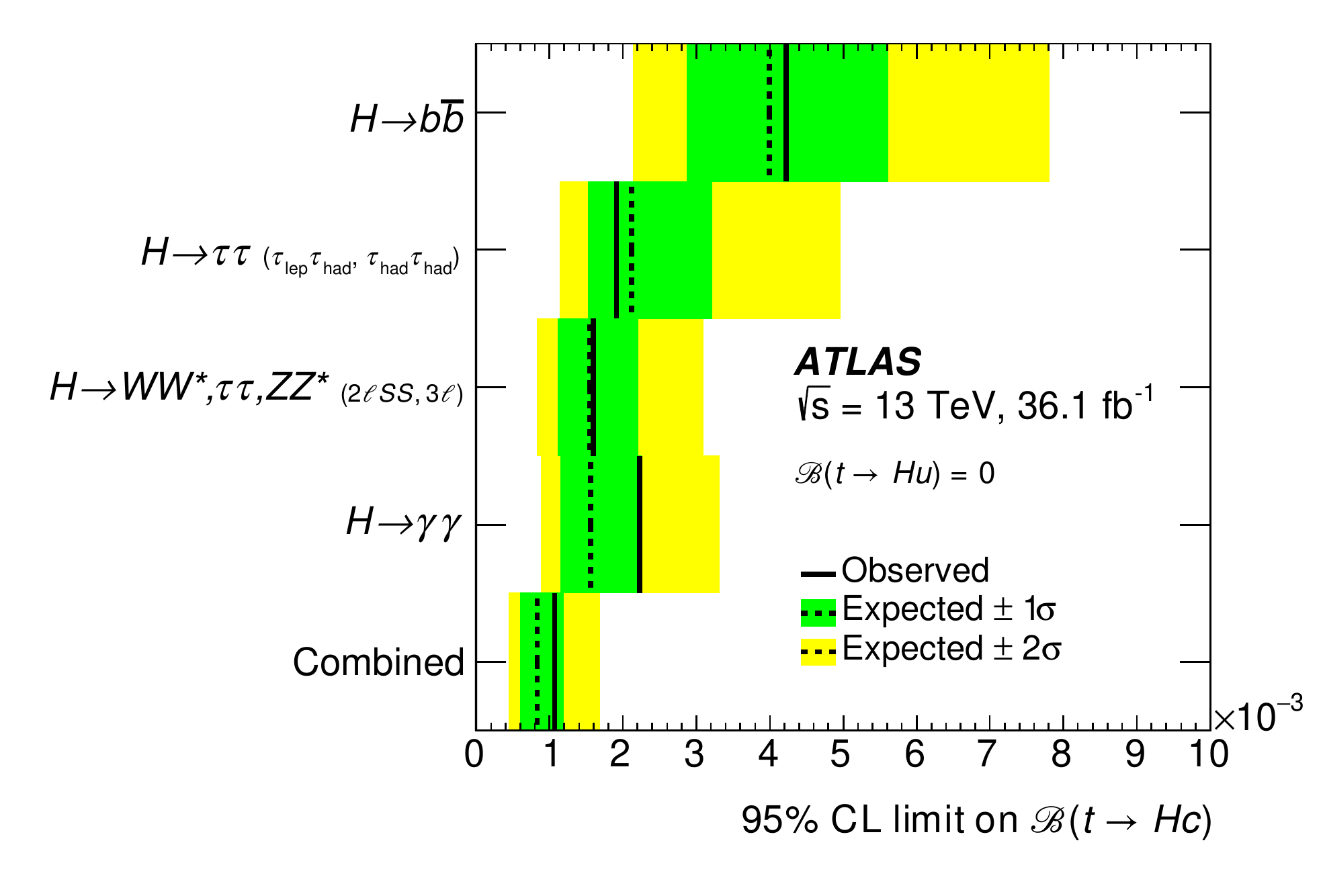}}
\caption{\small {95\% CL upper limits on $\BR(t\to Hc)$ for the individual searches as well as their
combination, assuming $\BR(t\to Hu)=0$. The observed limits (solid lines) are compared with the
expected (median) limits under the background-only
hypothesis (dotted lines). The surrounding shaded bands correspond to the 68\% and 95\% CL intervals around the expected limits,
denoted by $\pm 1\sigma$ and $\pm 2\sigma$, respectively.
}}
\label{fig:limits_combo_1D_hc}
\end{center}
\end{figure*}
 
\begin{figure*}[h!]
\begin{center}
\subfloat[]{\includegraphics[width=0.7\textwidth]{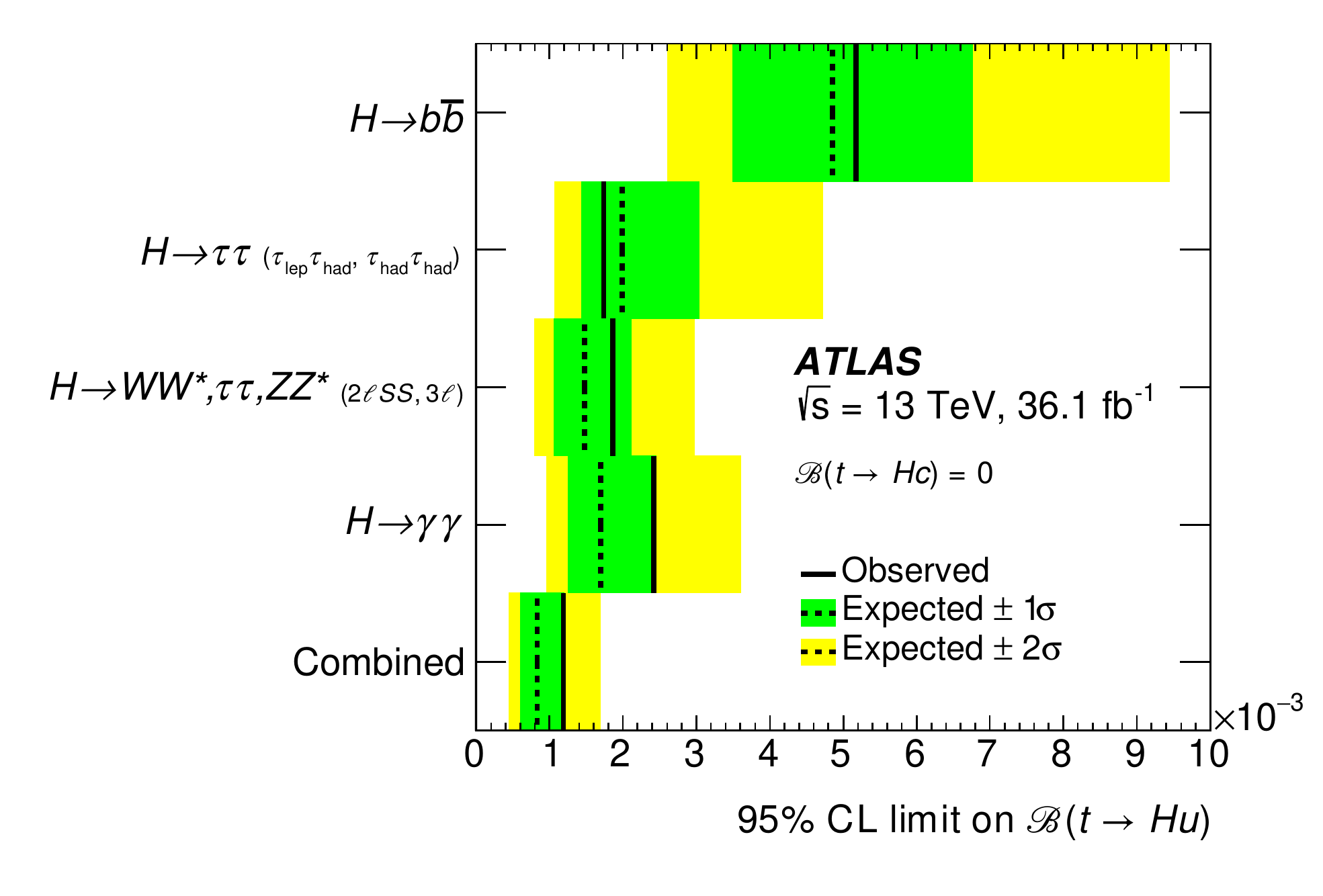}}
\caption{\small {95\% CL upper limits on $\BR(t\to Hu)$ for the individual searches as well as their
combination, assuming $\BR(t\to Hc)=0$. The observed limits (solid lines) are compared with the
expected (median) limits under the background-only
hypothesis (dotted lines). The surrounding shaded bands correspond to the 68\% and 95\% CL intervals around the expected limits,
denoted by $\pm 1\sigma$ and $\pm 2\sigma$, respectively.
}}
\label{fig:limits_combo_1D_hu}
\end{center}
\end{figure*}
 
A similar set of results can be obtained by simultaneously varying both branching ratios in the likelihood function.
Figure~\ref{fig:limits_combo_2D}(a) shows the 95\% CL upper limits on the branching ratios in the $\BR(t\to Hu)$ versus $\BR(t\to Hc)$ plane.
The small differences between the limiting values (on the $x$- and $y$-axes) of the branching ratio limits obtained in the two-dimensional scan and
those reported in Table~\ref{tab:limits_summary}, result from slightly different choices in the $\HML$ search
regarding the final discriminant, which in the two-dimensional case should be common to both signals, and its binning.
The corresponding upper limits on the couplings in the $|\lamHu|$ versus $|\lamHc|$ plane are shown in Figure~\ref{fig:limits_combo_2D}(b).

\begin{figure*}[t!]
\begin{center}
\subfloat[]{\includegraphics[width=0.49\textwidth]{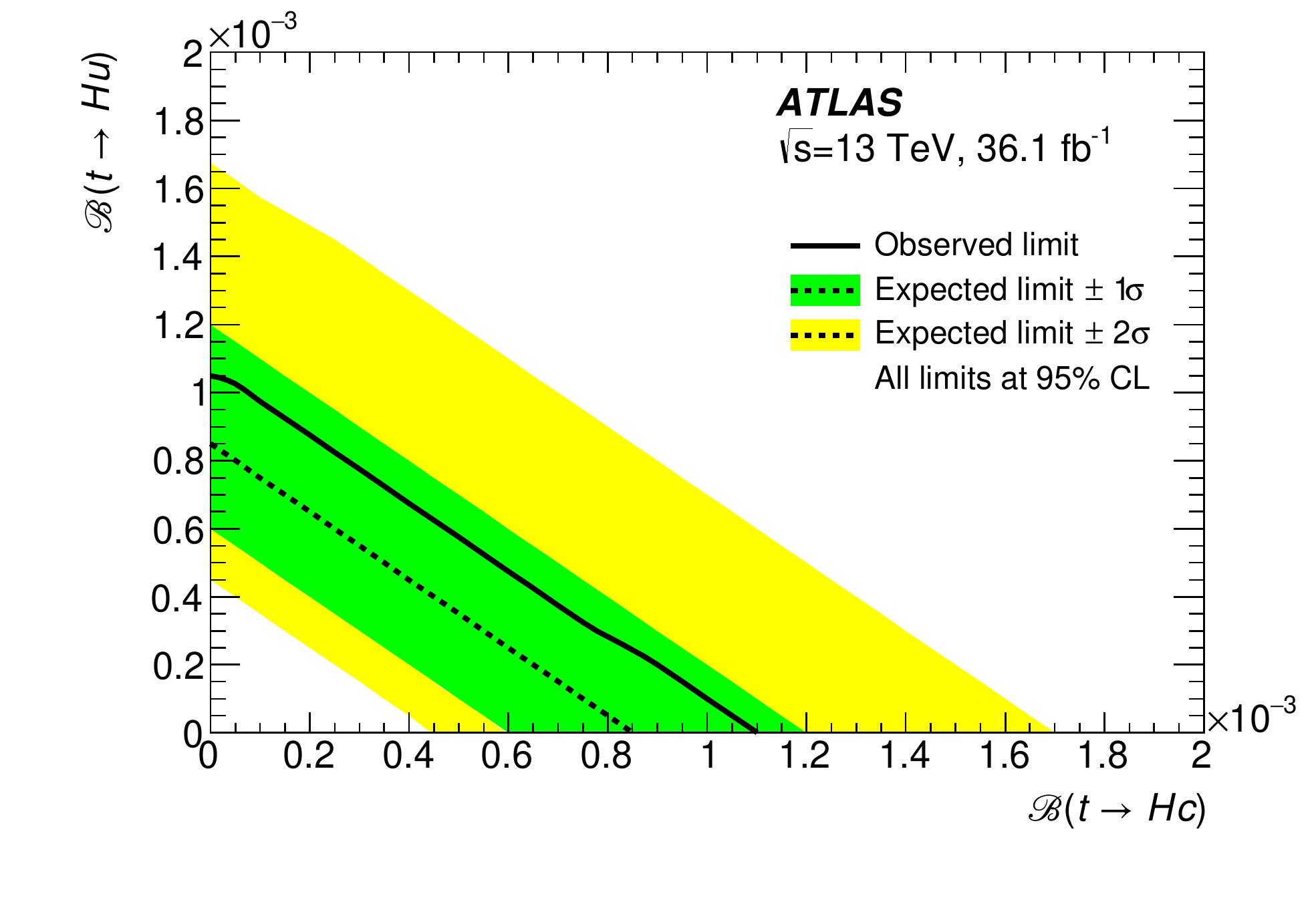}}
\subfloat[]{\includegraphics[width=0.49\textwidth]{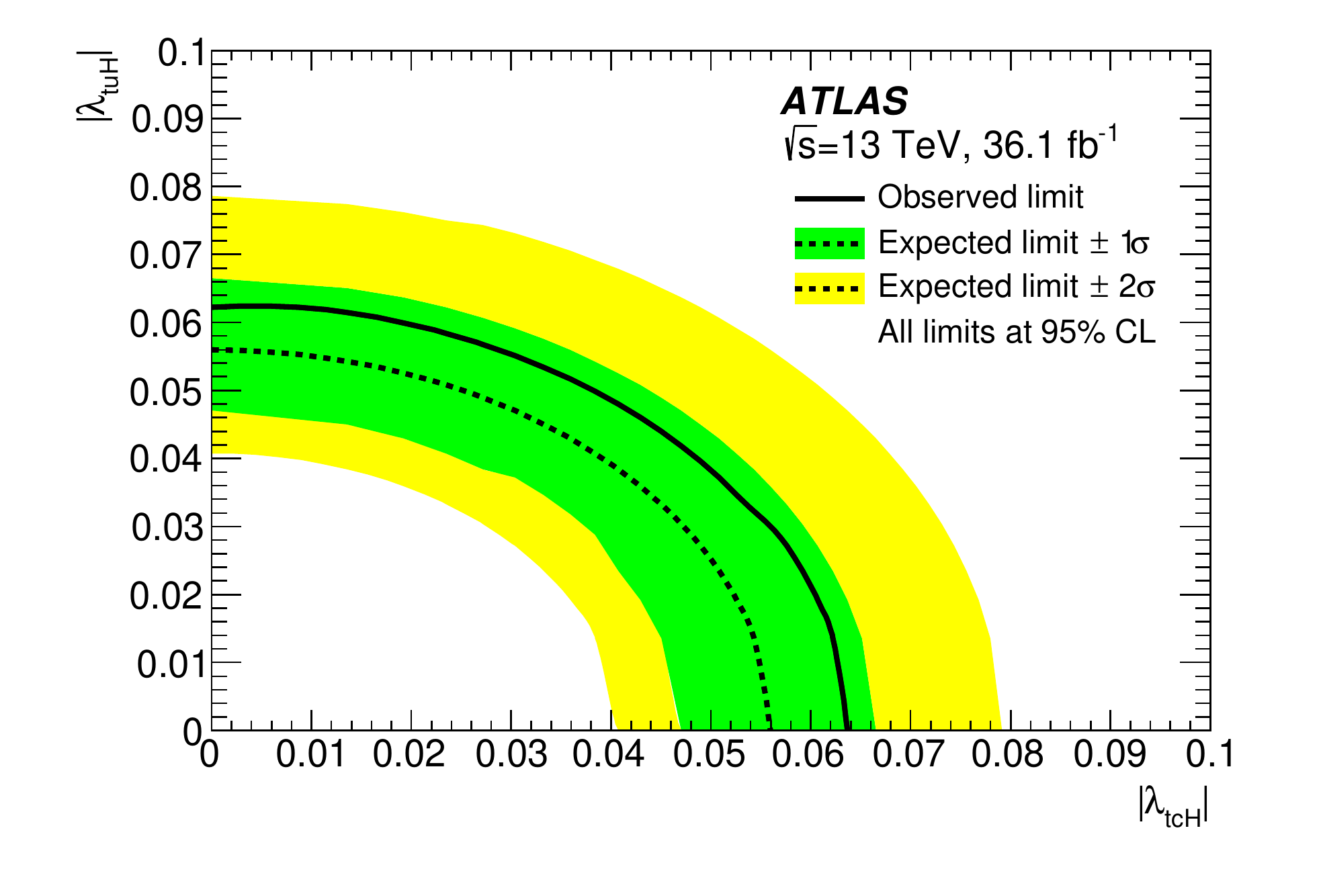}}
\caption{\small {95\% CL upper limits (a) on the plane of $\BR(t\to Hu)$ versus $\BR(t\to Hc)$ and (b) on the plane
of $|\lamHu|$ versus $|\lamHc|$ for the combination of the searches. The observed limits (solid lines) are compared with the expected (median) limits under the background-only hypothesis (dotted lines). The surrounding shaded bands correspond to the 68\% and 95\% CL intervals around the expected limits,
denoted by $\pm 1\sigma$ and $\pm 2\sigma$, respectively.}}
\label{fig:limits_combo_2D}
\end{center}
\end{figure*}
 
 
\FloatBarrier
 
\section{Conclusion}
\label{sec:conclusion}
 
A search for flavour-changing neutral-current decays of a top quark into an up-type quark ($q=u, c$) and the Standard Model Higgs boson, $t\to Hq$, is presented.
The search is based on a dataset of $pp$ collisions at $\sqrt{s}=13~\tev$ recorded in 2015 and 2016 with the ATLAS detector at the
CERN Large Hadron Collider and corresponding to an integrated luminosity of 36.1 fb$^{-1}$.
Two complementary analyses are performed to search for top-quark pair events in which one top quark decays into $Wb$ and the other top quark decays into $Hq$,
and target the $H \to b\bar{b}$ and $H \to \tau^+\tau^-$  decay modes, respectively.
The $\Hbb$ search selects events with one isolated electron or muon from the $W \to \ell\nu$ decay, and multiple jets, with several
of them being identified with high purity as originating from the hadronisation of $b$-quarks.
The $\Htautau$ search selects events with either one or two hadronically decaying $\tau$-lepton candidates, as well as multiple jets.
Both searches employ multivariate techniques to discriminate between the signal and the background on the basis of their different kinematics.
No significant excess of events above the background expectation is found, and 95\% CL upper limits on the $t\to Hq$ branching ratios are derived.
In the case of the $\Hbb$ search, the observed (expected) 95\% CL upper limits on the $t\to Hc$ and $t\to Hu$ branching ratios
are $4.2 \times 10^{-3}\,(4.0 \times 10^{-3})$ and $5.2 \times 10^{-3}\,(4.9 \times 10^{-3})$, respectively.
In the case of the $\Htautau$ search, the observed (expected) 95\% CL upper limits on the $t\to Hc$ and $t\to Hu$ branching ratios
are $1.9 \times 10^{-3}\,(2.1 \times 10^{-3})$ and $1.7 \times 10^{-3}\,(2.0 \times 10^{-3})$, respectively.
The combination of these searches with ATLAS searches in diphoton and multilepton final states
yields observed (expected) 95\% CL upper limits on the $t\to Hc$ and $t\to Hu$ branching ratios of $1.1 \times 10^{-3}$ ($8.3 \times 10^{-4}$)
and $1.2 \times 10^{-3}$ ($8.3 \times 10^{-4}$), assuming $\BR(t\to Hu)=0$ and $\BR(t\to Hc)=0$ respectively.
The corresponding combined observed (expected) upper limits on the $|\lambda_{tcH}|$ and $|\lambda_{tuH}|$ couplings are 0.064 (0.055) and 0.066 (0.055), respectively.
 
\section*{Acknowledgements}
 
We thank CERN for the very successful operation of the LHC, as well as the
support staff from our institutions without whom ATLAS could not be
operated efficiently.
 
We acknowledge the support of ANPCyT, Argentina; YerPhI, Armenia; ARC, Australia; BMWFW and FWF, Austria; ANAS, Azerbaijan; SSTC, Belarus; CNPq and FAPESP, Brazil; NSERC, NRC and CFI, Canada; CERN; CONICYT, Chile; CAS, MOST and NSFC, China; COLCIENCIAS, Colombia; MSMT CR, MPO CR and VSC CR, Czech Republic; DNRF and DNSRC, Denmark; IN2P3-CNRS, CEA-DRF/IRFU, France; SRNSFG, Georgia; BMBF, HGF, and MPG, Germany; GSRT, Greece; RGC, Hong Kong SAR, China; ISF and Benoziyo Center, Israel; INFN, Italy; MEXT and JSPS, Japan; CNRST, Morocco; NWO, Netherlands; RCN, Norway; MNiSW and NCN, Poland; FCT, Portugal; MNE/IFA, Romania; MES of Russia and NRC KI, Russian Federation; JINR; MESTD, Serbia; MSSR, Slovakia; ARRS and MIZ\v{S}, Slovenia; DST/NRF, South Africa; MINECO, Spain; SRC and Wallenberg Foundation, Sweden; SERI, SNSF and Cantons of Bern and Geneva, Switzerland; MOST, Taiwan; TAEK, Turkey; STFC, United Kingdom; DOE and NSF, United States of America. In addition, individual groups and members have received support from BCKDF, CANARIE, CRC and Compute Canada, Canada; COST, ERC, ERDF, Horizon 2020, and Marie Sk{\l}odowska-Curie Actions, European Union; Investissements d' Avenir Labex and Idex, ANR, France; DFG and AvH Foundation, Germany; Herakleitos, Thales and Aristeia programmes co-financed by EU-ESF and the Greek NSRF, Greece; BSF-NSF and GIF, Israel; CERCA Programme Generalitat de Catalunya, Spain; The Royal Society and Leverhulme Trust, United Kingdom.
 
The crucial computing support from all WLCG partners is acknowledged gratefully, in particular from CERN, the ATLAS Tier-1 facilities at TRIUMF (Canada), NDGF (Denmark, Norway, Sweden), CC-IN2P3 (France), KIT/GridKA (Germany), INFN-CNAF (Italy), NL-T1 (Netherlands), PIC (Spain), ASGC (Taiwan), RAL (UK) and BNL (USA), the Tier-2 facilities worldwide and large non-WLCG resource providers. Major contributors of computing resources are listed in Ref.~\cite{ATL-GEN-PUB-2016-002}.
 

\clearpage
\appendix
\part*{Appendix}
\addcontentsline{toc}{part}{Appendix}
\section{Pre-fit and post-fit event yields in the $\Hbb$ search}
\label{sec:prepostfit_yields_Hbb_appendix}
 
Table~\ref{tab:Hbb_Prefit_Yields_Unblind} presents the observed and predicted yields in each of the analysis regions
for the $\Hbb$ search before the fit to data.
Tables~\ref{tab:Hbb_Postfit_Yields_Unblind_Hc} and~\ref{tab:Hbb_Postfit_Yields_Unblind_Hu} present the observed and predicted yields
in each of the analysis regions after the fit to the data under the signal-plus-background hypothesis, assuming
$\Hc$ and $\Hu$ as signal, respectively.
 
\begin{table}[htbp]
\caption{
$\Hbb$ search: Predicted and observed yields in each of the analysis regions considered.
The prediction is shown before the fit to data. Also shown are the signal expectations for
$\Hc$ and $\Hu$ assuming $\BR(t\to Hc)=1\%$ and $\BR(t\to Hu)=1\%$ respectively.
The quoted uncertainties are the sum in quadrature of statistical and systematic uncertainties of the yields,
excluding the normalisation uncertainty of the $\ttbin$ background, which is determined via a likelihood fit to data.
}
\small
\begin{center}
\begin{tabular}{l*{3}{c}}
\hline\hline
& 4j, 2b & 4j, 3b & 4j, 4b \\
\hline
$\Hc$ & $ 1990 \pm 190 $ &   $ 1260 \pm 190 $ &   $ 24.8 \pm 9.5 $ \\
$\Hu$ & $ 1950 \pm 190 $ &   $ 1110 \pm 170 $ &   $ 19 \pm 16 $ \\
\hline
$\ttbar$+light-jets & $ 87000 \pm 11000 $ &   $ 4300 \pm 1200 $ &   $ 10.2 \pm 9.6 $ \\
$\ttcin$ & $ 8300 \pm 4300 $ &   $ 1050 \pm 640 $ &   $ 3.2 \pm 3.3 $ \\
$\ttbin$ & $ 3620 \pm 440 $ &   $ 2900 \pm 580 $ &   $ 95 \pm 33 $ \\
$t\bar{t}V$ & $ 176 \pm 31 $ &   $ 34.8 \pm 6.9 $ &   $ 2.84 \pm 0.74 $ \\
$t\bar{t}H$ & $ 61.7 \pm 9.2 $ &   $ 48.7 \pm 8.3 $ &   $ 5.1 \pm 1.0 $ \\
$W$+jets & $ 5400 \pm 2400 $ &   $ 280 \pm 130 $ &   $ 3.3 \pm 1.8 $ \\
$Z$+jets & $ 2120 \pm 960 $ &   $ 115 \pm 55 $ &   $ 2.4 \pm 1.4 $ \\
Single top & $ 7100 \pm 1300 $ &   $ 400 \pm 120 $ &   $ 7.8 \pm 6.0 $ \\
Diboson & $ 267 \pm 97 $ &   $ 17.2 \pm 6.5 $ &   $ 0.58 \pm 0.27 $ \\
Multijet & $ 7800 \pm 3400 $ &   $ 930 \pm 360 $ &   $ 31 \pm 17 $ \\
\hline
Total background & $ 120000 \pm 15000 $ &   $ 10000 \pm 2000 $ &   $ 162 \pm 44 $ \\
\hline
Data & 120572  & 11275  & 176  \\
\hline\hline
\end{tabular}
\vspace{0.2cm}
 
\begin{tabular}{l*{3}{c}}
\hline\hline
& 5j, 2b & 5j, 3b & 5j, $\geq$4b \\
\hline
$\Hc$ & $ 1260 \pm 240 $ &   $ 1010 \pm 190 $ &   $ 26.2 \pm 8.8 $ \\
$\Hu$ & $ 1160 \pm 240 $ &   $ 930 \pm 160 $ &   $ 23 \pm 12 $ \\
\hline
$\ttbar$+light-jets & $ 41300 \pm 9100 $ &   $ 3200 \pm 900 $ &   $ 13 \pm 11 $ \\
$\ttcin$ & $ 5900 \pm 3100 $ &   $ 1320 \pm 760 $ &   $ 21 \pm 17 $ \\
$\ttbin$ & $ 3040 \pm 250 $ &   $ 4300 \pm 760 $ &   $ 310 \pm 83 $ \\
$t\bar{t}V$  &   $ 175 \pm 29 $ &   $ 67 \pm 12 $ &   $ 9.1 \pm 2.0 $ \\
$t\bar{t}H$  &   $ 81.3 \pm 9.5 $ &   $ 103 \pm 15 $ &   $ 18.4 \pm 3.5 $ \\
$W$+jets  &   $ 2400 \pm 1100 $ &   $ 186 \pm 89 $ &   $ 7.3 \pm 3.9 $ \\
$Z$+jets  &   $ 780 \pm 350 $ &   $ 83 \pm 39 $ &   $ 6.1 \pm 3.8 $ \\
Single top  &   $ 2990 \pm 780 $ &   $ 350 \pm 110 $ &   $ 16.6 \pm 7.6 $ \\
Diboson  &   $ 125 \pm 56 $ &   $ 13.7 \pm 6.3 $ &   $ 0.89 \pm 0.47 $ \\
Multijet  &   $ 3700 \pm 1500 $ &   $ 500 \pm 230 $ &   $ 3.8 \pm 4.9 $ \\
\hline
Total background &  $ 60000 \pm 11000 $ &   $ 10100 \pm 1900 $ &   $ 405 \pm 98 $ \\
\hline
Data & 58557  & 11707  & 466  \\
\hline\hline
\end{tabular}
\vspace{0.2cm}
 
\begin{tabular}{l*{3}{c}}
\hline\hline
& $\geq$ 6j, 2b & $\geq$6j, 3b & $\geq$6j, $\geq$4b \\
\hline
$\Hc$ & $ 760 \pm 250 $ &   $ 690 \pm 210 $ &   $ 60 \pm 60 $ \\
$\Hu$ & $ 680 \pm 240 $ &   $ 570 \pm 180 $ &   $ 36 \pm 40 $ \\
\hline
$\ttbar$+light-jets  &   $ 22900 \pm 8100 $ &   $ 2400 \pm 910 $ &   $ 14 \pm 18 $ \\
$\ttcin$  &   $ 5300 \pm 3000 $ &   $ 1800 \pm 1100 $ &   $ 29 \pm 23 $ \\
$\ttbin$  &   $ 3270 \pm 510 $ &   $ 7300 \pm 1300 $ &   $ 1100 \pm 240 $ \\
$t\bar{t}V$  &   $ 229 \pm 41 $ &   $ 154 \pm 30 $ &   $ 30.8 \pm 6.9 $ \\
$t\bar{t}H$  &   $ 140 \pm 18 $ &   $ 262 \pm 39 $ &   $ 71 \pm 14 $ \\
$W$+jets  &   $ 1360 \pm 630 $ &   $ 200 \pm 100 $ &   $ 15.4 \pm 8.2 $ \\
$Z$+jets  &   $ 410 \pm 200 $ &   $ 63 \pm 32 $ &   $ 5.1 \pm 4.0 $ \\
Single top  &   $ 1510 \pm 560 $ &   $ 360 \pm 160 $ &   $ 34 \pm 20 $ \\
Diboson  &   $ 93 \pm 47 $ &   $ 18.5 \pm 9.6 $ &   $ 2.1 \pm 1.2 $ \\
Multijet  &   $ 1920 \pm 820 $ &   $ 780 \pm 360 $ &   $ 43 \pm 29 $ \\
\hline
Total background &  $ 37100 \pm 9600 $ &   $ 13400 \pm 2600 $ &   $ 1360 \pm 290 $ \\
\hline
Data & 35886  & 14877  & 1335  \\
\hline\hline
\end{tabular}
 
\end{center}
\label{tab:Hbb_Prefit_Yields_Unblind}
\end{table}
 
\begin{table}[htbp]
\caption{
$\Hbb$ search: Predicted and observed yields in each of the analysis regions considered.
The background prediction is shown after the fit to data under the signal-plus-background hypothesis
(assuming $\Hc$ as signal).
The quoted uncertainties are the sum in quadrature of statistical and systematic uncertainties of the yields,
computed taking into account correlations among nuisance parameters and among processes.
}
\small
\begin{center}
\begin{tabular}{l*{3}{c}}
\hline\hline
& 4j, 2b & 4j, 3b & 4j, 4b \\
\hline
$\Hc$  &   $ -30 \pm 470 $ &   $ -20 \pm 300 $ &   $ -0.4 \pm 5.9 $ \\
\hline
$\ttbar$+light-jets  &    $ 82900 \pm 4200 $ &   $ 4900 \pm 500 $ &   $ 16 \pm 12 $ \\
$\ttcin$  &   $ 11400 \pm 4800 $ &   $ 1360 \pm 550 $ &   $ 5.9 \pm 4.2 $ \\
$\ttbin$  &   $ 4270 \pm 590 $ &   $ 3400 \pm 350 $ &   $ 110 \pm 17 $ \\
$t\bar{t}V$  &    $ 174 \pm 28 $ &   $ 35.0 \pm 5.9 $ &   $ 2.69 \pm 0.55 $ \\
$t\bar{t}H$  &   $ 62.6 \pm 7.8 $ &   $ 47.3 \pm 6.3 $ &   $ 4.68 \pm 0.69 $ \\
$W$+jets  &   $ 4800 \pm 1800 $ &   $ 260 \pm 100 $ &   $ 2.9 \pm 1.3 $ \\
$Z$+jets  &    $ 1870 \pm 730 $ &   $ 102 \pm 41 $ &   $ 1.9 \pm 1.0 $ \\
Single-top  &  $ 6360 \pm 980 $ &   $ 393 \pm 96 $ &   $ 7.6 \pm 5.2 $ \\
Diboson  &   $ 242 \pm 84 $ &   $ 16.3 \pm 5.7 $ &   $ 0.50 \pm 0.22 $ \\
Multijet  &   $ 9000 \pm 3500 $ &   $ 820 \pm 240 $ &   $ 29 \pm 16 $ \\
\hline
Total & $ 121100 \pm 2200 $ &   $ 11290 \pm 280 $ &   $ 181 \pm 23 $ \\
\hline
Data & 120572  & 11275  & 176  \\
\hline\hline
\end{tabular}
\vspace{0.2cm}
 
\begin{tabular}{l*{3}{c}}
\hline\hline
& 5j, 2b & 5j, 3b & 5j, $\geq$4b \\
\hline
$\Hc$  &   $ -20 \pm 300 $ &   $ -10 \pm 240 $ &   $ -0.4 \pm 6.2 $ \\
\hline
$\ttbar$+light-jets  &   $ 38000 \pm 3100 $ &   $ 3480 \pm 460 $ &   $ 15.8 \pm 9.5 $ \\
$\ttcin$  &   $ 8300 \pm 3400 $ &   $ 2000 \pm 760 $ &   $ 39 \pm 18 $ \\
$\ttbin$  &   $ 3410 \pm 470 $ &   $ 4900 \pm 460 $ &   $ 356 \pm 29 $ \\
$t\bar{t}V$  &   $ 168 \pm 26 $ &   $ 65 \pm 10 $ &   $ 8.2 \pm 1.4 $ \\
$t\bar{t}H$  &   $ 81.1 \pm 8.9 $ &   $ 99 \pm 12 $ &   $ 16.6 \pm 2.3 $ \\
$W$+jets  &   $ 2080 \pm 820 $ &   $ 169 \pm 68 $ &   $ 6.0 \pm 2.8 $ \\
$Z$+jets  &   $ 700 \pm 270 $ &   $ 74 \pm 30 $ &   $ 5.6 \pm 3.2 $ \\
Single-top  &   $ 2560 \pm 590 $ &   $ 322 \pm 90 $ &   $ 13.3 \pm 5.8 $ \\
Diboson  &   $ 111 \pm 48 $ &   $ 12.5 \pm 5.4 $ &   $ 0.76 \pm 0.39 $ \\
Multijet  &   $ 3380 \pm 950 $ &   $ 560 \pm 230 $ &   $ 3.6 \pm 4.8 $ \\
\hline
Total & $ 58800 \pm 1400 $ &   $ 11690 \pm 360 $ &   $ 465 \pm 29 $ \\
\hline
Data & 58557  & 11707  & 466  \\
\hline\hline
\end{tabular}
\vspace{0.2cm}
 
\begin{tabular}{l*{3}{c}}
\hline\hline
& $\geq$6j, 2b & $\geq$6j, 3b & $\geq$6j, $\geq$4b \\
\hline
$\Hc$  &   $ -10 \pm 180 $ &   $ -10 \pm 160 $ &   $ -1 \pm 14 $ \\
\hline
$\ttbar$+light-jets  &   $ 20100 \pm 2500 $ &   $ 2560 \pm 490 $ &   $ 21 \pm 23 $ \\
$\ttcin$  &   $ 7800 \pm 3300 $ &   $ 3000 \pm 1100 $ &   $ 59 \pm 25 $ \\
$\ttbin$  &   $ 3390 \pm 480 $ &   $ 7510 \pm 760 $ &   $ 1106 \pm 83 $ \\
$t\bar{t}V$  &   $ 213 \pm 34 $ &   $ 145 \pm 24 $ &   $ 27.0 \pm 4.8 $ \\
$t\bar{t}H$  &   $ 134 \pm 15 $ &   $ 240 \pm 30 $ &   $ 61.6 \pm 8.8 $ \\
$W$+jets  &   $ 1200 \pm 470 $ &   $ 183 \pm 75 $ &   $ 12.5 \pm 5.7 $ \\
$Z$+jets  &   $ 350 \pm 150 $ &   $ 56 \pm 24 $ &   $ 3.5 \pm 2.2 $ \\
Single-top  &   $ 1220 \pm 400 $ &   $ 310 \pm 120 $ &   $ 27 \pm 14 $ \\
Diboson  &   $ 82 \pm 40 $ &   $ 16.7 \pm 8.2 $ &   $ 1.70 \pm 0.90 $ \\
Multijet  &   $ 1540 \pm 530 $ &   $ 860 \pm 340 $ &   $ 37 \pm 26 $ \\
\hline
Total &   $ 36000 \pm 1300 $ &   $ 14880 \pm 500 $ &   $ 1360 \pm 72 $ \\
\hline
Data & 35886  & 14877  & 1335  \\
\hline\hline
\end{tabular}
\end{center}
\label{tab:Hbb_Postfit_Yields_Unblind_Hc}
\end{table}
 
\begin{table}[htbp]
\caption{
$\Hbb$ search: Predicted and observed yields in each of the analysis regions considered.
The background prediction is shown after the fit to data under the signal-plus-background hypothesis (assuming $\Hu$ as signal).
The quoted uncertainties are the sum in quadrature of statistical and systematic uncertainties of the yields,
computed taking into account correlations among nuisance parameters and among processes.
}
\small
\begin{center}
\begin{tabular}{l*{3}{c}}
\hline\hline
& 4j, 2b & 4j, 3b & 4j, 4b \\
\hline
$\Hu$  &   $ 40 \pm 550 $ &   $ 20 \pm 320 $ &   $ 0.4 \pm 5.3 $ \\
\hline
$\ttbar$+light-jets  &   $ 82700 \pm 4400 $ &   $ 4860 \pm 530 $ &   $ 15 \pm 12 $ \\
$\ttcin$  &   $ 11500 \pm 5100 $ &   $ 1400 \pm 580 $ &   $ 5.8 \pm 4.2 $ \\
$\ttbin$  &   $ 4260 \pm 590 $ &   $ 3400 \pm 350 $ &   $ 110 \pm 17 $ \\
$t\bar{t}V$  &   $ 173 \pm 28 $ &   $ 34.8 \pm 5.8 $ &   $ 2.68 \pm 0.54 $ \\
$t\bar{t}H$  &   $ 62.4 \pm 7.7 $ &   $ 47.1 \pm 6.2 $ &   $ 4.66 \pm 0.68 $ \\
$W$+jets  &   $ 4800 \pm 1900 $ &   $ 260 \pm 100 $ &   $ 2.9 \pm 1.4 $ \\
$Z$+jets  &   $ 1880 \pm 740 $ &   $ 103 \pm 42 $ &   $ 1.9 \pm 1.0 $ \\
Single-top  &   $ 6380 \pm 990 $ &   $ 392 \pm 96 $ &   $ 7.5 \pm 5.2 $ \\
Diboson  &   $ 243 \pm 85 $ &   $ 16.3 \pm 5.7 $ &   $ 0.50 \pm 0.22 $ \\
Multijet  &   $ 9000 \pm 3500 $ &   $ 810 \pm 240 $ &   $ 29 \pm 16 $ \\
\hline
Total &   $ 121000 \pm 2300 $ &   $ 11290 \pm 290 $ &   $ 181 \pm 23 $ \\
\hline
Data & 120572  & 11275  & 176  \\
\hline\hline
\end{tabular}
\vspace{0.2cm}
 
\begin{tabular}{l*{3}{c}}
\hline\hline
& 5j, 2b & 5j, 3b & 5j, $\geq$4b \\
\hline
$\Hu$  &   $ 20 \pm 330 $ &   $ 20 \pm 270 $ &   $ 0.4 \pm 6.6 $ \\
\hline
$\ttbar$+light-jets  &   $ 37800 \pm 3400 $ &   $ 3450 \pm 500 $ &   $ 15.8 \pm 9.7 $ \\
$\ttcin$  &   $ 8400 \pm 3700 $ &   $ 2000 \pm 800 $ &   $ 39 \pm 19 $ \\
$\ttbin$  &   $ 3400 \pm 470 $ &   $ 4920 \pm 460 $ &   $ 356 \pm 29 $ \\
$t\bar{t}V$  &   $ 168 \pm 26 $ &   $ 65 \pm 10 $ &   $ 8.2 \pm 1.4 $ \\
$t\bar{t}H$  &   $ 81.0 \pm 8.9 $ &   $ 99 \pm 12 $ &   $ 16.6 \pm 2.3 $ \\
$W$+jets  &   $ 2100 \pm 840 $ &   $ 169 \pm 69 $ &   $ 6.0 \pm 2.8 $ \\
$Z$+jets  &   $ 710 \pm 280 $ &   $ 74 \pm 30 $ &   $ 5.5 \pm 3.2 $ \\
Single-top  &   $ 2570 \pm 600 $ &   $ 320 \pm 90 $ &   $ 13.4 \pm 5.8 $ \\
Diboson  &   $ 112 \pm 48 $ &   $ 12.5 \pm 5.5 $ &   $ 0.77 \pm 0.39 $ \\
Multijet  &   $ 3430 \pm 990 $ &   $ 560 \pm 230 $ &   $ 3.6 \pm 4.8 $ \\
\hline
Total &   $ 58800 \pm 1500 $ &   $ 11690 \pm 380 $ &   $ 465 \pm 29 $ \\
\hline
Data & 58557  & 11707  & 466  \\
\hline\hline
\end{tabular}
\vspace{0.2cm}
 
\begin{tabular}{l*{3}{c}}
\hline\hline
& $\geq$6j, 2b & $\geq$6j, 3b & $\geq$6j, $\geq$4b \\
\hline
$\Hu$ &   $ 10 \pm 190 $ &   $ 10 \pm 160 $ &   $ 1 \pm 10 $ \\
\hline
$\ttbar$+light-jets  &   $ 20000 \pm 2700 $ &   $ 2530 \pm 520 $ &   $ 20 \pm 24 $ \\
$\ttcin$  &   $ 7900 \pm 3600 $ &   $ 3000 \pm 1200 $ &   $ 58 \pm 26 $ \\
$\ttbin$  &   $ 3390 \pm 480 $ &   $ 7520 \pm 760 $ &   $ 1106 \pm 83 $ \\
$t\bar{t}V$  &   $ 213 \pm 34 $ &   $ 147 \pm 24 $ &   $ 27.0 \pm 4.8 $ \\
$t\bar{t}H$  &   $ 135 \pm 16 $ &   $ 241 \pm 30 $ &   $ 61.9 \pm 9.0 $ \\
$W$+jets  &   $ 1210 \pm 480 $ &   $ 184 \pm 76 $ &   $ 12.6 \pm 5.8 $ \\
$Z$+jets  &   $ 360 \pm 150 $ &   $ 57 \pm 24 $ &   $ 3.6 \pm 2.2 $ \\
Single-top  &   $ 1240 \pm 400 $ &   $ 320 \pm 120 $ &   $ 27 \pm 14 $ \\
Diboson  &   $ 83 \pm 40 $ &   $ 16.8 \pm 8.3 $ &   $ 1.71 \pm 0.91 $ \\
Multijet  &   $ 1530 \pm 530 $ &   $ 860 \pm 340 $ &   $ 37 \pm 26 $ \\
\hline
Total &   $ 36000 \pm 1400 $ &   $ 14880 \pm 530 $ &   $ 1360 \pm 73 $ \\
\hline
Data & 35886  & 14877  & 1335  \\
\hline\hline
\end{tabular}
\end{center}
\label{tab:Hbb_Postfit_Yields_Unblind_Hu}
\end{table}
\FloatBarrier
\section{Pre-fit and post-fit event yields in the $\Htautau$ search}
\label{sec:prepostfit_yields_Htautau_appendix}
 
Table~\ref{tab:Htautau_Prefit_Yields_Unblind} presents the observed and predicted yields in each of the analysis regions
for the $\Htautau$ search before the fit to data.
Tables~\ref{tab:Htautau_Postfit_Yields_Unblind_Hc} and~\ref{tab:Htautau_Postfit_Yields_Unblind_Hu} present the observed and predicted yields
in each of the analysis regions after the fit to the data under the signal-plus-background hypothesis, assuming
$\Hc$ and $\Hu$ as signal, respectively.
 
\begin{table}[htbp]
\caption{
$\Htautau$ search: Predicted and observed yields in each of the analysis regions considered.
The prediction is shown before the fit to data. Also shown are the signal expectations for
$\Hc$ and $\Hu$ assuming $\BR(t\to Hc)=1\%$ and $\BR(t\to Hu)=1\%$ respectively.
The contributions with real $\had$ candidates from $\ttbar$,  $\ttbar V$, $\ttbar H$, and single-top-quark backgrounds are combined into
a single background source referred to as ``Top (real $\had$)'', whereas the small contributions from
$Z\to \ell^+\ell^-$ ($\ell = e, \mu$) and diboson backgrounds are combined into ``Other''.
The quoted uncertainties are the sum in quadrature of statistical and systematic uncertainties of the yields,
excluding the normalisation uncertainty of the fake $\had$ background, which is determined via a likelihood fit to data.
}
\small
\begin{center}
\begin{tabular}{l*{4}{c}}
\hline\hline
& $\lephad$, 3j & $\lephad$, $\geq$4j & $\hadhad$, 3j &  $\hadhad$, $\geq$4j \\
\hline
$\Hc$  &   $ 89 \pm 14 $ &   $ 226 \pm 43 $ &   $ 46 \pm 14 $ &   $ 122 \pm 32 $ \\
$\Hu$  &   $ 100 \pm 17 $ &   $ 237 \pm 47 $ &   $ 32 \pm 10 $ &   $ 114 \pm 28 $ \\
\hline
Fake $\had$  &   $ 2828 \pm 78 $ &   $ 3200 \pm 100 $ &   $ 710 \pm 110 $ &   $ 500 \pm 62 $ \\
Top (real $\had$)  &   $ 3840 \pm 720 $ &   $ 3160 \pm 890 $ &   $ 113 \pm 72 $ &   $ 117 \pm 35 $ \\
$Z \to \tau\tau$  &   $ 420 \pm 140 $ &   $ 320 \pm 120 $ &   $ 283 \pm 99 $ &   $ 267 \pm 96 $ \\
Other  &   $ 168 \pm 56 $ &   $ 103 \pm 33 $ &   $ 8.9 \pm 2.5 $ &   $ 11.2 \pm 2.5 $ \\
\hline
Total background  &   $ 7260 \pm 730 $ &   $ 6770 \pm 880 $ &   $ 1120 \pm 120 $ &   $ 900 \pm 120 $ \\
\hline
Data  & 7259  & 6768  & 1119  & 894  \\
\hline\hline
\end{tabular}
 
\end{center}
\label{tab:Htautau_Prefit_Yields_Unblind}
\end{table}
 
\begin{table}[htbp]
\caption{
$\Htautau$ search: Predicted and observed yields in each of the analysis regions considered.
The background prediction is shown after the fit to data under the signal-plus-background hypothesis
(assuming $\Hc$ as signal).
The contributions with real $\had$ candidates from $\ttbar$,  $\ttbar V$, $\ttbar H$, and single-top-quark backgrounds are combined into
a single background source referred to as ``Top (real $\had$)'', whereas the small contributions from
$Z\to \ell^+\ell^-$ ($\ell = e, \mu$) and diboson backgrounds are combined into ``Other''.
The quoted uncertainties are the sum in quadrature of statistical and systematic uncertainties of the yields,
computed taking into account correlations among nuisance parameters and among processes.
}
\small
\begin{center}
\begin{tabular}{l*{4}{c}}
\hline\hline
& $\lephad$, 3j & $\lephad$, $\geq$4j & $\hadhad$, 3j &  $\hadhad$, $\geq$4j \\
\hline
$\Hc$  &   $ -4.2 \pm 8.2 $ &   $ -11 \pm 21 $ &   $ -2.4 \pm 4.3 $ &   $ -10 \pm 11 $ \\
\hline
Fake $\had$  &   $ 2290 \pm 680 $ &   $ 2640 \pm 880 $ &   $ 640 \pm 110 $ &   $ 440 \pm 100 $ \\
Top (real $\had$)  &   $ 4300 \pm 670 $ &   $ 3660 \pm 860 $ &   $ 147 \pm 84 $ &   $ 139 \pm 35 $ \\
$Z \to \tau\tau$  &   $ 500 \pm 100 $ &   $ 359 \pm 90 $ &   $ 320 \pm 79 $ &   $ 306 \pm 76 $ \\
Other  &   $ 178 \pm 45 $ &   $ 112 \pm 28 $ &   $ 9.6 \pm 2.6 $ &   $ 12.5 \pm 2.6 $ \\
\hline
Total  &   $ 7230 \pm 160 $ &   $ 6760 \pm 170 $ &   $ 1117 \pm 65 $ &   $ 893 \pm 45 $ \\
\hline
Data & 7259  & 6768  & 1119  & 894 \\
\hline\hline
\end{tabular}
\end{center}
\label{tab:Htautau_Postfit_Yields_Unblind_Hc}
\end{table}
 
\begin{table}[htbp]
\caption{
$\Htautau$ search: Predicted and observed yields in each of the analysis regions considered.
The background prediction is shown after the fit to data under the signal-plus-background hypothesis
(assuming $\Hu$ as signal).
The contributions with real $\had$ candidates from $\ttbar$,  $\ttbar V$, $\ttbar H$, and single-top-quark backgrounds are combined into
a single background source referred to as ``Top (real $\had$)'', whereas the small contributions from
$Z\to \ell^+\ell^-$ ($\ell = e, \mu$) and diboson backgrounds are combined into ``Other''.
The quoted uncertainties are the sum in quadrature of statistical and systematic uncertainties of the yields,
computed taking into account correlations among nuisance parameters and among processes.
}
\small
\begin{center}
\begin{tabular}{l*{4}{c}}
\hline\hline
& $\lephad$, 3j & $\lephad$, $\geq$4j & $\hadhad$, 3j &  $\hadhad$, $\geq$4j \\
\hline
$\Hu$  &   $ -5.7 \pm 8.6 $ &   $ -14 \pm 21 $ &   $ -2,0 \pm 2.8 $ &   $ -7.1 \pm 9.8 $ \\
\hline
Fake $\had$  &   $ 2270 \pm 680 $ &   $ 2620 \pm 880 $ &   $ 640 \pm 110 $ &   $ 440 \pm 100 $ \\
Top (real $\had$)  &   $ 4320 \pm 660 $ &   $ 3680 \pm 860 $ &   $ 148 \pm 84 $ &   $ 140 \pm 35 $ \\
$Z \to \tau\tau$  &   $ 470 \pm 100 $ &   $ 359 \pm 89 $ &   $ 321 \pm 79 $ &   $ 308 \pm 77 $ \\
Other  &   $ 177 \pm 44 $ &   $ 111 \pm 27 $ &   $ 9.7 \pm 2.6 $ &   $ 12.5 \pm 2.6 $ \\
\hline
Total  &   $ 7230 \pm 160 $ &   $ 6760 \pm 160 $ &   $ 1118 \pm 66 $ &   $ 892 \pm 45 $ \\
\hline
Data  & 7259  & 6768  & 1119  & 894  \\
\hline\hline
\end{tabular}
\end{center}
\label{tab:Htautau_Postfit_Yields_Unblind_Hu}
\end{table}
\FloatBarrier
\clearpage
 
\printbibliography

\clearpage 
 
\begin{flushleft}
{\Large The ATLAS Collaboration}

\bigskip

M.~Aaboud$^\textrm{\scriptsize 34d}$,    
G.~Aad$^\textrm{\scriptsize 99}$,    
B.~Abbott$^\textrm{\scriptsize 125}$,    
D.C.~Abbott$^\textrm{\scriptsize 100}$,    
O.~Abdinov$^\textrm{\scriptsize 13,*}$,    
B.~Abeloos$^\textrm{\scriptsize 129}$,    
D.K.~Abhayasinghe$^\textrm{\scriptsize 91}$,    
S.H.~Abidi$^\textrm{\scriptsize 164}$,    
O.S.~AbouZeid$^\textrm{\scriptsize 39}$,    
N.L.~Abraham$^\textrm{\scriptsize 153}$
% ...
\bigskip
\\

% ...
$^{13}$Institute of Physics, Azerbaijan Academy of Sciences, Baku; Azerbaijan.\\
% ...
$^{34}$$^{(a)}$Facult\'e des Sciences Ain Chock, R\'eseau Universitaire de Physique des Hautes Energies - Universit\'e Hassan II, Casablanca;$^{(b)}$Centre National de l'Energie des Sciences Techniques Nucleaires (CNESTEN), Rabat;$^{(c)}$Facult\'e des Sciences Semlalia, Universit\'e Cadi Ayyad, LPHEA-Marrakech;$^{(d)}$Facult\'e des Sciences, Universit\'e Mohamed Premier and LPTPM, Oujda;$^{(e)}$Facult\'e des sciences, Universit\'e Mohammed V, Rabat; Morocco.\\
% ...
$^{39}$Niels Bohr Institute, University of Copenhagen, Copenhagen; Denmark.\\
% ...
$^{91}$Department of Physics, Royal Holloway University of London, Egham; United Kingdom.\\
% ...
$^{99}$CPPM, Aix-Marseille Universit\'e, CNRS/IN2P3, Marseille; France.\\
$^{100}$Department of Physics, University of Massachusetts, Amherst MA; United States of America.\\
% ...
$^{125}$Homer L. Dodge Department of Physics and Astronomy, University of Oklahoma, Norman OK; United States of America.\\
% ...
$^{129}$LAL, Universit\'e Paris-Sud, CNRS/IN2P3, Universit\'e Paris-Saclay, Orsay; France.\\
% ...
$^{153}$Department of Physics and Astronomy, University of Sussex, Brighton; United Kingdom.\\
% ...
$^{164}$Department of Physics, University of Toronto, Toronto ON; Canada.\\
% ...
$^{a}$ Also at Borough of Manhattan Community College, City University of New York, NY; United States of America.\\
$^{b}$ Also at California State University, East Bay; United States of America.\\
$^{c}$ Also at Centre for High Performance Computing, CSIR Campus, Rosebank, Cape Town; South Africa.\\
$^{d}$ Also at CERN, Geneva; Switzerland.\\
$^{e}$ Also at CPPM, Aix-Marseille Universit\'e, CNRS/IN2P3, Marseille; France.\\
% ...
$^{*}$ Deceased

\end{flushleft}


\end{document}